\documentclass[acmsmall]{acmart}

\AtBeginDocument{%
  \providecommand\BibTeX{{%
    \normalfont B\kern-0.5em{\scshape i\kern-0.25em b}\kern-0.8em\TeX}}}

\setcopyright{acmcopyright}
\copyrightyear{2023}
\acmYear{2023}
\acmDOI{XXXXXXX.XXXXXXX}

\usepackage{tikz}
\usepackage{makecell}
\usepackage{gensymb}
\newcounter{num}

\newcommand{\find}[1]{ \begin{tcolorbox}
\textbf{Finding \refstepcounter{num}\thenum}: #1
\end{tcolorbox}}

\newcommand{\tool}[1]{\textsc{TenFuzz}}

\usepackage{cleveref}
\usepackage{soul}
\usepackage{framed}
\usepackage{graphicx}
\usepackage{subfigure}
\usepackage{multirow}
\usepackage{url}
\usepackage{tcolorbox}
\usepackage{diagbox}
\usepackage{tikz,listings,caption} 
\usepackage{adjustbox}
\usepackage{threeparttable}
\usepackage{microtype}
\usepackage{enumitem}
\usepackage{subfigure}
\usepackage{graphbox}

\acmJournal{JACM}
\acmVolume{37}
\acmNumber{4}
\acmArticle{111}
\acmMonth{8}

\definecolor{dkgreen}{rgb}{0,0.6,0}
\definecolor{gray}{rgb}{0.5,0.5,0.5}
\definecolor{mauve}{rgb}{0.58,0,0.82}

\lstdefinestyle{Code}{language=Python,
    frame=tb,
    language=Java,
    aboveskip=3mm,
    belowskip=3mm,
    showstringspaces=false,
    columns=flexible,
    basicstyle={\small\ttfamily},
    numbers=none,
    numberstyle=\tiny\color{gray},
    keywordstyle=\color{blue},
    commentstyle=\color{dkgreen},
    stringstyle=\color{mauve},
    breaklines=true,
    breakatwhitespace=true,
    tabsize=3
}

\lstdefinestyle{Python}{
    language = Python,
    basicstyle = \scriptsize\ttfamily,          
    rulesepcolor= \color{gray},            
    breaklines = true,                
    numbers = left,
    numbersep=-6pt,
    numberstyle = \scriptsize\color{gray}, 
    keywordstyle = \color{blue},            
    commentstyle =\color{green!100},       
    stringstyle = \color{red!100},          
    frame = lines,                  
    frameround = tttt,
    morekeywords={with, self},
    showspaces = false,                 
    columns = fixed,                    
    morekeywords = {as},               
    deletendkeywords = {compile},
    escapeinside={<@}{@>}
}

\lstdefinestyle{customc}{
	belowcaptionskip=1\baselineskip,
	breaklines=true,
	frame=l,
	xleftmargin=\parindent,
	language=C,
	showstringspaces=false,
	basicstyle=\footnotesize\ttfamily,
	keywordstyle=\bfseries\color{green!40!black},
	commentstyle=\itshape\color{purple!40!black},
	identifierstyle=\color{blue},
	stringstyle=\color{orange},
	tabsize=2,
}

\usepackage[normalem]{ulem}
\newcommand{\del}[1]{{}}
\newcommand{\add}[1]{{{#1}}}
\newcommand{\ddel}[1]{{}}
\newcommand{\aadd}[1]{{{#1}}}

\begin{document}

\title{Toward Understanding Deep Learning Framework Bugs}

\author{Junjie Chen}
\affiliation{%
    \department{College of Intelligence and Computing}
  \institution{Tianjin University}
  \city{Tianjin}
  \country{China}
  \postcode{300350}
}
\email{junjiechen@tju.edu.cn}

\author{Yihua Liang}
\authornotemark[1]
\affiliation{%
    \department{College of Intelligence and Computing}
  \institution{Tianjin University}
  \city{Tianjin}
  \country{China}
  \postcode{300350}
}
\email{liangyihua@tju.edu.cn}

\author{Qingchao Shen}
\authornote{Equal contribution.}
\affiliation{%
    \department{School of New Media and Communication}
  \institution{Tianjin University}
  \city{Tianjin}
  \country{China}
  \postcode{300350}
}
\email{qingchao@tju.edu.cn}

\author{Jiajun Jiang}
\authornote{Corresponding author.}
\affiliation{%
    \department{College of Intelligence and Computing}
  \institution{Tianjin University}
  \city{Tianjin}
  \country{China}
}
\email{jiangjiajun@tju.edu.cn}

\author{Shuochuan Li}
\affiliation{%
    \department{College of Intelligence and Computing}
  \institution{Tianjin University}
  \city{Tianjin}
  \country{China}
  \postcode{300350}
}
\email{lishuochuan@tju.edu.cn}

\renewcommand{\shortauthors}{Chen et al.}

\begin{abstract}
DL frameworks are the basis of constructing all DL programs and models, and thus their bugs could lead to the unexpected behaviors of any DL program or model relying on them.
Such \add{a} wide effect demonstrates the necessity and importance of guaranteeing DL frameworks' quality.
Understanding the characteristics of DL framework bugs is a fundamental step for this quality assurance task, facilitating \del{to design} \add{designing} effective bug detection and debugging approaches.
Hence, in this work we conduct the most large-scale study on 1,000 bugs from four popular and diverse DL frameworks (i.e., TensorFlow, PyTorch, MXNet, and DL4J).
By analyzing the root causes and symptoms of DL framework bugs associated with 5 components decomposed from DL frameworks, as well as measuring test coverage achieved by three state-of-the-art testing techniques,
we obtain \ddel{13}\aadd{12} major findings for the comprehensive understanding of DL framework bugs and the current status of existing DL framework testing practice, and then provide a series of actionable guidelines for better DL framework bug detection and debugging. Finally, based on the guidelines\add{,} we design and implement a prototype DL-framework testing tool, called \tool{}, which is evaluated to be effective and finds 3 unknown bugs on the latest TensorFlow framework in a preliminary study, indicating the significance of our guidelines.
\end{abstract}



\setcopyright{acmlicensed}
\acmJournal{TOSEM}
\acmYear{2023} \acmVolume{32} \acmNumber{6} \acmArticle{135} \acmMonth{11} \acmPrice{15.00}\acmDOI{10.1145/3587155}

\maketitle

\section{Introduction}
\label{sec:intro}

In recent years, Deep Learning (DL) systems have become one of the most popular types of software systems and have been widely used in many domains, such as autonomous driving~\cite{chen2015deepdriving}, 
aircraft collision avoidance~\cite{julian2016policy}, and software engineering~\cite{ferreira2019software,tian2022learning,yang2021semi,kang2021apirecx}.
However, like traditional software, DL systems also contain bugs, which could lead to huge economic losses or even threaten human lives.
For example, in 2018, an Uber autonomous \del{driving} car killed a pedestrian in Arizona~\cite{news1} and a Tesla Model S in autopilot mode crashed into a fire truck parked with light flashing on a California freeway~\cite{news2}.
Therefore, guaranteeing the quality of DL systems is critical.

A DL system typically involves three levels~\cite{yan2021exposing}: the production level (i.e., DL models), program level (i.e., DL programs used for training DL models), and framework level (i.e., DL frameworks, also called DL libraries in some existing work~\cite{wang2020deep}, used by developers for implementing DL programs).
Bugs \del{in} \add{at} any level could affect the overall quality of the DL system.
Hence, it is necessary to ensure DL systems' quality at all these levels.
Over the years, a lot of research has focused on the production level by designing various DL model testing metrics~\cite{DBLP:conf/icse/SADL,DBLP:conf/kbse/deepgauge,DBLP:journals/corr/abs-1806-07723}, proposing various adversarial input generation methods~\cite{DBLP:journals/corr/GoodfellowSS14,DBLP:conf/iclr/KurakinGB17a,shen2022natural,you2023DRFuzz}, or prioritizing/selecting test inputs for improving DL model testing~\cite{wang2021prioritizing,chen2020practical, DBLP:conf/kbse/ZhangWJYC22}, as well as the program level by studying the characteristics of DL program bugs~\cite{zhang2018empirical,islam2019comprehensive,humbatova2020taxonomy,wang2022empirical} or designing bug detection and diagnosis methods~\cite{yan2021exposing,DBLP:conf/icse/ZhangZMS21,wardat2021deeplocalize}.
However, there is less attention on the framework level.
Actually, DL frameworks are the basis of constructing all DL programs and models, and thus their bugs could produce much wider \del{effect} \add{effects} than the bugs in a specific DL program or model.
Therefore, it is very essential to put more effort in ensuring the quality of DL frameworks, and this work does focus on the framework level.

Indeed, DL frameworks' quality has begun to receive attention recently, and some DL framework testing techniques have been proposed~\cite{pham2019cradle,wang2020deep,guo2020audee,zhang2021predoo}.
Although they have been demonstrated to be effective to detect some new bugs in their experiments, they tend to treat the DL framework under test as a black box and lack \del{the}\add{a} comprehensive understanding of the DL framework bug characteristics (such as root causes and bug distribution).
Such a lack could limit their performance and hinder the design of more effective bug detection techniques.
Moreover, it could limit the development of DL framework bug diagnosis techniques since this kind of tasks require much more sufficient understanding of detected bugs.
That is, understanding the characteristics of DL framework bugs comprehensively is the fundamental task in the area of DL framework quality assurance, which is also the goal of our work.

In the literature, some studies on investigating DL bug characteristics have been conducted~\cite{zhang2018empirical,islam2019comprehensive,humbatova2020taxonomy}, but almost all of them target DL program bugs rather than DL framework bugs.
Due to the significant differences between DL programs and DL frameworks, their bug characteristics are different.
Specifically, a DL program is to invoke the APIs provided by a DL framework for implementing the desired neural network structure, and thus DL program bugs actually refer to those caused by the incorrect usage of the DL framework rather than the bugs inside the DL framework code (which are DL framework bugs).
Regarding DL framework bug characteristics, Jia et al.~\cite{jia2020empirical} made the only one attempt till now, but it is still not enough to comprehensively understand bugs in the family of DL frameworks due to its small scale and limited study points (e.g., studying only one DL framework from three aspects).
More details on the differences between our work and these existing studies can be found in Section~\ref{sec:related_work}.
Hence, in this work \textbf{we conduct a comprehensive study to facilitate the sufficient understanding of DL framework bugs}.

Specifically, we used four popular DL frameworks (in terms of the number of forks in their GitHub repositories) in the study, including TensorFlow~\cite{tensorflow} from Google, PyTorch~\cite{pytorch} from Facebook, MXNet~\cite{mxnet} from Apache, and Deeplearning4j (DL4J)~\cite{dl4j} from Eclipse, as the experimental subjects.
In particular, they have great diversity, e.g., involving both static and dynamic computational graphs, various programming languages for implementations, and different development organizations, which facilitates the generalizability of our conclusions.
In total, we studied 1,000 real bugs collected from their bug repositories and manually labeled them according to a systematic process (to be presented in Section~\ref{sec:methodology}).
To our best knowledge, our study is the most large-scale one for investigating DL framework bugs.
Based on the 1,000 bugs from the four DL frameworks, our study aims to address the following five research questions:

\begin{itemize} 

\item \textbf{RQ1: What are the root causes of DL framework bugs and their distribution?}
The root causes are helpful to understand the nature of DL framework bugs, which facilitates the detection, localization, and fixing of bugs.
Also, it is interesting to investigate the root causes specific to DL framework bugs and explore whether the conclusions on prevalent root causes between DL framework bugs and other software bugs are consistent or not.

\item \textbf{RQ2: What are the symptoms of DL framework bugs and their distribution?}
The symptoms are helpful to understand the consequences of DL framework bugs, which facilitates to triage them and assess their impacts.
Also, we analyzed \del{in} \add{at} which stages of the DL pipeline we can observe these symptoms.
The results can guide the improvement of test oracles for more effective testing of DL frameworks.

\item \textbf{RQ3: What is the relationship between root causes and symptoms of DL framework bugs?}
After investigating the root causes and symptoms of DL framework bugs individually, it can obtain more comprehensive information about the bugs by associating them to study which root cause is more likely to produce a specific bug symptom.


\item \textbf{RQ4: Which levels in DL frameworks are more fragile to bugs?}
In general, a DL framework consists of five levels (to be introduced in Section~\ref{sec:background}) and the fragility of different levels may be different for different kinds of DL framework bugs.
Identifying bug-prone levels for each kind of bugs can make the testing and debugging practice more targeted.

\item \textbf{RQ5: Do the bugs of different DL frameworks have commonality?}
We investigated whether there is some relationship among the bugs of different DL frameworks.
It is helpful to guide the design of more general testing and debugging techniques for DL frameworks.
Also, it may improve the testing and debugging practice of a DL framework by drawing the experience from other DL frameworks. 

\end{itemize}


In our study, we decomposed a DL framework into 5 levels, and identified 13 root causes and 6 symptoms of DL framework bugs through systematic manual analysis.
By studying each aspect individually and associating different aspects together, we obtained \ddel{11} \aadd{10} major findings. 
Inspired by these findings, we further conducted a preliminary experiment to investigate the state-of-the-art DL framework testing techniques in terms of test coverage on each level of DL frameworks, and obtained additional 2 findings about the current status of existing approaches. Based on the empirical results and all the findings, we further provided a series of actionable guidelines for future DL framework testing and debugging. Furthermore, to evaluate the usefulness of our guidelines, we have designed and developed a prototype DL-framework testing tool, called \tool{}, and evaluated its effectiveness on the latest version of \add{the} TensorFlow framework. The results showed that it successfully detected 6 bugs, \del{in} \add{of} which 3 bugs are previously unknown ones and have been confirmed by the maintainers, demonstrating the significance of our guidelines for future research.


To sum up, our work makes the following major contributions:

\begin{itemize} 
    \item We conduct the most large-scale comprehensive study on DL framework bugs based on 1,000 real bugs from four popular and diverse DL frameworks.
    
    \item We provide a classification of root causes and symptoms of DL framework bugs, and associate them with each other as well as each level of DL frameworks.
    
    \item We conduct a preliminary experiment to investigate the current status of the state-of-the-art DL framework testing techniques, further confirming the necessity of developing more effective testing approaches.
    
    \item We provide a series of actionable guidelines for future DL framework testing and debugging practice according to our findings.
    
    \item We design and implement a prototype DL-framework testing tool and conduct a preliminary study with it. The result demonstrates the significance of our findings and guidelines.
    
    
\end{itemize}


\section{Deep Learning Frameworks}
\label{sec:background}
DL frameworks are the basis for implementing DL programs and building DL models.
To complete a prediction task, developers have to implement a DL program by invoking the APIs provided by a DL framework, and then a model can be built by executing the DL program with training data.
The core functionalities of a DL program include determining the structure of a neural network (e.g., selecting proper layers and setting their order) and configuring the training process (e.g., setting the optimizer and loss function).
All the detailed implementations under the invoked APIs for these DL functionalities are \textit{inside the used DL framework}.
Besides implementing various DL functionalities, \add{DL frameworks also have the other two typical characteristics:
On the one hand,}
DL frameworks are the bridge between DL functionalities and various hardware, and thus they also implement some strategies to support DL functionalities on different hardware.
\add{On the other hand, DL is still a fast-growing area and thus DL frameworks are frequently updated to incorporate the rapid advancement in DL algorithms.}
Therefore, DL frameworks are definitely important for DL development and very complicated especially compared with widely-studied DL programs.

\begin{figure}
    \centering
    \includegraphics[width=0.95\linewidth]{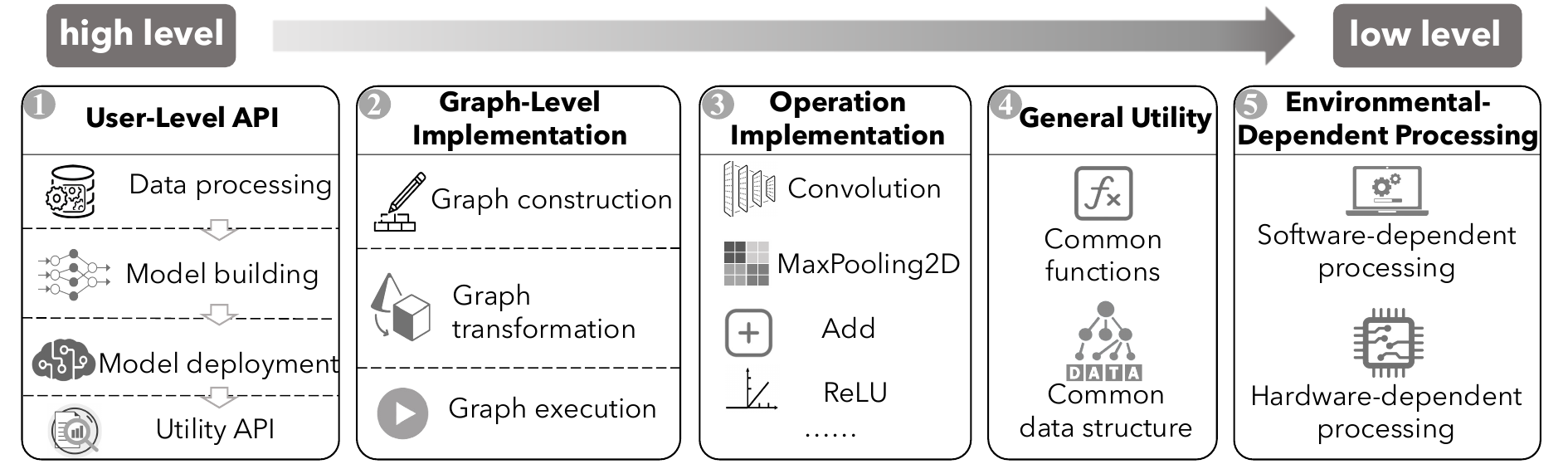}
    \caption{Architecture of DL frameworks}
    \label{fig:DLFcomponent}
\end{figure}


By referring to the existing work~\cite{DBLP:journals/corr/AbadiABBCCCDDDG16}
and understanding the functionality of DL frameworks,
a DL framework can be decomposed into a general five-level architecture as shown in Figure~\ref{fig:DLFcomponent}.
The five levels are \textbf{User-Level API}, \textbf{Graph-Level Implementation}, \textbf{Operation Implementation}, \textbf{General Utility}, and \textbf{Environment-Dependent Processing}, where User-Level API is the highest level that can be directly accessed by users to implement their DL programs while Environment-Dependent Processing is the lowest level that is related to the underlying infrastructure.

\textcircled{1} \textbf{User-Level API}. 
This level contains a large number of high-level APIs,
which \del{aims} \add{aim} to provide convenience for users to use DL frameworks to conduct their DL tasks.
According to the workflow of DL, this level can be further divided into four components 
1) \textit{Data-Processing API} aims to process the input data to make them meet the corresponding requirement of a DL model, e.g., image resizing and text tokenization.
2) \textit{Model-Building API} aims to construct a model structure and search for a group of optimal parameters for the model to make it well fit the training data via a given optimization target (e.g., a loss function).
For example, APIs for various layers and loss functions, as well as various optimizers (e.g., Adam) belong to it.
3) \textit{Model-Deployment API} aims to integrate a built DL model into an existing production environment to make practical prediction.
Typically, it involves the processing (such as model quantization) that makes a DL model work in a specific environment.
4) \textit{Utility API}:
There are many utility APIs across the whole workflow of DL, which provide some auxiliary functionalities to facilitate the DL process, e.g., model visualization and checkpointing.

\textcircled{2} \textbf{Graph-Level Implementation}. 
After implementing a DL program based on these user-level APIs, the follow-up process is mainly based on a static or dynamic computational graph.
A computational graph is a directed graph, in which each node represents an operation (e.g., convolution operation). 
An operation can feed its outputs to another operation through an edge, and the values that flow along an edge are tensors.
This level contains all the computational-graph-level implementations in DL frameworks.
According to the functionalities on a computational graph, this level can be divided into three components:
1) \textit{Graph Construction} aims to create a computational graph and obtain subgraphs via partitioning a graph for distributed execution (especially for a static graph).
2) \textit{Graph Transformation} is responsible for graph optimization (e.g., common subexpression elimination and operation fusion) to improve computation performance, and graph conversion (e.g., converting to the ONNX format).
3) \textit{Graph Execution} aims to execute the graph in a runtime environment, including local and distributed execution.
For example, the execution process involves data propagation and gradient computation.
Please note that the functionalities of these components are conducted in order for static computational graphs, but are mostly intertwined for dynamic computational graphs. 

\textcircled{3} \textbf{Operation Implementation}.
As presented above, each node in a graph is an operation.
This level contains all the detailed implementations for these operations.
An operation takes zero or more tensors as input and produces zero or more tensors as output.
There are a large number of operations implemented in DL frameworks, such as convolution operations, pooling operations, batch normalization operations, mathematical operations (e.g., log), and array manipulation operations (e.g., shuffle).

\textcircled{4} \textbf{General Utility}.
To facilitate the implementations of the above levels, there are many general utilities in DL frameworks, including common data structures and common functions (such as type conversion and padding functions).
This level includes all these general utilities. 
Please note that this level is different from \textit{Utility API} in User-Level API that focuses on facilitating the process of training and deployment for users.

\textcircled{5} \textbf{Environment-Dependent Processing}.
This level is the lowest one, which aims to support the functionalities of DL frameworks in different environments.
A typical example is the memory allocation strategies on different devices, which aim to achieve high efficiency on different devices by considering their corresponding characteristics.
That is, this level contains all the implementations establishing connections between the functionalities of DL frameworks and environments, including both hardware environments (e.g., GPU) and software environments (e.g., operating systems).

\section{Methodology}
\label{sec:methodology}

\subsection{Data Collection}
\label{sec:data}
In the study, we selected four popular DL frameworks as our subjects\ddel{ according to the number of forks in the GitHub repository}, i.e., TensorFlow~\cite{tensorflow} from Google, PyTorch~\cite{pytorch} from Facebook, MXNet~\cite{mxnet} from Apache, and DL4J~\cite{dl4j} from Eclipse. 
\ddel{As presented in the existing work~\cite{10.1145/2627508.2627515}, the number of forks is an important indicator of the popularity of a project, and thus we used it as the metric of DL framework selection.}\aadd{Following the existing work~\cite{10.1145/2627508.2627515,han2020empirical}, we used the number of forks and the number of projects using the DL framework in GitHub to measure the popularity of a DL framework, which is used for DL framework selection.
We found that Top-6 DL frameworks are the same, i.e., TensorFlow, PyTorch, Keras, Caffe, MXNet, and DL4J, regardless of the used metrics.}
Although Caffe~\cite{caffe} and Keras~\cite{keras} \ddel{have more forks} \aadd{are more popular} than MXNet and DL4J \aadd{in terms of both metrics}, Caffe has stopped its update for a relatively long time (i.e., more than 12 months before the accessing date in October, 2021)
while Keras is only a front end and has to run on top of some other DL frameworks (e.g., TensorFlow).
\aadd{More specifically, Keras can be considered as a ``User-Level API'' in the general five-level architecture. 
In actual, our work studied more comprehensive DL frameworks in order to sufficiently understand DL framework bugs at various levels.}
Hence, we did not use them in our study.
All the four DL frameworks are built with the above five-level architecture, but they are also diverse, e.g., involving different programming languages for implementations, different development organizations, and different types of computational graphs.
\ddel{Indeed, one metric may not guarantee the representativeness of our selection.
Hence, we further adopted the number of projects in GitHub using the DL framework to measure the popularity of the corresponding DL framework following the existing work~\cite{han2020empirical}.
We found that Top-6 DL frameworks are also TensorFlow, PyTorch, Keras, Caffe, MXNet, and DL4J in terms of this metric.
Due to the above mentioned reason for Caffe and Keras, we will also select the four DL frameworks (i.e., TensorFlow, PyTorch, MXNet, and DL4J) as the subjects in terms of this metric.
Therefore, that confirms the representativeness of our selection to some degree.}

Since our study aims to investigate the characteristics of DL framework bugs, we collected \textit{closed} and \textit{merged} pull requests that are responsible to fix bugs from the corresponding GitHub repositories of the four DL frameworks following the existing work~\cite{islam2019comprehensive,garcia2020comprehensive,dlcompiler}.
On the one hand, the bugs involved in these pull requests have been accepted and fixed by developers;
On the other hand, these pull requests tend to contain more comprehensive information, e.g., code changes, links to related issues, and discussions among developers, which is helpful to understand the bugs.
In fact, not all of such pull requests are responsible to fix bugs, e.g., some of them aim to add new features or update documents.
Hence, we further identified \textit{bug-fixing pull requests} through keyword searching in the tags and titles of pull requests.
Following the existing work\cite{islam2019comprehensive,garcia2020comprehensive,dlcompiler}, we adopted several bug-relevant keywords, including \textit{fix, defect, error, bug, issue, mistake, correct, fault}, and \textit{flaw}.

\begin{table}[t]
\centering
\caption{Statistical Information on Our Dataset}
\vspace{-2mm}
\begin{adjustbox}{max width=0.95\columnwidth,center}
\begin{tabular}{l||r|r|r|c|r|c|c}
\toprule
\textbf{Framework} & 
\textbf{\#SLOC} &
\textbf{\#PR} &
\textbf{\#Bug} &
\textbf{Duration} &
\textbf{\#Fork} &
\textbf{Language} &
\textbf{Organization}
\\
\midrule
TensorFlow &3,090,623 &319 &250 & 2020/08-2021/10& 85.3k & C++, Python & Google \\
PyTorch &1,762,228 &307 &250 & 2018/06-2021/10& 13.5k& C++, Python &  Facebook \\
MXNet &455,363 &359 &250 &2020/03-2021/10& 6.9k  & C++, Python & Apache\\
DL4J & 1,041,927 &265 &250 &2018/12-2021/10& 4.9k & Java, C++ &  Eclipse\\
\midrule
\textit{Total} &6,320,141 &1,250 &1,000 &2018/06-2021/10 &110.6k &-  & -\\
\bottomrule
\end{tabular}
\end{adjustbox}
\label{tab:frame_info}
\end{table}

It is quite time-consuming to manually analyze bugs, and thus it is unaffordable for us to collect all bugs for manual inspection.
Following the existing studies~\cite{JIA2021110935,islam2019comprehensive, di2017comprehensive}, we collected bugs in a specific duration (i.e., 2018/06 - 2021/10 in our study).
However, different DL frameworks have different numbers of bugs within the same period, which may affect the analysis and conclusions across different DL frameworks, and thus we further balanced the number of studied bugs for each DL framework by selecting the same number of bugs for them.
Specifically, among the four DL frameworks, PyTorch has the smallest number of bug-fixing pull requests within this period, and thus we first manually analyzed all these pull requests and finally obtained 250 bugs for PyTorch.
The detailed process for manual analysis will be presented in Section~\ref{sec:classification}.
Then, for each of the other three DL frameworks, we analyzed the bug-fixing pull requests in the reversed chronological order within this period like the existing study~\cite{perez2017prevalence}, until the same number of bugs as PyTorch (i.e., 250) were identified.
\add{That is, we did not manually analyze all the collected pull requests for each of the other three DL frameworks (except PyTorch), but analyzed a subset of bug-fixing pull requests corresponding to the identified 250 bugs.}
\del{Indeed, more recent bugs can better reflect the current characteristics of DL framework bugs, which is more helpful to improve the current DL frameworks.}
\add{More recent bugs may be more relevant to the characteristics of the current versions of DL frameworks.
Specifically, DL is a fast-growing area and thus the characteristics of DL frameworks tend to be frequently updated, such as incorporating the rapid advancement in DL.
Therefore, paying more attention to recent bugs could be more helpful to improve the current DL frameworks.}
Meanwhile, the involved duration for the 250 bugs of each DL framework is at least 14 months (i.e., 2020/08 - 2021/10 for TensorFlow), which can also support the generalizability of our conclusions to a large extent.
\add{Please note that in our study each bug is uniquely different from the other 249 bugs for each DL framework.
In general, developers merge a bug-fixing pull request after they carefully verify that the target bugs have been correctly fixed by the pull request.
Also, many bug-fixing pull requests have the related issue reports, and we have checked that all the related issue reports in our study are indeed different.
We also regard it as a potential threat in our study since it is also possible that developers make mistakes when verifying the bugs fixed by a pull request, but this threat may be not serious according to the above analysis.
}

Table~\ref{tab:frame_info} shows the statistical information of our dataset, where each column presents
the number of source lines of code (SLOC), \del{the number of analyzed bug-fixing pull requests (PR)} \add{the number of pull requests (PR) that were manually analyzed by us for obtaining the 250 bugs (rather than the total number of bug-fixing pull requests collected initially)}, \del{the number of bugs identified from these pull requests} \add{the number of identified bugs (i.e., 250 for each DL framework)}, the duration for the identified bugs, the number of forks in the GitHub repository (accessed in October, 2021),
the used major programming languages, and the development organization, respectively.
In total, we collected 1,000 bugs from the four popular DL frameworks after manually analyzing 1,250 bug-fixing pull requests.
The involved duration for the collected bugs ranges from 14 months to 40 months across different DL frameworks.
\add{To our best knowledge, our study is the most large-scale in this area.}
In particular, we have released our dataset at our project homepage: \textbf{\url{https://github.com/DLFrameworkBug/DLFrameworkBugsData}}, to facilitate the replication of our study and promote the future research in this area.

\subsection{Classification and Labeling Process}
\label{sec:classification}
In the study, for each bug, we labeled its \textit{root cause}, the \textit{symptom} that the bug exhibits, the \textit{stage} of the DL pipeline (to be introduced in Section~\ref{sec:symptom_distribution}) in which the bug symptom is observed, and the \textit{level} of the DL framework in which the bug occurs.
To label the root cause and \add{the} symptom of each bug, we first adopted the general taxonomies of root causes and symptoms from the existing work~\cite{garcia2020comprehensive,islam2019comprehensive,MLBugs,tan2014bug,humbatova2020taxonomy} as the \textit{initial} taxonomies, and then adapted them to DL framework bugs.
Specifically, following the general open-coding scheme~\cite{lune2017qualitative}, two authors
went through all the pull requests to determine the root-cause and symptom categories of our collected DL framework bugs based on the initial general taxonomies by adding DL-framework-specific categories (e.g., Environment Incompatibility) on demand and removing irrelevant categories. 

Regarding the levels of DL frameworks, we have introduced them in Section~\ref{sec:background}.
To label the level of the DL framework in which each bug occurs, we divided all the source files of each DL framework into the five levels by understanding the functionality of each source file based on the source code, comments, and documents for the source file and the corresponding folder.
This task was conducted by two authors together.
We have also released our classification results for the source files of each DL framework at our project homepage to facilitate the replication of our study.

As mentioned above, based on the prepared root-cause and symptom categories as well as the classification results for the source files of each DL framework, two authors \textit{independently} labeled each pull request via an open-coding scheme following the existing studies~\cite{islam2019comprehensive,dlcompiler}.
Specifically, to label the level of the DL framework in which a bug occurs, the two authors identified the bug-fixing code changes in the pull request and labeled it according to which source files the bug-fixing code changes lie in.
To label the other aspects, the two authors carefully understood the bug-fixing code changes, the descriptions about the bugs in the related issues, and the discussions among developers in the pull request. 
During the labeling process, we adopted the Cohen's Kappa coefficient~\cite{vieira2010cohen} to measure the inter-rater agreement between them following the existing work~\cite{islam2019comprehensive,dlcompiler}.
The Cohen's Kappa coefficient was just nearly 35\% for the first 5\% of labeling results, and thus we conducted a training session about labeling.
Then, the Cohen's Kappa coefficient reached 80\% for the first 10\% of labeling results (including the first 5\%).
Through further discussion on these inconsistencies, the Cohen's Kappa coefficients were over 95\% in all the subsequent labeling studies (i.e., labeling 20\%$\sim$100\% of bugs with the interval of 10\%).
For the inconsistencies in each labeling study \add{(including the training session)}, the two authors discussed with the third author until all the bugs were labeled consistently.

In particular, we provided an example to illustrate how a bug was manually labeled in more detail at our project homepage. 
During the careful labeling process, we came across slight inaccuracy in the prepared root-cause and symptom categories, and thus we further improved them to obtain better classification results.
Moreover, we filtered out the pull requests that are actually irrelevant to bug fixing.
There are some pull requests where more than one bugs were fixed, and we treated each of them as an individual bug following the existing work~\cite{garcia2020comprehensive,dlcompiler}.

\section{Results and Analysis}
\label{sec:results}


\subsection{RQ1: Root Causes}
\label{sec:root_causes}
\subsubsection{Root Cause Classification Results}
\label{sec:root_causes_classification}

Based on the above classification and labeling process, we identified the following 13 root causes of DL framework bugs.
The first four root causes involve the characteristics of DL frameworks \aadd{(meaning that there is at least one sub-category specific to the characteristics of DL frameworks in each of the four root causes)}, while the others are common categories.
\aadd{Besides the four root causes involving the characteristics of DL frameworks, we also newly added the root cause of Dependent Module Issue over the general taxonomy of traditional software.
This is because the number of this category of bugs is non-negligible for DL frameworks, and thus we did not treat these bugs as Others.
The remaining categories were adapted from the general taxonomy of traditional software.}
In the study, we not only discussed these root causes very relevant to DL frameworks, but also investigated whether these common root causes have different distributions and characteristics between DL frameworks and traditional software.

\begin{figure}
    \centering
    \subfigure[{\footnotesize Type Issue (PyTorch\#8230)}]{
        \label{fig:ex_type}
        \includegraphics[width=0.4750\linewidth,trim=0 170 410 165,clip]{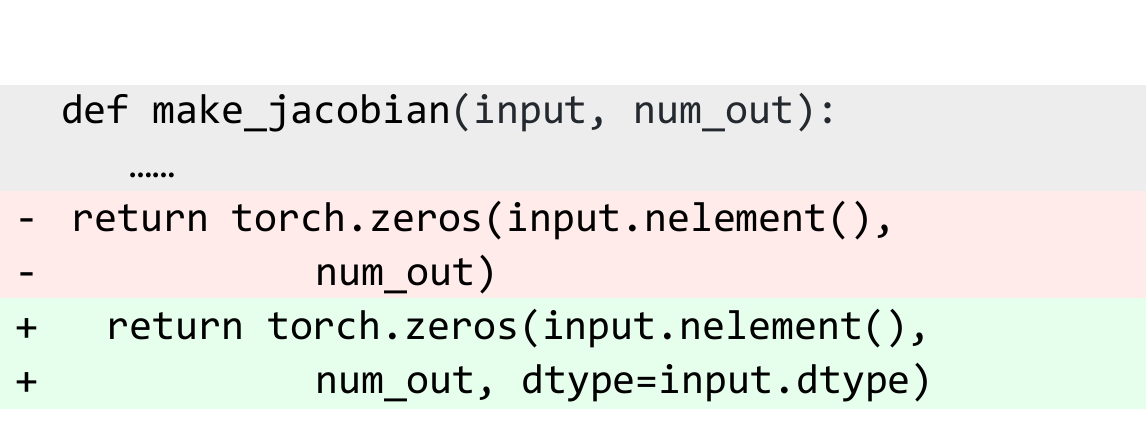}
    }
    \subfigure[{\footnotesize Tensor Shape Misalignment (PyTorch\#66214)}]{
        \label{fig:ex_shape}
        \includegraphics[width=0.475\linewidth,trim=0 170 410 165,clip]{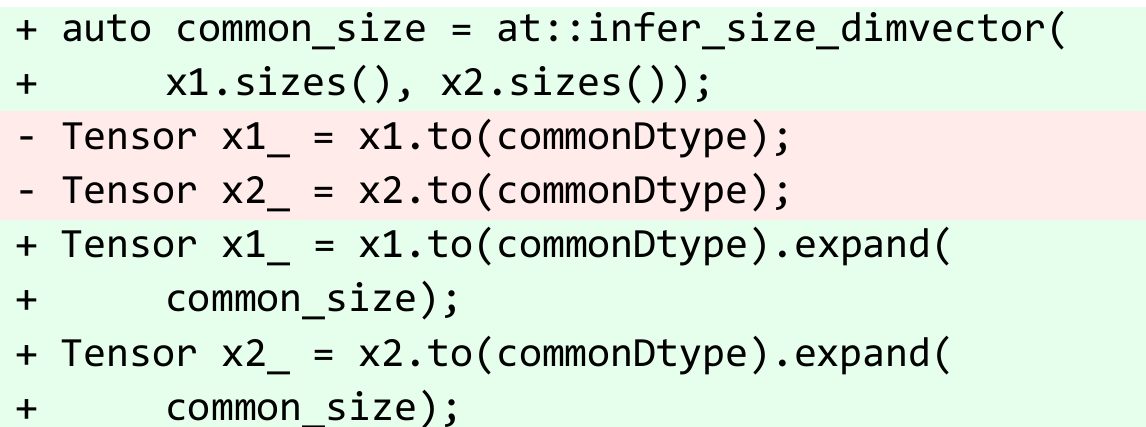}
    }
    
    \subfigure[{\footnotesize Incorrect Algorithm Impl. (PyTorch\#35416)}]{
        \label{fig:ex_algo}
        \includegraphics[width=0.475\linewidth,trim=0 115 410 110,clip]{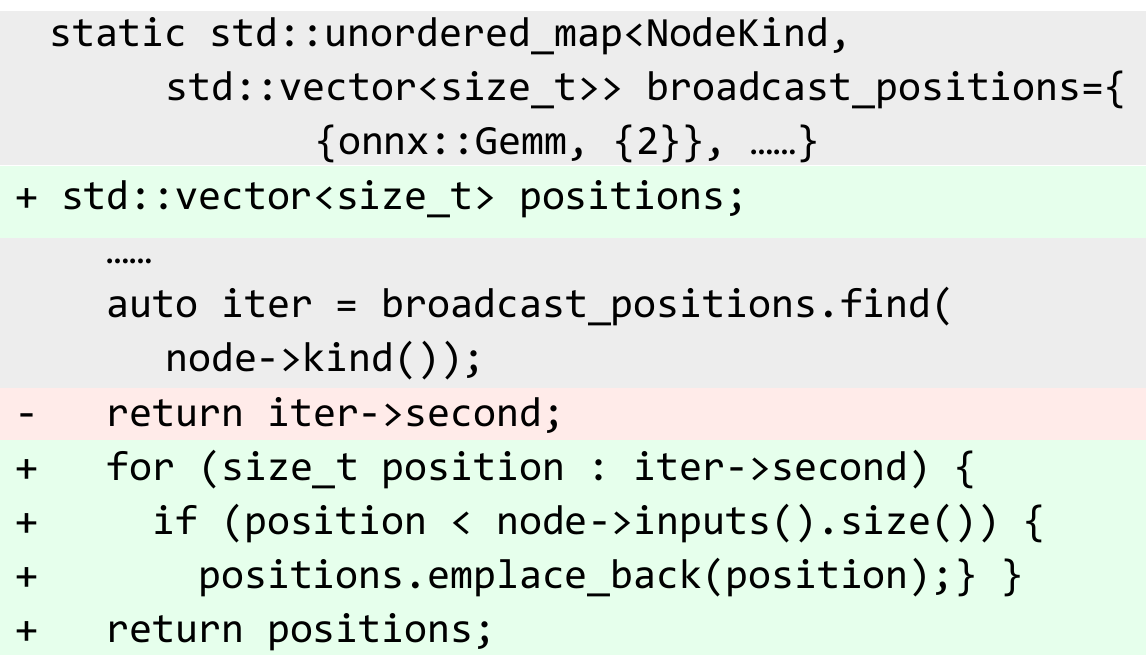}
    }
     \subfigure[{\footnotesize Environment Incompatibility (MXNet\#19236)}]{
        \label{fig:ex_envi}
        \includegraphics[width=0.475\linewidth,trim=0 115 410 110,clip]{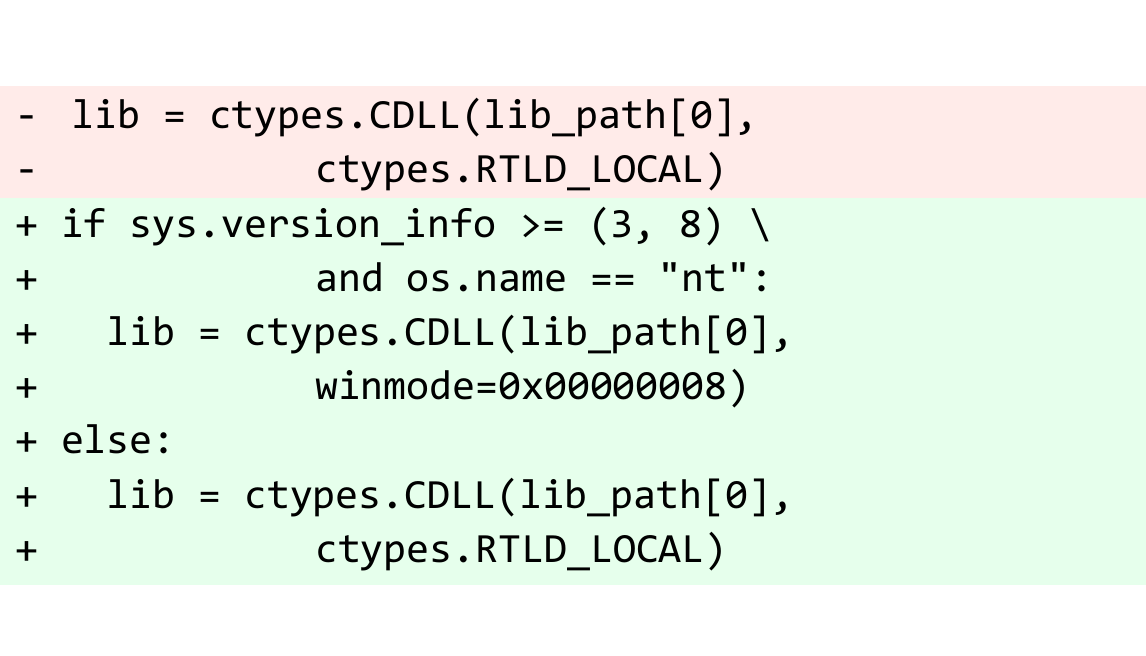}
    }
    
    \subfigure[{\footnotesize API Incompatibility (PyTorch\#59008)}]{
        \label{fig:ex_apicom}
        \includegraphics[width=0.475\linewidth,trim=0 230 410 230,clip]{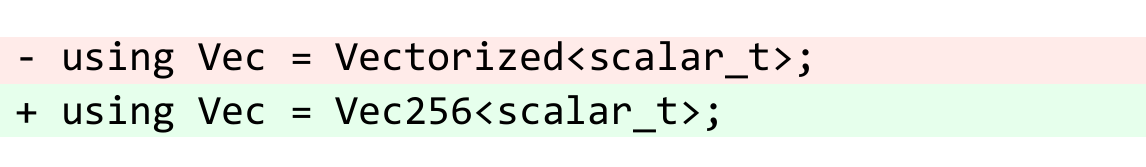}
    }
    \subfigure[{\footnotesize API Misuse (TensorFlow\#44066)}]{
        \label{fig:ex_apim}
        \includegraphics[width=0.475\linewidth,trim=0 230 410 230,clip]{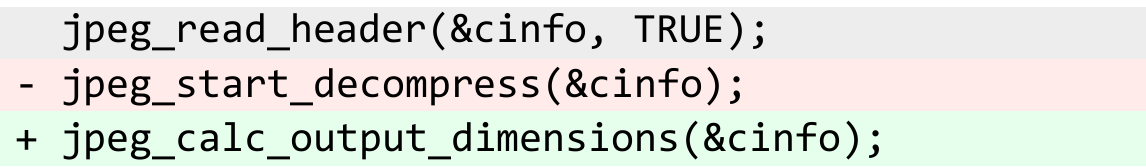}
    }    

    \subfigure[{\footnotesize Incorrect Assignment (TensorFlow\#42032)}]{
        \label{fig:ex_assign}
        \includegraphics[width=0.475\linewidth,trim=0 150 410 150,clip]{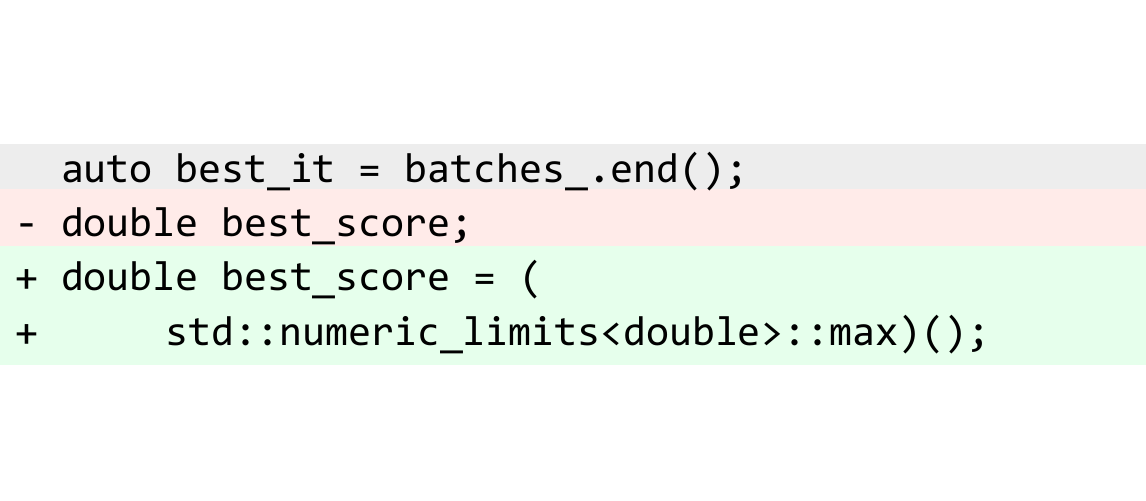}
    }  
    \subfigure[{\footnotesize Incorrect Exception Handling (PyTorch\#27398)}]{
        \label{fig:ex_excep}
        \includegraphics[width=0.475\linewidth,trim=0 150 410 150,clip]{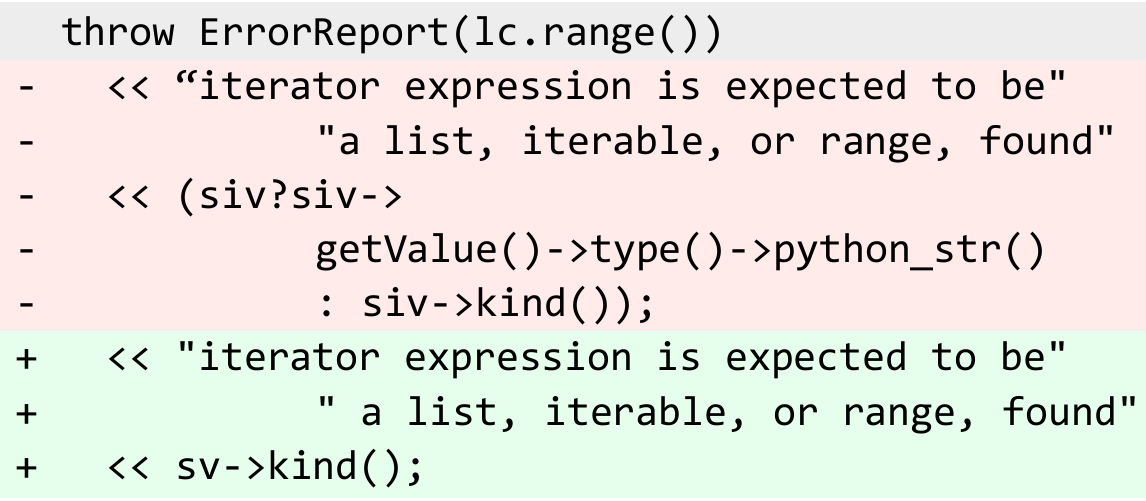}
    } 
    
    \subfigure[{\footnotesize Misconfiguration (MXNet\#20570)}]{
        \label{fig:miscon}
        \includegraphics[width=0.475\linewidth,trim=0 130 410 130,clip]{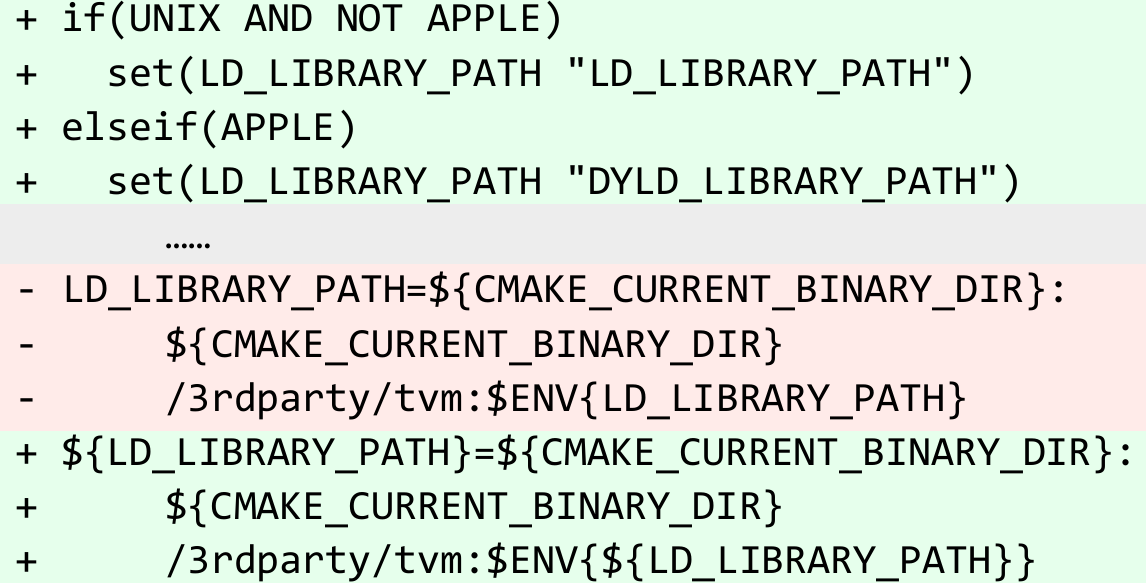}
    }  
    \subfigure[{\footnotesize Numerical Issue (MXNet\#17937)}]{
        \label{fig:ex_nume}
        \includegraphics[width=0.475\linewidth,trim=0 130 410 130,clip]{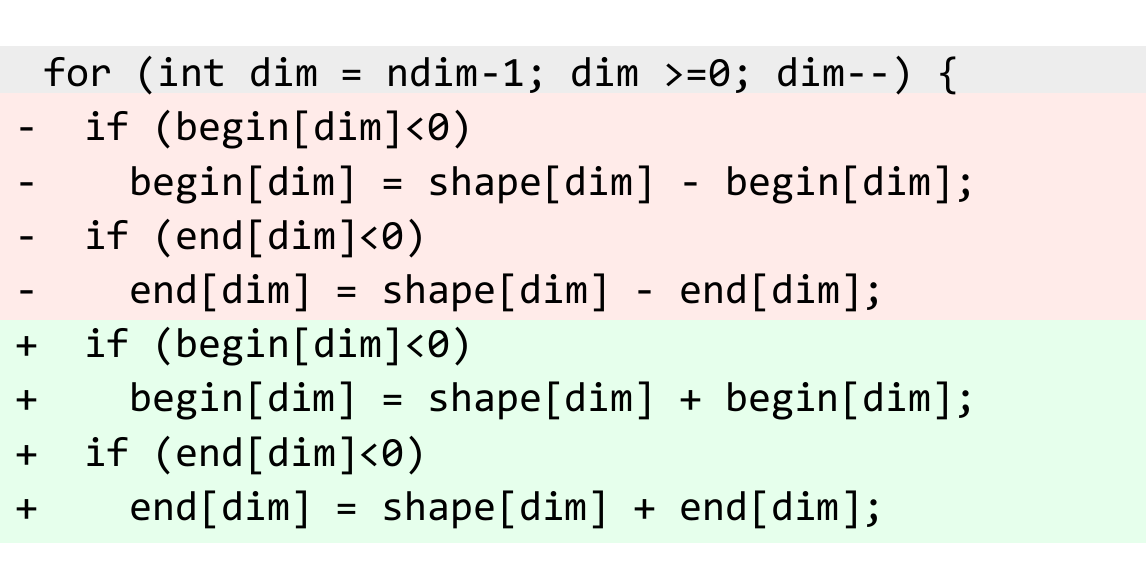}
    }  
    
    \subfigure[{\footnotesize Concurrency Issue(TensorFlow\#50382)}]{
        \label{fig:ex_concu}
        \includegraphics[width=0.475\linewidth,trim=0 221 410 222,clip]{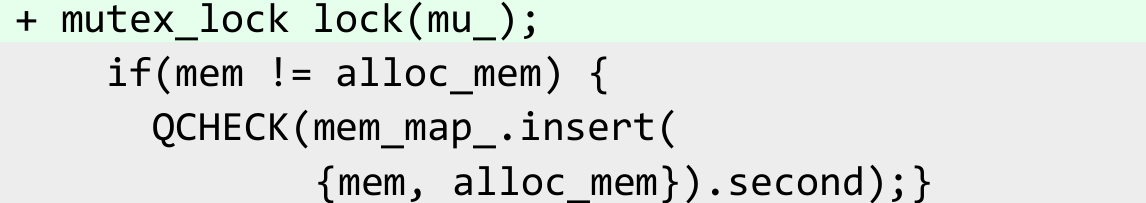}
    }  
    \subfigure[{\footnotesize Dependent Module Issue (DL4J\#9113)}]{
        \label{fig:ex_depend}
        \includegraphics[width=0.475\linewidth,trim=0 221 410 222,clip]{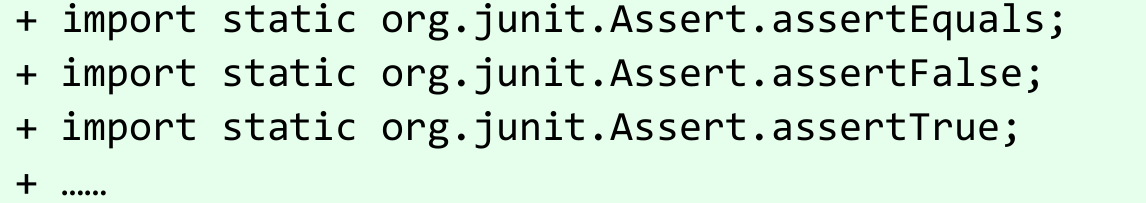}
    }  
    \vspace{-3mm}
    \caption{\add{Bug Examples \ddel{in}\aadd{of Different} Root Causes}}
    \label{fig:examples}
\end{figure}

\textcircled{1}
\textbf{Type Issue}.
This kind of bugs involves type-related problems, such as type conversion and type checking.
Different from traditional software, tensors are quite widely-used in the development of a DL framework.
A tensor is a multi-dimensional matrix consisting of elements with some data type.
According to this characteristic of DL frameworks, we divide this root cause into two sub-categories.
1) \textit{Tensor type issue}, which refers to the bugs caused by the data types of tensors.
2) \textit{Traditional data type issue}, which refers to the bugs caused by the types of traditional variables.
\add{For example, Figure~\ref{fig:ex_type} shows the patch\footnote{\add{https://github.com/pytorch/pytorch/pull/8230}} for an example bug belonging to this category. By default, the function {\tt torch.zeros()} returns a matrix with type {\tt float32}, while the expected type of the returned matrix should be consistent with that of the fed {\tt input}.}

\textcircled{2}\textbf{Tensor Shape Misalignment}.
This kind of bug occurs due to tensor shape mismatching in shape-related operations, e.g., tensor shape inference and transformation.
Specifically, tensor shape describes the number of elements in each dimension.
\add{For example, when computing the \textit{cosine} value between two tensors, their dimensions (i.e., tensor shapes) should be broadcastable to a common shape, but the buggy code requires that their original dimensions must be the same. The patch\footnote{\add{https://github.com/pytorch/pytorch/pull/66214}} in Figure~\ref{fig:ex_shape} fixes it by calculating the common size of the two tensors.}

\textcircled{3}\textbf{Incorrect Algorithm Implementation}.
This kind of bugs is caused by the problematic implementation logic \add{(rather than the lack of implementation)} of an algorithm, which tends to involve a number of statements or blocks.
According to the functionality of an algorithm, we divide this root cause into two sub-categories.
1) \textit{Incorrect DL-specific algorithm implementation}: there are a large number of algorithms with DL-specific functionalities in a DL framework (such as operation fusion and gradient computation algorithms), and this sub-category of bugs occurs due to the incorrect implementation logic of these algorithms.
2) \textit{Incorrect DL-irrelevant algorithm implementation}: This sub-category of bugs occurs due to the incorrect implementation logic of the algorithms with DL-irrelevant functionalities (such as memory allocation algorithms).
\add{Figure~\ref{fig:ex_algo} shows a patch\footnote{\add{https://github.com/pytorch/pytorch/pull/35416}} from the method {\tt getBroadcastPositions()} in PyTorch to fix an incorrect DL-specific algorithm implementation for model transformation. 
Specifically, when transforming {\tt Torch.mm} to {\tt Gemm} in ONNX, the output of the method should satisfy certain constraints (i.e., {\tt position $<$ node$\rightarrow$inputs().size()}), while the buggy code missed the checking and produced incorrect results.}

\textcircled{4}\textbf{Environment Incompatibility}.
This kind of bugs occurs due to neglecting some characteristics (e.g., the endianness for an architecture) of a specific environment (e.g., hardware or operating systems).
\add{For example, the lack of implementation for supporting the hardware is one reason for this kind of bugs.}
This root cause is common in DL frameworks since DL frameworks are required to work on various environments, such as CPU and GPU with various architectures.
\add{Figure~\ref{fig:ex_envi} shows an example patch\footnote{\add{https://github.com/apache/incubator-mxnet/pull/19236}}, which fixes the compatibility issue when deploying MXNet on different versions of the system.}

\textcircled{5}\textbf{API Incompatibility}.
This kind of bugs contains two sub-categories.
1) \textit{Internal incompatibility} refers to the API compatibility issues within a DL framework caused by API evolution;
2) \textit{External incompatibility} refers to the API compatibility issues between a DL framework and third-party libraries (such as NumPy and HIPIFY) caused by the update of the latter.
\add{Please note that we did not consider the bugs in the third-party libraries depended by DL frameworks, but consider the bugs caused by the incompatibility between DL frameworks and the third-party libraries.
This is because this latter kind of bugs are fixed by modifying the DL-framework code (rather than the third-party library code) to replace the invoked incompatible API with a compatible one.}
\add{Taking the patch\footnote{\add{https://github.com/pytorch/pytorch/pull/59008}} shown in Figure~\ref{fig:ex_apicom} as an example, the version 1.9 of PyTorch changed the class {\tt Vectorized} to {\tt Vec256}, while the use of this class was not consistently updated, causing an API incompatibility issue.}

\textcircled{6}\textbf{API Misuse}.
This kind of bugs occurs due to the misunderstanding of the APIs invoked by the DL framework, which contains three \add{main} sub-categories.
1) \textit{Condition missing/redundancy} means that developers miss to add (or redundantly add) a condition check for an API; 
2) \textit{API missing/redundancy} means that developers miss to use (or redundantly use) an API;
3) \textit{Wrong API} means that developers use a wrong name/argument/receiver of an API.
By taking an API call {\tt a.b(x,y)} as an example, {\tt a} is the receiver, {\tt b} is the API name, and {\tt x} and {\tt y} are the arguments.
\add{For example, developers misused APIs {\tt jpeg\_start\_decompress} and {\tt jpeg\_calc\_output\_dimensions} as shown in Figure~\ref{fig:ex_apim}\footnote{\add{https://github.com/tensorflow/tensorflow/pull/44066}}, where the former API may cause over 50x performance decline on certain images due to the \textit{decompression} operation in it.}
\add{In fact, the incorrect implementation of an ``iteration'' over API calls is also a potential reason for this kind of bugs, but it does not occur in our study.
This conclusion is almost consistent with the existing work~\cite{amann2018systematic_api}, which shows that only 2 among 165 API misuses for traditional software are caused by this reason.}

\textcircled{7}\textbf{Incorrect Assignment}.
This kind of bugs occurs when a variable is incorrectly assigned or lacks initialization,
\add{such as the example\footnote{\add{https://github.com/tensorflow/tensorflow/pull/42032}} shown in Figure~\ref{fig:ex_assign}, where the variable {\tt score} was uninitialized before use.}

\textcircled{8}\textbf{Incorrect Exception Handling}.
This kind of bugs occurs due to incorrect exception handling, which includes three scenarios:
1) \textit{Missing exception}, i.e., a DL framework does not throw\add{/handle} an exception when it should;
2) \textit{Spurious exception}, i.e., a DL framework throws an exception when it should not;
3) \textit{Wrong exception message}, i.e., a DL framework produces incorrect/imprecise exception messages for an exception,
\add{such as the example\footnote{\add{https://github.com/pytorch/pytorch/pull/27398}} shown in Figure~\ref{fig:ex_excep}.}

\textcircled{9}\textbf{Misconfiguration}.
This kind of bugs is caused by incorrect configurations in a DL framework, such as configurations in Bazel files and various Shell configuration scripts.
\add{For example, the default path configuration of dynamic shared libraries is ``LD\_LIBRARY\_PATH'' for building MXNet, which works well on Linux systems. However, for MacOS, it will cause build failures since this path changes to ``DYLD\_LIBRARY\_PATH''. As a result, the configuration file should be updated for explicitly setting the path for different systems\footnote{\add{https://github.com/apache/incubator-mxnet/pull/20570}} (see Figure~\ref{fig:miscon}).}

\textcircled{10}\textbf{Numerical Issue}.
This kind of bugs is caused by incorrect numerical computations, such as dividing by 0, \aadd{overflow/underflow,} using wrong operators or operands \add{e.g., a computation should use ``+'' but use ``$\times$'' or a computation should be ``i+1'' but wrongly write as ``i+2''}, and missing operands.
\add{For example, developers misused the operators ``+'' and ``-'' as shown in Figure~\ref{fig:ex_nume}, which caused the incorrect computation of batch size\footnote{\add{https://github.com/apache/incubator-mxnet/pull/17937}}.}

\textcircled{11}\textbf{Concurrency Issue}.
This kind of bugs is caused by incorrect operations on concurrency-oriented structures, such as threads, shared memory, and race conditions.
\add{For example, the variable {\tt mem\_map\_} shown in Figure~\ref{fig:ex_concu} records the allocated memory resources and can be updated by different threads\footnote{\add{https://github.com/tensorflow/tensorflow/pull/50382}}. Therefore, ensuring the mutually exclusive access to it is critical. To fix the bug, a {\tt mutex\_lock} was introduced.}

\textcircled{12}\textbf{Dependent Module Issue}.
This kind of bugs occurs due to missing to import dependent modules or importing wrong modules,
\add{such as the example patch\footnote{\add{https://github.com/deeplearning4j/deeplearning4j/pull/9113}} for DL4J shown in Figure~\ref{fig:ex_depend}.}

\textcircled{13}\textbf{Others}.
Each bug in this root cause is unusual and cannot be assigned to any other root causes.

\subsubsection{Root Cause Distribution}
\label{sec:root_causes_distribution}

\begin{figure}[t]
    \centering
    \includegraphics[width=\linewidth]{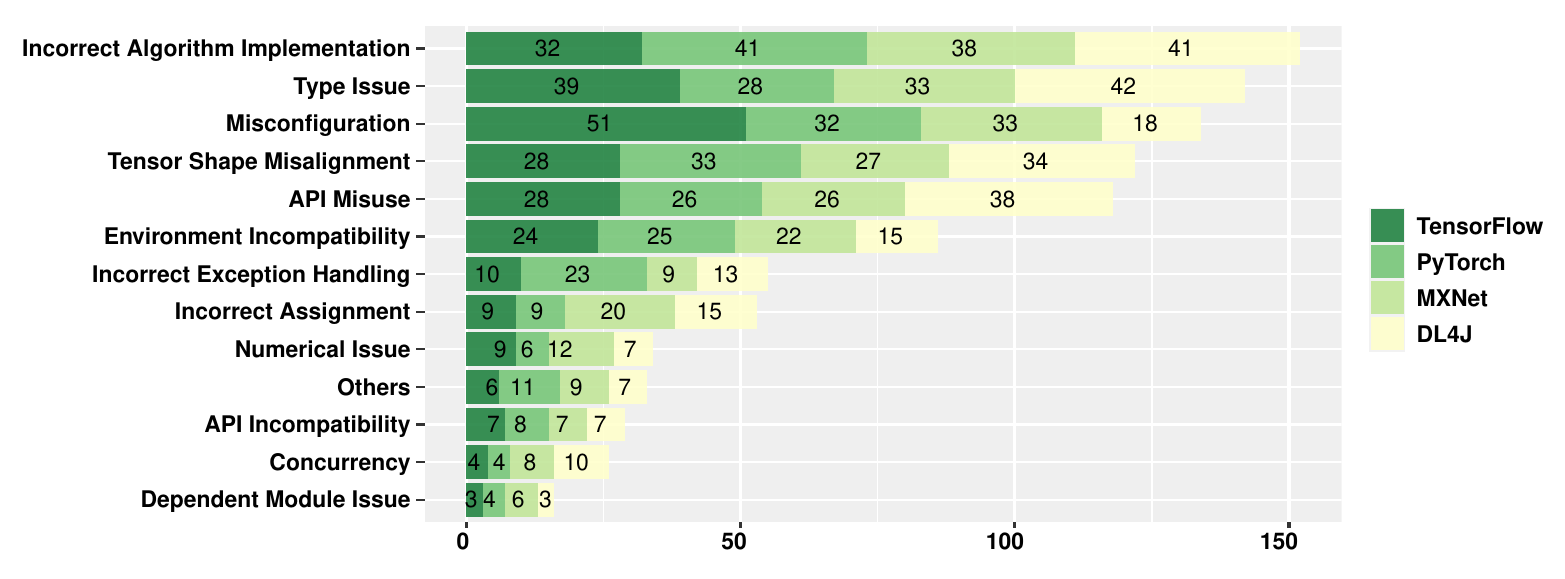}
    \vspace{-2mm}
    \caption{Bug Distribution by Root Causes}
    \label{fig:rootcause_distribute}
\end{figure}


Figure~\ref{fig:rootcause_distribute} shows the bug distribution by the identified root causes.
From this figure, the four root causes involving the characteristics of DL frameworks (i.e., Incorrect Algorithm Implementation, Type Issue, Tensor Shape Misalignment, and Environment Incompatibility) are indeed prevalent, all of which are ranked within Top-6 (among 13 root causes) and account for 50.2\% of bugs in total.
Among all these root causes, Incorrect Algorithm Implementation is the most prevalent one.
It accounts for 152 bugs in total, including 32 in TensorFlow, 41 in PyTorch, 38 in MXNet, and 41 in DL4J.
The reason mainly lies in that deep learning is a fast-growing area and thus DL frameworks have to be frequently updated to incorporate the rapid advancement in DL algorithms.
Moreover, hardware (especially DL-related hardware) is also rapidly developed and thus DL frameworks are required to provide the corresponding implementations to support these new features in hardware.
Regardless of supporting advanced DL algorithms or new hardware features, the corresponding implementations in DL frameworks tend to involve complicated code logic, and thus it is very likely for them to incur various technical debts.
Through further analysis, we found that about 79.61\% of this kind of bugs (121 out of 152) occur in the implementations of DL-specific algorithms, significantly outnumbering the bugs in the implementations of DL-irrelevant algorithms.

\find{Regarding the root causes involving DL framework characteristics, all of them are prevalent, accounting for 50.2\% of bugs in total.
Among them, the most prevalent root cause is Incorrect Algorithm Implementation (especially in DL-specific algorithms).}

Type Issue is the second most prevalent one among all the root causes. 
It accounts for 142 bugs in total, including 39 in TensorFlow, 28 in PyTorch, 33 in MXNet, and 42 in DL4J.
Through further investigation, nearly 70.42\% of this kind of bugs (100 out of 142) are caused by tensor types rather than traditional data types.
This is because all the DL operations depend on tensors, and meanwhile tensor type is an important property in a tensor and \add{is} usually involved in various operations.
In particular, type conversion, especially implicit type conversion, tends to incur Type Issue bugs in DL frameworks, which deserves more attention in practice.

\find{Type Issue is the second most prevalent root cause, which accounts for 14.20\% of DL framework bugs and mainly occurs on tensor types. }


In addition, there are common categories of root causes between DL frameworks and traditional software, and some of them are also notable.
Besides the four root causes involving DL framework characteristics, the remaining two root causes ranked within Top-6 are Misconfiguration and API Misuse.
In particular, Misconfiguration is the third \add{most} prevalent one among all the root causes, which accounts for 134 bugs in total.
The phenomenon is different from the existing studies on traditional software bugs where Misconfiguration bugs either are ignored by them~\cite{tan2014bug,sun2016toward} or account for only a small percentage among all the studied bugs~\cite{ocariza2013empirical,MLBugs}.
For example, as shown in the existing study~\cite{MLBugs}, only 5.7\% of bugs are caused by Misconfiguration in traditional machine learning systems, which is ranked at $9^{\textit{th}}$ position among their identified 11 root causes.

The reason why DL frameworks contain many Misconfiguration bugs may lie in that, there are a large number of configuration files/options for compilation, installation, and ensuring compatibility of DL frameworks due to their complex implementations involving multiple programming languages as well as the large number of dependent third-party libraries and hardware/software environments.
\add{Through further analysis, we found that 22.39\% (30 out of 134) bugs are caused by the configuration files/options for compilation, 41.04\% (55 out of 134) are caused by the configuration files/options for installation, and 36.57\% (49 out of 134) are caused by the configuration files/options for ensuring compatibility of DL frameworks.}


\begin{table}[t]

\caption{Distribution of API Misuse Bugs}
\vspace{-2mm}
\begin{adjustbox}{width=0.8\columnwidth,center}
\begin{threeparttable}
\small
\begin{tabular}{@{}l||c|c|cccc@{}|c}
\toprule
\multirow{2}{*}{\textbf{Framework}} &
  \multirow{2}{*}{\textbf{\begin{tabular}[c]{@{}c@{}}Condition Missing\\or Redundancy\end{tabular}}} &
  \multirow{2}{*}{\textbf{\begin{tabular}[c]{@{}c@{}}API Missing\\ or Redundancy\end{tabular}}} &
  \multicolumn{4}{c|}{\textbf{Wrong API}} & \multirow{2}{*}{\textbf{\textit{Total}}} \\ \cline{4-7} 
                                &   &    & \textbf{Receiver} & \textbf{Name} & \textbf{Args} & \textit{Sum} \\ 
                                \midrule
TensorFlow   & 3 & 4 & 4 & 7 & 10 & 21 & 28\\
PyTorch      & 3 & 5 & 2 & 9 & 7 & 18 & 26\\
MXNet        & 3 & 3 & 6 & 5 & 9 & 20 & 26 \\
DL4J         & 4 & 9 & 4 & 13 & 8 & 25 & 38 \\ \midrule
\textit{Total}        & 13 & 21 & 16 & 34 & 34 & 84 & 118 \\ \bottomrule
\end{tabular}
\end{threeparttable}
\end{adjustbox}
\label{tab:api_misuse}
\end{table}

API Misuse is another prevalent root cause for DL framework bugs.
Indeed, this root cause is also common in traditional software, but it is unknown whether they manifest in the same way or not.
To further investigate it, we then show its distribution in the three subcategories in Table~\ref{tab:api_misuse},
following the existing work~\cite{amann2018systematic_api,zhang2018code_api}.
From Table~\ref{tab:api_misuse}, 71.19\% of API Misuse bugs (84 out of 118) are due to using wrong APIs, significantly outnumbering those caused by the other two subcategories.
However, as demonstrated by the existing studies on MuBench~\cite{amann2016mubench,amann2018systematic_api} (one of the most widely-studied benchmarks in the area of API misuse, including 90 API misuses from Java projects), API Missing/Redundancy is the most prevalent subcategory.
That is, while API Misuse is a common root cause for both DL frameworks and traditional software, they manifest in a significantly different way.
The result indicates that in DL frameworks, developers may usually confuse different API usage scenarios, especially for a set of similar APIs, calling for new API misuse detection methods that can clearly distinguish those similar APIs.

\find{Regarding the common categories between DL frameworks and traditional software, Misconfiguration and API Misuse are two most notable root causes, but those bugs in DL frameworks have different characteristics and distributions with those in traditional software, calling for different bug detection strategies.}

\subsection{RQ2: Symptoms}
\label{sec:symptoms}

\subsubsection{Symptom Classification Results}
\label{sec:symptoms_classification}
According to the aforementioned classification and labeling process, we identified the following 6 symptoms of DL framework bugs.

\textcircled{1}\textbf{Crash}.
This symptom refers to that a DL framework terminates unexpectedly during running, such as terminating with an error message like ``out of memory'' or ``null pointer''.
\add{For example, the bug shown in Figure~\ref{fig:ex_shape} made PyTorch crash during computing {\it cosine} similarity due to inconsistent tensor shapes.}

\begin{figure}

    \centering
    \subfigure[{\footnotesize Incorrect Functionality (TensorFlow\#46509)}]{
        \label{fig:ex_function}
        \includegraphics[width=0.475\linewidth,trim=0 160 400 160,clip]{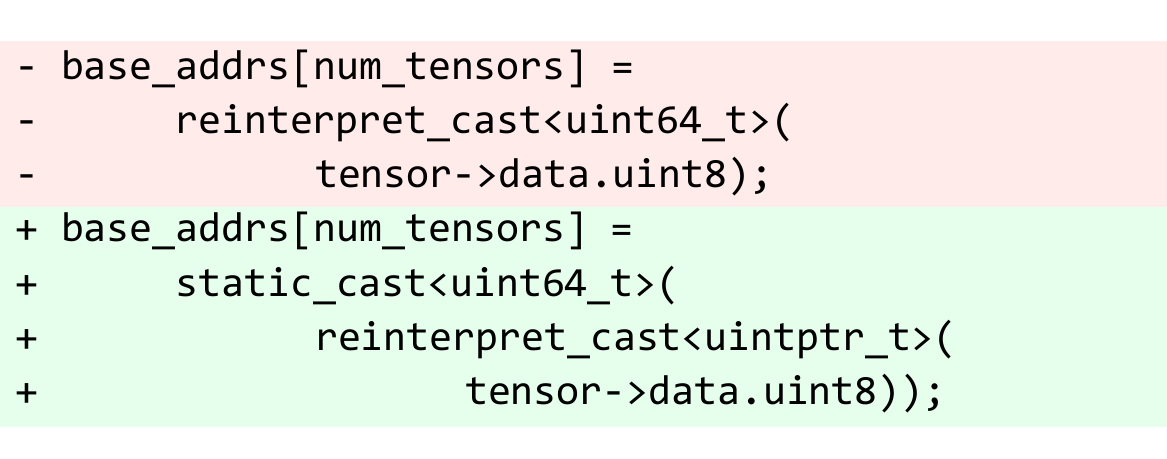}
    }
    \subfigure[{\footnotesize Hang (MXNet\#19349)}]{
        \label{fig:ex_hang}
        \includegraphics[width=0.475\linewidth,trim=0 160 400 160,clip]{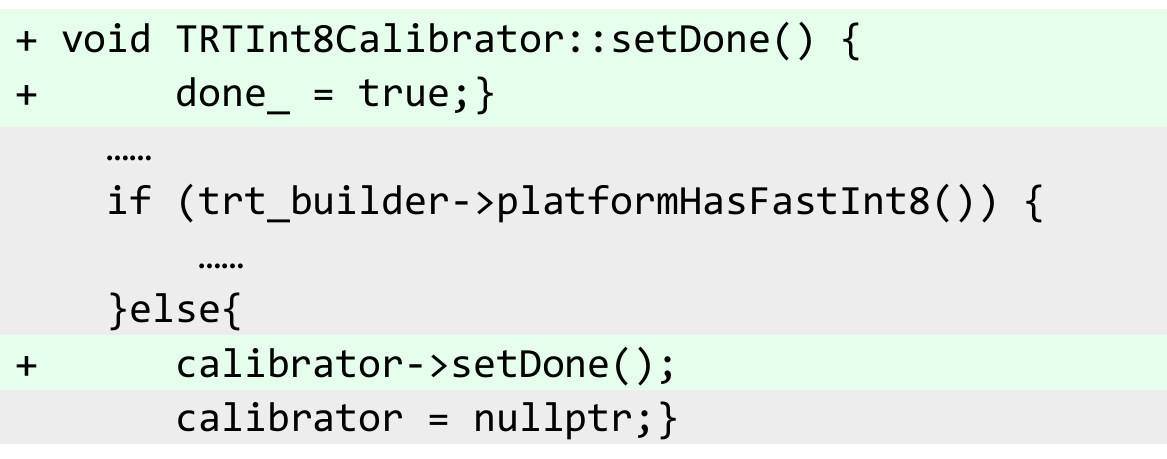}
    }
    \vspace{-3mm}
    \caption{\add{Bug Examples \ddel{in}\aadd{of Different} Symptoms}}
    \label{fig:examples2}
\end{figure}

\vspace{0.5mm}
\textcircled{2}\textbf{Incorrect Functionality}.
This symptom refers to that a DL framework behaves incorrectly but does not crash, such as producing unexpected prediction results, unexpected model structures, or incorrect intermediate states.
\add{Figure~\ref{fig:ex_function} presents such an example, which produces incorrect intermediate results due to an unsafe pointer conversion to {\tt uint64\_t} since GCC prevents using addresses at 0x80000000 or above\footnote{\add{https://github.com/tensorflow/tensorflow/pull/46509}}.}


\vspace{0.5mm}
\textcircled{3}\textbf{Build Failure}.
This symptom refers to that a DL framework fails to be installed,
\add{such as the previously introduced example shown in Figure~\ref{fig:miscon}.}

\vspace{0.5mm}
\textcircled{4}\textbf{Poor Performance}.
This symptom refers to that the spent time or consumed resource (such as memory) is much larger than expectation during the usage of a DL framework.
\add{As mentioned above, the API misuse shown in Figure~\ref{fig:ex_apim} may cause about 50x decline of running efficiency.}


\vspace{0.5mm}
\textcircled{5}\textbf{Hang}.
This symptom refers to that a DL program written on top of a DL framework cannot terminate within a long period of time (due to a DL framework bug).
\add{Figure~\ref{fig:ex_hang} presents such an example from MXNet\footnote{\add{https://github.com/apache/incubator-mxnet/pull/19349}}. Specifically, before nullifying the pointer {\tt calibrator}, the flag {\tt done\_ } denoting the finish of the process was not properly set, causing that it cannot correctly terminate.}


\vspace{0.5mm}
\textcircled{6}\textbf{Unreported}.
We cannot identify the symptoms for some bugs
after carefully reading the corresponding pull requests, including the related issues, discussions, and code changes.

\begin{figure}[t]
    \centering
    \includegraphics[width=\linewidth]{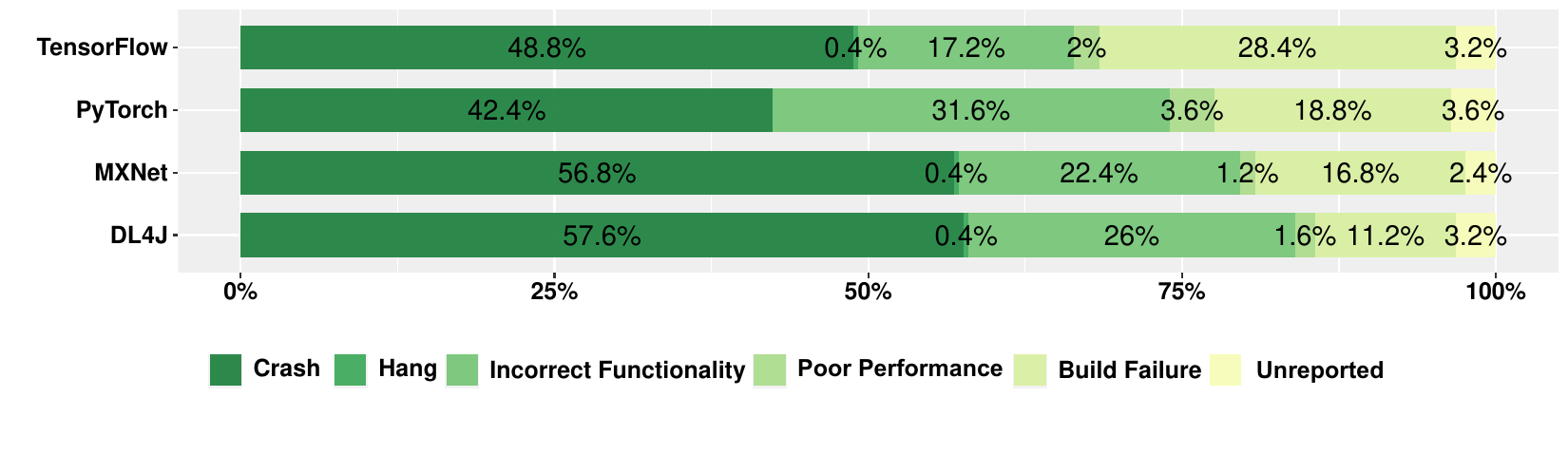}
    \vspace{-6mm}
    \caption{Bug Distribution by Symptoms}
    \label{fig:symptom_distribute}
\end{figure}

\subsubsection{Symptom Distribution}
\label{sec:symptom_distribution}
Figure~\ref{fig:symptom_distribute} shows the bug distribution of the identified symptoms.
We found that Crash is the most common symptom.
The number of bugs exhibiting this symptom is 122, 106, 142, and 144 for TensorFlow, PyTorch, MXNet, and DL4J respectively, and the total number is 514.
The detection of this kind of bugs has an explicit test oracle, and thus automated test input generation (that tends to suffer from the test oracle problem but does not here) has a great potential to facilitate the detection of the large percentage of Crash bugs.
Also, Crash bugs occur with error messages, which can provide hints for the bugs, and thus designing effective debugging techniques based on those informative messages is beneficial for such a large percentage of Crash bugs.

\find{Crash is the most common symptom for DL framework bugs, which accounts for 51.40\% of bugs.}

Incorrect Functionality takes the second place, which accounts for 243 bugs in total, including 43 in TensorFlow, 79 in PyTorch, 56 in MXNet, and 65 in DL4J.
The major challenge for detecting this kind of bugs lies in the test oracle problem since the symptom is not as obvious as Crash.
Specifically, a DL framework is used by developers for implementing a DL program and then building a DL model, but it is difficult to determine the correctness of the DL program/model due to its complexity.
Hence, the adverse impact of Incorrect Functionality bugs is severe due to the weak observability. 
Through analyzing the large number of studied Incorrect Functionality bugs, we found that they were often detected by checking whether the prediction results of the built model, the model structure, or some intermediate states (e.g., the calculation results of some operations) are as expected.
More specifically, among the 243 bugs, 115 of them produced incorrect intermediate states, \aadd{105 of them produced incorrect prediction results, and 23 of them caused incorrect model structures.}
\aadd{The 115 bugs producing incorrect intermediate states include} \ddel{including} 18 in TensorFlow, 46 in PyTorch, 17 in MXNet, and 34 in DL4J. 
We further investigated the possible reason why PyTorch produces more incorrect intermediate state bugs, and found that most of such bugs were introduced for optimizing the ONNX-related component in PyTorch, which involves many parameters and thus may be easy to be incorrectly modified. 
\aadd{Since most of Incorrect Functionality bugs in PyTorch produce incorrect intermediate states, this also causes that PyTorch has more Incorrect Functionality bugs than the other three DL frameworks.}\ddel{Besides, another 105 bugs produced incorrect prediction results, while the remaining 23 bugs caused incorrect model structures.} Therefore, \del{Deciding}\add{deciding} which information should be observed and how to determine its expected result are important but indeed challenging to effectively detect this kind of bugs.

\find{Incorrect Functionality is the second most common symptom for DL framework bugs, accounting for 24.30\% of bugs. 
Defining effective test oracles deserves much more attention for the detection of this kind of bugs.}

\begin{table}[]
\centering
\caption{Bug Distribution by Symptoms in Each Stage.}
\vspace{-2mm}
\begin{adjustbox}{width=0.90\columnwidth,center}
\begin{tabular}{l||c|c|c|c|c|c}
\toprule
\textbf{\diagbox{Symptoms}{Stages}}
& 
  \textbf{Installation} &
  \textbf{Preprocessing} &
  \textbf{Training} &
  \textbf{Deployment} &
  \textbf{Utility Operation} &
  \textbf{\textit{Total}}\\ \midrule
Crash &2 &20 &350 &65 &77 & 514\\
Incorrect Functionality &4 &9 &154 &28 &48 & 243\\
Build Failure &166 &0 &13 &5 &4 & 188\\
Poor Performance &1 &2 &14 &2 &2 & 21\\
Hang &0 &0 &2 &1 &0 & 3\\
Unreported &0 &1 &18 &9 &3 & 31\\\midrule
\textit{Total} &173 &32 &551 &110 &134 & 1,000\\
\bottomrule
\end{tabular}
\end{adjustbox}
\label{tab:symptom_stage}
\end{table}

\add{In addition, both Poor Performance and Hang are rare for DL framework bugs, indicating that functional bugs are generally more common than performance-related bugs for DL frameworks.
It is also consistent with the existing studies on bugs of other software~\cite{dlcompiler,garcia2020comprehensive,di2017comprehensive,islam2019comprehensive}.
Besides, there are some DL framework bugs whose symptoms are not provided in the pull requests (including the related issue reports).
This is also as expected, because different developers may have different styles to describe bugs.}

Based on the symptoms, we then analyzed when we can observe these bugs.
From \textit{the view of DL framework users}, we classified the DL pipeline into five stages following the existing work~\cite{islam2019comprehensive,9113719,hapke2020building}:
\textcircled{1}\textbf{Installation}: the stage of installing the DL framework;
\textcircled{2}\textbf{Preprocessing}: the stage of preprocessing the dataset used for model building;
\textcircled{3}\textbf{Training}: the stage of training and validating a model;
\textcircled{4}\textbf{Deployment}: the stage of deploying the built model to a device;
\textcircled{5}\textbf{Utility Operation}: the stage of conducting auxiliary operations, e.g., model visualization.
\aadd{To determine the stage in which a bug can be observed, we carefully understood the description about the bug in the related issue, the discussion among developers in the pull request, and the bug-fixing code changes.}
Table~\ref{tab:symptom_stage} shows the bug distribution according to the stage in which the bugs with each symptom were observed.
From the view of DL framework users, DL framework bugs are mainly observed at the stage of Training (i.e., 55.10\%).
The Training stage tends to be time-consuming due to heavy numerical computation based on a large amount of training data.
As presented in the existing study~\cite{yan2021exposing}, the typical training time ranges from a few minutes to several days.
Hence, the bugs observed at this stage, especially those Incorrect Functionality bugs (account for 27.95\% of bugs observed at this stage), may be manifested after hours or even days into the training process.
This is very harmful to the efficiency of both testing and debugging.
In particular, the training process for exposing the bugs has to be repeated several times to validate whether a fix is correct, and meanwhile the training process involves randomness that further aggravates the difficulty of bug reproduction\aadd{~\cite{yan2021exposing}}.
Hence, the large percentage of bugs observed at the Training stage suggests the urgent need of speeding up the process of exposing bugs at this stage.

\find{About 55.10\% of DL framework bugs are observed at the Training stage.
It could lead to lengthy testing and debugging for them, especially the large number of Incorrect Functionality bugs without halfway crashes, due to the costly and non-deterministic training process.}

\subsection{RQ3: Relationship between Root Causes and Symptoms}
\label{sec:rootcause_symptom}

Table~\ref{tab:rootcause_symptom} presents the number of each kind of bugs \add{(including each sub-category of a root cause)} exhibiting each symptom.
Here, Crash and Incorrect Functionality are the most common symptoms for all the root causes (except Misconfiguration for both, and Environment Incompatibility, API Incompatibility, Dependent Module Issue for the latter).
The result indicates designing effective test oracles targeting the two symptoms is helpful to detect a wide variety of DL framework bugs.
As presented before, Crash has an explicit test oracle, while the test oracle problem is the major challenge for detecting Incorrect Functionality bugs.
Currently, differential testing has been adopted as the test oracle for the latter~\cite{pham2019cradle,wang2020deep}, but it could lead to false positives and false negatives due to the \textit{randomness} in DL (which is different from traditional software).
Hence, more precise test oracles are still desirable.

\add{
Intuitively, Type Issue and Tensor Shape Misalignment usually lead to Crash rather than Incorrect Functionality.
However, from Table~\ref{tab:rootcause_symptom}, the two root causes can also lead to many Incorrect Functionality bugs.
For better understanding this phenomenon, we thus present two examples.
The first one is a TensorFlow bug caused by Type Issue\footnote{\add{https://github.com/tensorflow/tensorflow/pull/46509}} (shown in Figure~\ref{fig:ex_function}).
This bug occurs due to unsafe type conversion from pointer to {\tt uint64\_t} in Ethos-U kernel, since GCC prevents using addresses at 0x80000000 or above.
Such unsafe conversion causes that the value becomes a wrong value, and thus leads to Incorrect Functionality.
The patch is to first convert the pointer to {\tt uintptr\_t} and then convert {\tt uintptr\_t} to {\tt uint64\_t}.
The second example is a PyTorch bug caused by Tensor Shape Misalignment\footnote{\add{https://github.com/pytorch/pytorch/pull/38583}}.
This bug occurs due to performing Conv2d non-zero padding in wrong dimensions, leading to wrong data values for subsequent calculation.
Hence, it also leads to Incorrect Functionality.}

Regarding the symptoms of Build Failure and Poor Performance, they can be produced by some specific root causes.
Specifically, among the 188 bugs exhibiting the symptom of Build Failure, 65.43\% are produced by Misconfiguration and 13.30\% are produced by Environment Incompatibility.
Among the 21 bugs exhibiting the symptom of Poor Performance, 52.38\% are produced by Incorrect Algorithm Implementation or API Misuse.
\add{Among the 6 Poor Performance bugs caused by API Misuse, all of them are caused by \textit{Wrong API}. In particular, Figure~\ref{fig:ex_apim} shows an example for further illustrating how Wrong API leads to Poor Performance.}
Therefore, when a bug occurs with the two symptoms, developers can first check these highly relevant root causes to speed up the debugging process.

\find{The symptom of Build Failure is highly relevant to the root causes of Misconfiguration and Environment Incompatibility, while the symptom of Poor Performance is highly relevant to the root causes of Incorrect Algorithm Implementation and API Misuse.}

\begin{table}[]

\centering
\caption{\add{Bug Distribution by Symptom for Each Sub-Root-Cause}}
\vspace{-2mm}
\label{tab:rootcause_symptom}
\begin{adjustbox}{max width=0.99\textwidth,center}
\begin{threeparttable}
\begin{tabular}{l|l||c|c|c|c|c|c|c}
\toprule
\multicolumn{2}{c||}{\textbf{\diagbox{Root Cause}{Symptom}}} & 
\textbf{Crash} & 
\textbf{Incorrect Functionality} &
\textbf{Build Failure} & 
\textbf{Poor Performance} &
\textbf{Hang} & 
\textbf{Unreported} &
\textbf{Total} \\
\midrule
& DL-related   & 75 & 32  & 3  & 4 &0  & 7 \\
\multirow{-2}{*}{IAI} & DL-unrelated & 19 & 8  & 0  & 1 & 0 & 3  & \multirow{-2}{*}{152} \\\cline{1-9}
& Tensor Type Issue  & 68  & 23  & 1   & 2   & 0 & 6  \\
\multirow{-2}{*}{TI}  & Conventional Type Issue & 27 & 9  & 3 & 1  & 2  & 0 &\multirow{-2}{*}{142} \\\cline{1-9} 
MC &Misconfiguration &8 &3 &123 &0 &0 &0 &134\\\cline{1-9}
TSM &Tensor Shape Misalignment &80 &39 &1 &0 &0 &2 &122\\\cline{1-9}

& Condition M/R
  & 6  & 5  & 2  &0  &0 &  0  \\
& API M/R & 13   & 7     & 0     &0    &  0   & 1    \\
& Wrong API Args  & 18  & 9  & 1& 1  &0  & 5 \\
& Wrong API Name & 14 & 11 & 3  & 3 & 0 & 3\\
\multirow{-5}{*}{AM} & Wrong API Receiver & 9 & 4 & 1 & 2 &  0  & 0 &\multirow{-5}{*}{118} \\\cline{1-9}
EI &Environment Incompatibility &49 &8 &25 &1 &1 &2 &86\\\cline{1-9}
&Missing Exception & 12 & 9   & 0 & 0 &0 &0 \\
& Spurious Exception& 2  & 1   & 0   &0  & 0    &0   \\
\multirow{-3}{*}{IEH} & Wrong Exception Message & 18 & 13 &0 & 0 &0 &0 & \multirow{-3}{*}{55} \\\cline{1-9}
IA &Incorrect Assignment &27 &22 &3 &1 &0 &0 &53\\\cline{1-9}
NI &Numerical Issue &11 &21 &1 &1 &0 &0 &34\\\cline{1-9}
Others &Others &14 &9 &8 &1 &0 &1 &33\\\cline{1-9}
& External \aadd{Incompatibility}& 10  & 3 & 7   & 1  & 0 &0  \\
\multirow{-2}{*}{AI}  & Internal \aadd{Incompatibility} & 7  & 1  &0 &0  & 0  & 0 &\multirow{-2}{*}{29}\\\cline{1-9}
CI &Concurrency Issue &17 &6 &0 &2 &0 &1 &26\\\cline{1-9}
DMI &Dependent Module Issue &10 &0 &6 &0 &0 &0 &16\\
\bottomrule
\multicolumn{2}{l||}{\textbf{Total}} &514 &243 &188 &21 &3 &31 &1000\\\cline{1-9}
\end{tabular}

\begin{tablenotes}
\item {\footnotesize 
    \textbf{IAI}: Incorrect Algorithm Implementation; 
    \textbf{TI}: \del{Tensor} Type Issue; 
    \textbf{MC}: Misconfiguration;
    \textbf{TSM}: Tensor Shape Misalignment;
    \textbf{AM}: API Misuse; 
    \textbf{Condition M/R}: Condition missing or redundancy;
    \textbf{API M/R}: API missing or redundancy;
    \textbf{EI}: Environment Incompatibility;
    \textbf{IEH}: Incorrect Exception Handling};
    \textbf{IA}: Incorrect Assignment;
    \textbf{NI}: Numerical Issue;
    \textbf{AI}: API Incompatibility;
    \textbf{CI}: Concurrency Issue;
    \textbf{DMI}: Dependent Module Issue
\end{tablenotes}
\end{threeparttable}
\end{adjustbox}
\end{table}


\subsection{RQ4: Bug-Occurring Levels}
\label{sec:level}
We analyzed the distribution of each kind of bugs over different levels of DL frameworks in Table~\ref{tab:root_cause_comp}.
\del{Here, we excluded Misconfiguration bugs or the bugs caused by external configuration files, since our five-level architecture of DL frameworks does not contain configuration files and almost all the Misconfiguration bugs occur in configuration files.}
From this table, the number of bugs occurring at the level of Operation Implementation (i.e., 260) is the largest. 
It is reasonable since this level includes hundreds or even thousands of algorithms, which implement the complicated functionalities of DL models (e.g., gradient calculation), and always involve a significant amount of source code.
In contrast, the number of bugs occurring at the level of Environment-Dependent Processing is the smallest since this level involves the least amount of code.
\add{Also, among the five levels, Graph-Level Implementation and Operation Implementation are more core than the other three, since both the training process and the inference process in DL are based on a static or dynamic computational graph and each node in the graph is an operation.
From Table~\ref{tab:root_cause_comp}, 435 bugs occur at the two levels.}
The results indicate that it is urgent to conduct more extensive testing on the level of Operation Implementation (such as improving its test coverage) in terms of bug detection, in order to sufficiently improve the reliability of DL frameworks.


\find{
The level of Operation Implementation contains the most bugs, accounting for 30.77\% of the bugs, while the level of Environment-Dependent Processing contains the fewest bugs.
}

Further, the bugs caused by the four root causes \del{(i.e., TI, TSM, IAI and EI in Table~\ref{tab:root_cause_comp})} involving DL framework characteristics \add{(i.e., IAI, TI, TSM and EI in Table~\ref{tab:root_cause_comp})} are chiefly distributed \del{in} \add{at} the level of Operation Implementation, while the common root-cause categories of bugs are chiefly distributed \del{in} \add{at} the level of User-Level API.
The results indicate that the level of User-Level API is more similar to traditional software and the level of Operation Implementation is more specific to DL frameworks.
Therefore, it is likely to apply existing testing and debugging techniques to the former level, which may facilitate to ensure its quality to a large degree, while new testing and debugging techniques targeting DL framework characteristics are desirable for the latter level.
Regarding the level of Graph-Level Implementation, Incorrect Algorithm Implementation (IAI) and Tensor Shape Misalignment (TSM) are two major causes, since it involves many DL-specific algorithms (such as various graph transformation algorithms) and takes tensors as the basic elements of computational graphs.
In particular, 46.85\% (82 out of 175) bugs in this level occur in Graph Transformation.
As expected, Environment Incompatibility is the major cause for the bugs \del{in} \add{at} the level of Environment-Dependent Processing.
Such different bug distribution characteristics in terms of root causes call for different testing and debugging techniques for different levels.

\find{
Different levels of DL frameworks involve different major root causes.
DL-specific bugs are chiefly distributed \del{in} \add{at} the level of Operation Implementation, while traditional categories of bugs are chiefly distributed \del{in} \add{at} the level of User-Level API.
}

\begin{table}[t]
\centering
\caption{\add{Bug Distribution by Root Causes in Levels}}
\vspace{-2mm}
\begin{adjustbox}{width=0.99\columnwidth,center}
\begin{threeparttable}
\begin{tabular}{l||c|c|c|c|c|c|c|c|c|c|c|c|c}
\toprule
\textbf{\diagbox[width=16em]{Components}{Root Causes}} & 
\textbf{\underline{IAI}}& \textbf{\underline{TI}}&     \textbf{\underline{TSM}}&   \textbf{AM}&      \textbf{\underline{EI}}&       \textbf{IEH}& \textbf{IA}&  \textbf{NI}&  \textbf{Others}&   \textbf{AI}& \textbf{CI}&   \textbf{DMI}&
  \textbf{\textit{Total}}\\
                                               
\midrule                                                                                     
User-Level API                     &29 &35    &16   &41     &7       &17 &17 &4     &8    &8 &6   &5     &193 \\
Graph-Level API                    &51 &21    &31   &20     &15      &9  &7  &4     &5    &3 &7   &2     &175\\
Operation Implementation           &50 &43    &47   &27     &18      &14 &19 &16    &11   &3 &8   &4     &260\\
General Implementation             &18 &37    &28   &24     &8       &14 &7  &9     &3    &5 &4   &1     &158\\
Environment-Dependent Processing   &4  &6     &0    &6      &34      &1  &3  &0     &3    &1 &1   &0     &59\\
\bottomrule
\end{tabular}
\begin{tablenotes}
\item[*] {\footnotesize \textbf{IAI}: Incorrect Algorithm Implementation; \textbf{TI}: \del{Tensor} \add{Type} Issue; \textbf{TSM}: Tensor Shape Misalignment; \textbf{AM}: API Misuse; \textbf{EI}: Environment Incompatibility; \textbf{IEH}: Incorrect Exception Handling; \textbf{IA}: Incorrect Assignment; \textbf{NI}: Numerical Issue; \textbf{AI}: API Incompatibility; \textbf{CI}: Concurrency Issue; \textbf{DMI}: Dependent Module Issue}
\item[*]{\footnotesize  \add{We excluded Misconfiguration bugs and the bugs caused by external configuration files in this table, since our five-level architecture of DL frameworks does not contain configuration files and almost all the Misconfiguration bugs occur in configuration files.}}
\item[*]{\footnotesize \aadd{We added the underline for the root causes involving DL framework characteristics.}}
\end{tablenotes}
\end{threeparttable}
\end{adjustbox}
\label{tab:root_cause_comp}
\end{table}

\subsection{RQ5: Bug Commonality}
\label{sec:commonality}
\begin{figure}
\centering
\subfigure[Symptom]{
\includegraphics[width=0.3\linewidth]{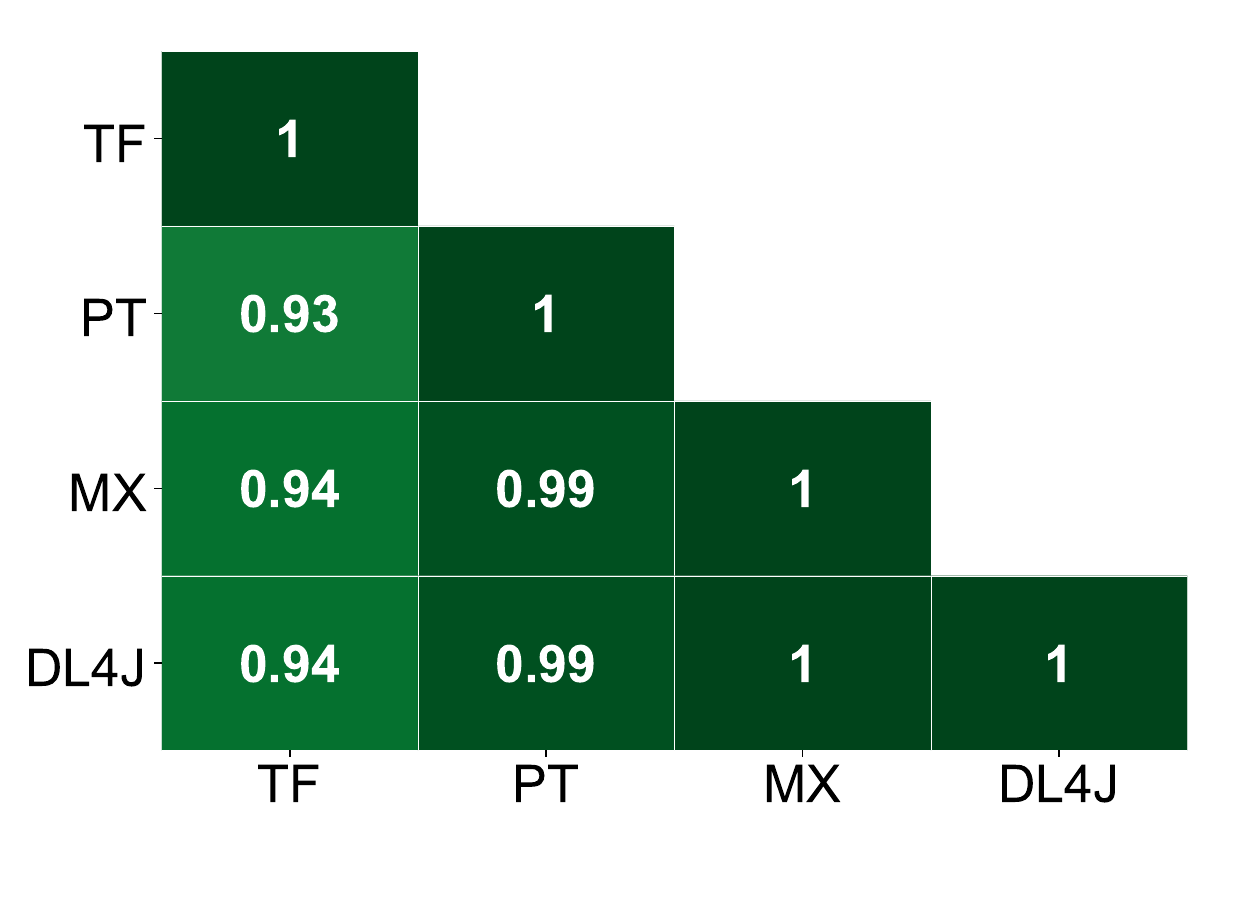} 
}
\subfigure[Root Cause]{
\includegraphics[width=0.3\linewidth]{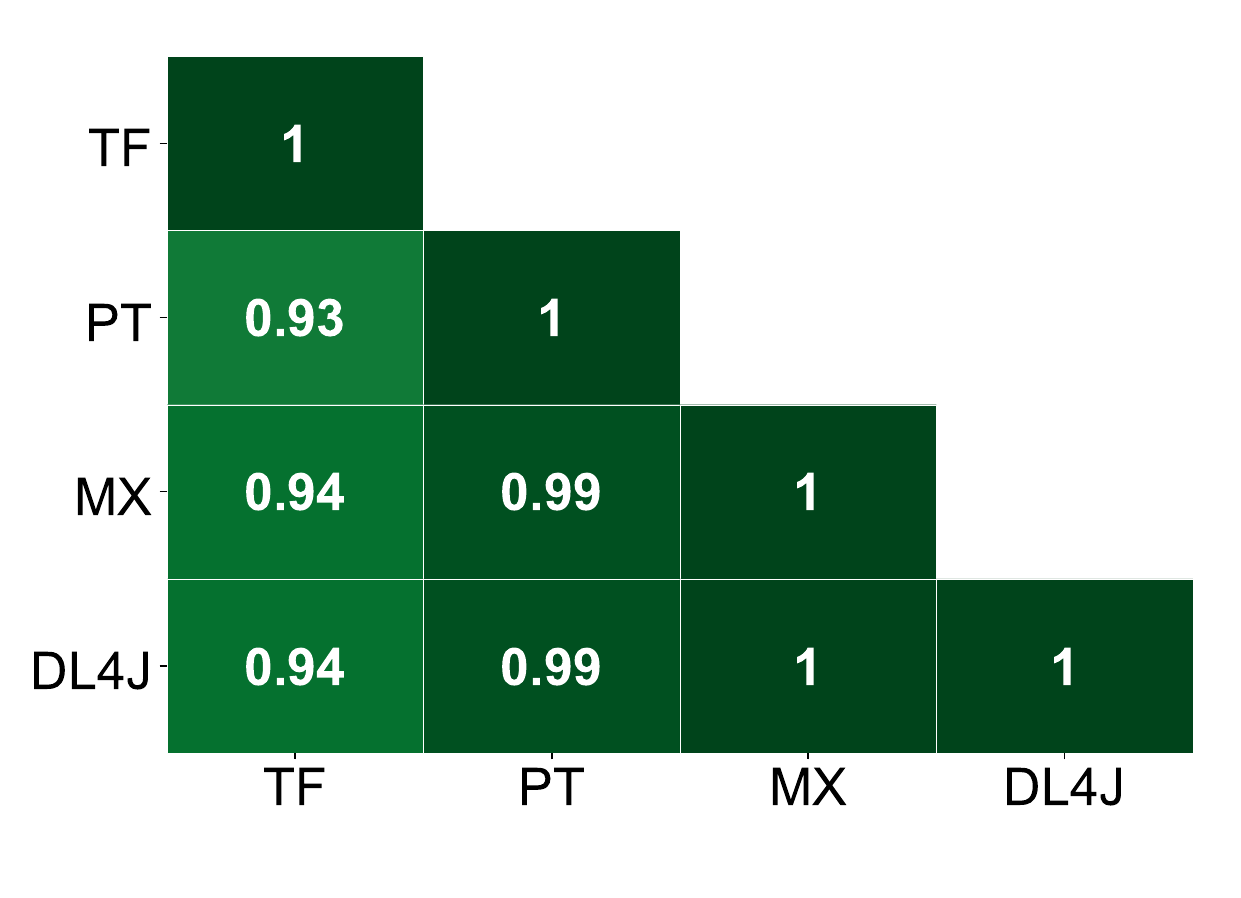} 
}
\subfigure[Component]{
\includegraphics[width=0.3\linewidth]{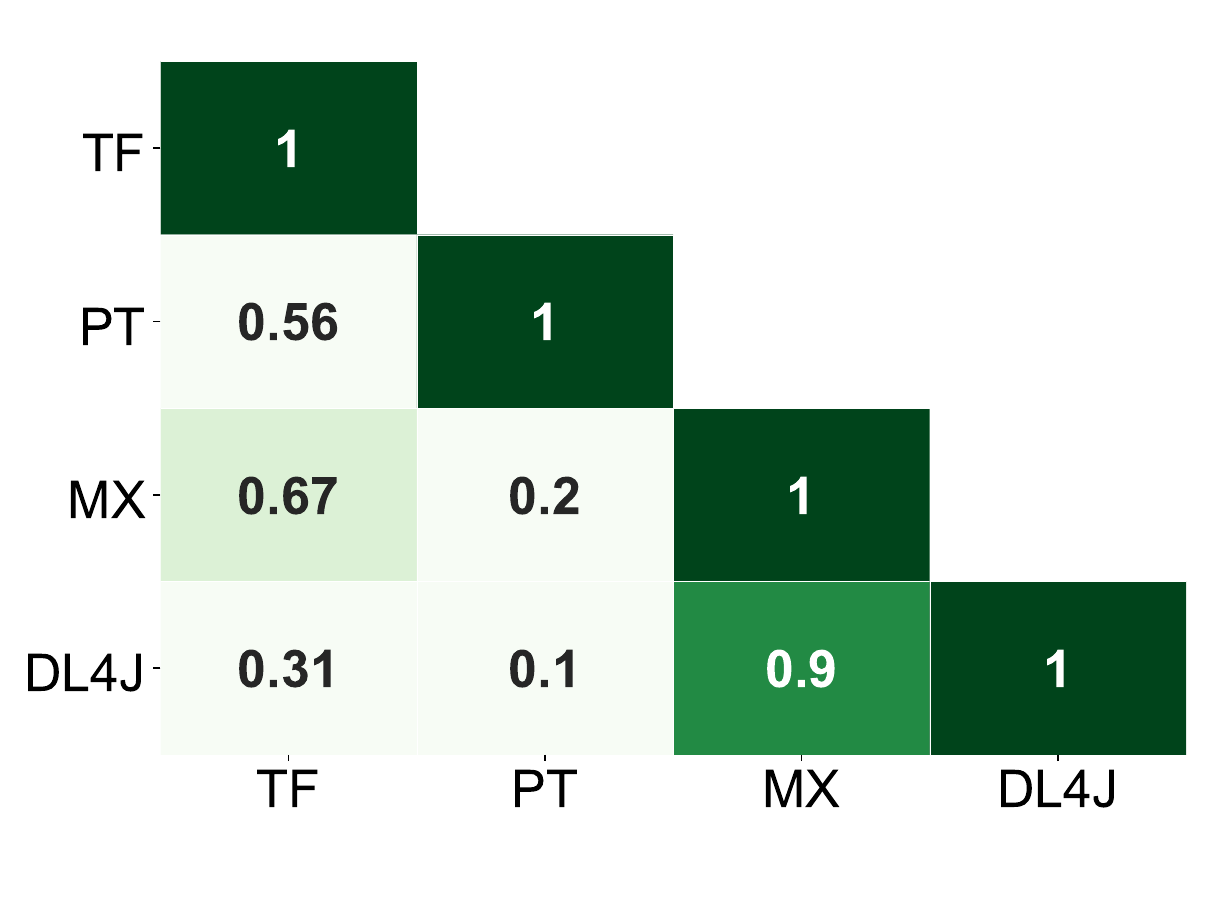}
}
\DeclareGraphicsExtensions.
\caption{Spearman Correlation across DL Frameworks}
\label{fig:spearman}
\end{figure}

In order to measure the bug commonality across different DL frameworks,
following \add{the} existing work~\cite{lou2019history,zhong2021understanding},
we calculated the Spearman correlation between each pair of DL frameworks in terms of bug distributions from the perspectives of root causes, symptoms and components respectively.
Spearman’s correlation coefficient is a statistical measure of the strength of a monotonic relationship between two paired variables~\cite{inbook}.
Figure~\ref{fig:spearman} shows the correlation results, where [$0.8$,$1.0$] indicates the very strong correlation, [$0.6$,$0.8$) indicates the strong correlation, [$0.4$, $0.6$) indicates the moderate correlation, while [$0$, $0.4$) indicates the weak or no correlation.
From this figure, when considering the bug distributions over different root causes and symptoms, all the correlation coefficients are larger than 0.8.
The results show that regardless of root causes or symptoms, the four DL frameworks share a high degree of the commonality, demonstrating the generality of our findings in the study and the potential of developing general testing and debugging techniques for various DL frameworks. 
\add{For example, from Figure~\ref{fig:rootcause_distribute}, Top-5 root causes for all the four DL frameworks are the same, i.e., Incorrect Algorithm Implementation, Type Issue, Misconfiguration, Tensor Shape Misalignment, and API Misuse.}
\add{Also, from Figure~\ref{fig:symptom_distribute}, Crash and Incorrect Functionality are more common than the other three symptoms for all the four DL frameworks.}
However, these frameworks tend to display diverse bug distributions over different components as shown in the figure. The major reason lies in the different designs and implementations of DL frameworks. For example, DL4J implements many User-Level APIs to bridge the gap between different frameworks (e.g., it can load models trained in Keras) so as to be compatible with the others, while PyTorch takes many efforts for constructing a uniform model structure (e.g., ONNX) for convenient optimization.
Please note that it does not contradict with the conclusions in Findings~9 and 10. 
Through a more fine-grained analysis and comparison, although the four DL frameworks do not consistently share the same bug distribution over different components, the bugs in Operation Implementation level still take a \del{significant} \add{significantly} large portion for each individual DL framework. Particularly, MXNet shares considerable commonalities (with at least moderate correlation) with other frameworks except for PyTorch. In summary, the results suggest that future research aiming to detect DL-framework bugs should consider the commonalities of bug symptoms and root causes so as to make the approaches have better generalizability over different frameworks.


\find{There is a significant commonality among the four DL frameworks in both root causes and symptoms.}

\section{Implications and The Application}

In this section, we explain the implications we have learned from our study to facilitate future research on DL framework bugs. In order to provide a more intuitive an targeted analysis, we have conducted an empirical study to investigate the \del{current} status of existing testing techniques. In the end, based on the implications, we have developed a prototype testing tool aiming at finding DL framework bugs, and evaluated its effectiveness in a preliminary experiment.

\subsection{Status of Existing Testing Techniques}
\label{sec:coverage}
To investigate the \del{current} status of existing testing techniques, we analyzed them in terms of test coverage on each DL-framework component.
Here, we studied three \add{typical} DL framework testing techniques, i.e., CRADLE~\cite{pham2019cradle}, LEMON~\cite{wang2020deep}, Audee~\cite{guo2020audee}.
\add{There are also some other DL framework testing techniques and more details about them can be found in Section~\ref{sec:related_work}.
Also, the selection of studied techniques may be a threat in this experiment, which will be discussed in Section~\ref{sec:dis}.}
All of \del{them} \add{the studied techniques} adopt differential testing as the test oracle.
Their main difference lies in the used test inputs:
CRADLE is the first technique, which takes real-world pre-trained DL models as test inputs;
LEMON and Audee adopt different search-based mutation strategies to generate mutated models based on pre-trained models as test inputs, where the former proposes to mutate the layers, neurons, and weights of pre-trained models and the latter proposes to mutate the parameters of layers, weights, and inputs (e.g., images).

Here, we used 8 pre-trained models widely-used in the existing studies~\cite{wang2020deep,guo2020audee}, involving different model structures and different sets of input data.
They are LeNet-5 trained on MNIST, LeNet-5 trained on Fashion-MNIST, AlexNet trained on Cifar10, MobileNetV2, ResNet-50, and VGG-16 trained on ImageNet, and two LSTM models trained on Sinewave and Price.
CRADLE uses the 8 models as test inputs directly, while LEMON and Audee produced 100 mutated models based on each pre-trained model respectively and the latter also produced mutated input data for each mutated model.
In total, there are 800 mutated models as test inputs for LEMON and Audee, respectively.
We 
measured the achieved test coverage (i.e., line, branch, and function coverage) by the three techniques respectively via \textit{Gcov} (for C code coverage collection)~\cite{gcov_tool} and \textit{Coverage.py} (for Python code coverage collection)~\cite{coverage_python}.
\ddel{Since collecting DL framework coverage is costly and different DL frameworks share a significant bug commonality as presented in Section~\ref{sec:commonality}, we used MXNet as the representative in this experiment since it shares the largest commonality with other three frameworks regarding bug distribution in different components}\aadd{Since collecting DL framework coverage is costly, we used MXNet and PyTorch as the representatives in this experiment. 
MXNet, TensorFlow, and DL4J share a significant bug commonality as presented in Section~\ref{sec:commonality}, and thus we used one of them (i.e., MXNet) as the representative\footnote{\aadd{Handling various environment/configuration/dependency issues for collecting code coverage of DL frameworks is challenging. 
We tried our best to handle these issues for collecting TensorFlow’s code coverage in order to make the experiments in Sections~\ref{sec:coverage} and~\ref{sec:application} use the same DL framework (i.e., TensorFlow) as the subject, but unfortunately failed due to a version incompatibility bug regarding Bazel~\cite{bazel} and Gcov.
In the future, we will try to solve this problem to obtain TensorFlow code coverage results for more sufficient evaluation.}}. 
To ensure the generality of conclusions, we also used PyTorch as another subject in this experiment.}
In particular, we also ran the equipped test suites in MXNet (version 1.9.0) \aadd{ and PyTorch (version 1.9.0) }and collected the achieved coverage to facilitate analysis.

We first measured the coverage results achieved by the three testing techniques together, and also compared them with the coverage result achieved by the equipped test suite, whose results are shown in Table~\ref{tab:basic_cov}.
We found that the line, branch, and function coverage achieved by these testing techniques are only 24.19\%, 4.58\% \ddel{and 23.38\%}\aadd{, 23.38\% on MXNet and 10.12\%, 1.30\%, 7.11\% on PyTorch} respectively, which are significantly smaller than those achieved by the equipped test suite (i.e., 70.51\%, 11.88\%, \ddel{and 38.99\%}\aadd{, 38.99\% on MXNet and 64.34\%, 23.60\%, 49.81\% on PyTorch}).
That is, the \del{current} \add{studied} testing techniques suffer from the low test coverage issue.
It is very harmful to the testing performance since test coverage is the first condition of bug detection according to the PIE theory~\cite{pie}.
Hence, improving test coverage is an important direction of designing new DL framework testing techniques.
Also, from Table~\ref{tab:basic_cov}, the \del{current} \add{studied} testing techniques tend to achieve relatively high test coverage on the components of User-Level API and Graph-Level Implementation (especially function coverage on the User-Level API component)\ddel{, but achieve low test coverage on} \aadd{compared with} the remaining three components.
The results suggest that focusing on the remaining three components could be more helpful to improve test coverage.

\find{The \del{current} \add{studied} DL framework testing techniques suffer from the low test coverage issue, especially on the components of Operation Implementation, General Utility, and Environment-Dependent Processing.}

\begin{table}[t]
\centering
\caption{\aadd{Coverage of Existing Testing Techniques and Equipped Test Suite}}
\begin{adjustbox}{max width=0.99\textwidth,center}
\begin{tabular}{l|l|l||c|c|c|c|c||r}
\toprule
\multicolumn{1}{l|}{Framework} & 
\multicolumn{2}{c||}{Component} &
  \textbf{User-Level API} &
  \textbf{Graph-Level Impl.} &
  \textbf{Operation Impl.} &
  \textbf{General Utility} &
  \textbf{Env.-Dep. Processing} &
  \textit{\textbf{Overall}} \\ \midrule
  {\multirow{6}{*}{MXNet}}            & 
\multirow{3}{*}{Test Suite} & Line   & 72.78\% & 68.75\% & 72.56\% & 65.58\% & 39.42\% & 70.51\% \\  
                           && Branch & 59.07\% & 29.72\% & 10.57\% & 17.11\% & 19.11\% & 11.88\% \\ 
                           && Function & 93.87\% & 62.23\% & 34.47\% & 51.54\% & 46.09\% & 38.99\% \\  \cmidrule(l){2-9}
\multicolumn{1}{c|}{} &\multirow{3}{*}{\begin{tabular}[c]{@{}c@{}}CRADLE+\\      LEMON+\\      Audee\end{tabular}} &
Line &  30.22\% &  38.37\% &  20.29\% &  17.79\% &  13.00\% &  24.19\% \\ 
& & Branch & 8.65\%  & 18.00\% & 4.05\%  & 3.81\%  & 7.57\%  & 4.58\%  \\ 
& & Function & 91.51\% & 37.54\% & 18.65\% & 20.56\% & 15.33\% & 23.38\% \\ \midrule

{\multirow{6}{*}{\aadd{PyTorch}}}            & 
\multirow{3}{*}{\aadd{Test Suite}} & \aadd{Line} & \aadd{64.43\%} & \aadd{67.06\%} & \aadd{70.51\%} & \aadd{55.97\%} & \aadd{48.25\%} & \aadd{64.34\%} \\
 &  & \aadd{Branch} & \aadd{54.86\%} & \aadd{21.82\%} & \aadd{25.17\%} & \aadd{24.00\%} & \aadd{13.20\%} & \aadd{23.60\%} \\
 &  & \aadd{Function} & \aadd{53.37\%} & \aadd{69.32\%} & \aadd{42.85\%} & \aadd{44.16\%} & \aadd{42.57\%} & \aadd{49.81\%} \\  \cmidrule(l){2-9}
\multicolumn{1}{c|}{} &\multirow{3}{*}{\begin{tabular}[c]{@{}c@{}}\aadd{CRADLE+}\\      \aadd{LEMON+}\\      \aadd{Audee}\end{tabular}} & 
\aadd{Line} & \aadd{15.04}\% & \aadd{7.71\%} & \aadd{4.39\%} & \aadd{15.34\%} & \aadd{2.96\%} & \aadd{10.12\%} \\
 &  & \aadd{Branch} & \aadd{7.00\%} & \aadd{0.90\%} & \aadd{0.50\%} & \aadd{3.43\%} & \aadd{0.55\%} & \aadd{1.30\%} \\
 &  & \aadd{Function} & \aadd{30.00\%} & \aadd{5.95\%} & \aadd{1.85\%} & \aadd{5.03\%} & \aadd{2.61\%} & \aadd{7.11\%} \\ \bottomrule
\end{tabular}
\end{adjustbox}
\label{tab:basic_cov}
\end{table}



\begin{figure}
\centering
\subfigure[Line coverage]{
\includegraphics[width=0.3\linewidth]{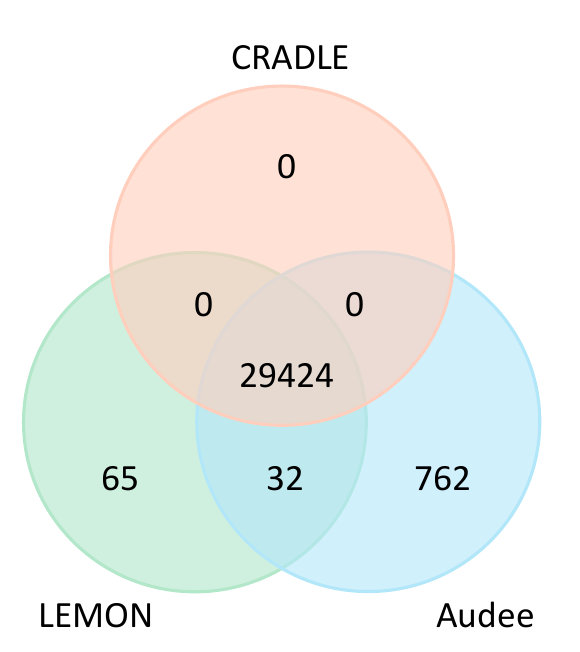} 
}
\subfigure[Branch coverage]{
\includegraphics[width=0.3\linewidth]{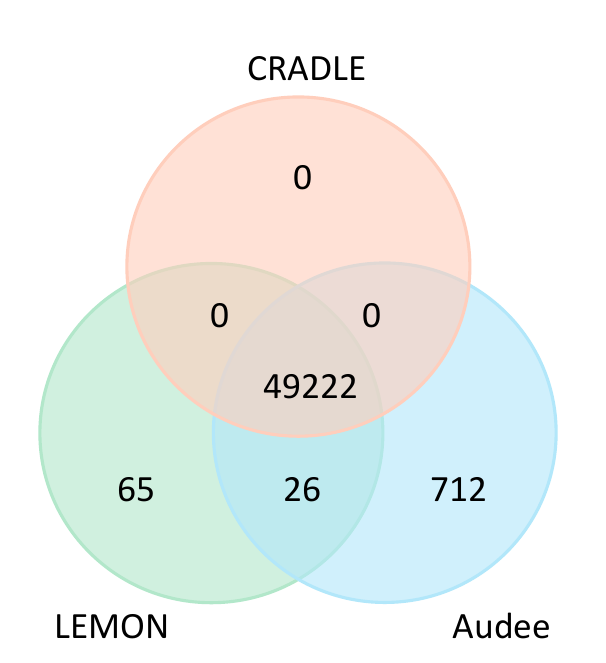}
}
\subfigure[Function coverage]{
\includegraphics[width=0.3\linewidth]{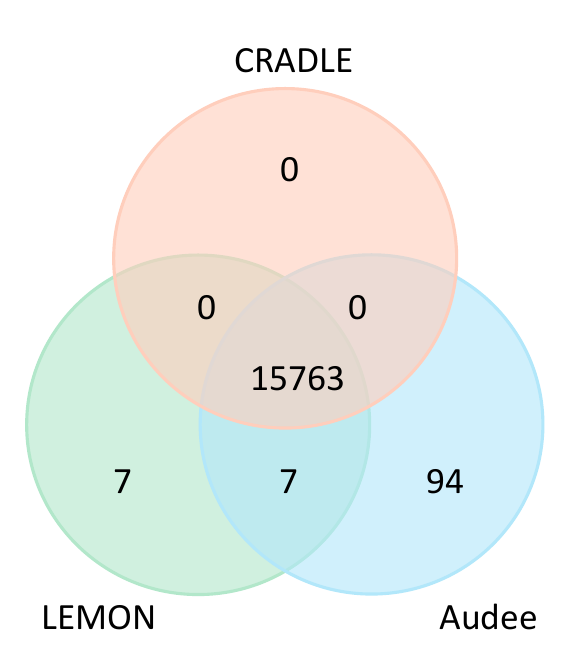}
}
\DeclareGraphicsExtensions.
\caption{Number of overlapping elements covered by different approaches \aadd{on MXNet}}
\label{fig:vn_unique_cov}
\end{figure}

\begin{figure}
\centering
\subfigure[\aadd{Line coverage}]{
\includegraphics[width=0.3\linewidth]{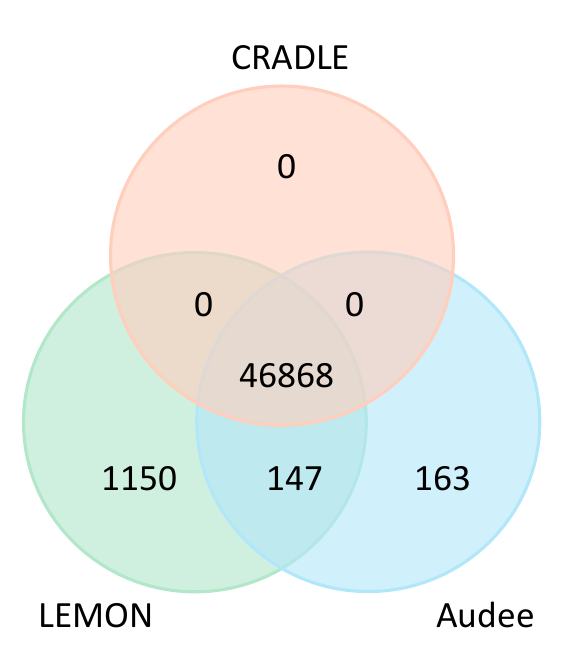} 
}
\subfigure[\aadd{Branch coverage}]{
\includegraphics[width=0.3\linewidth]{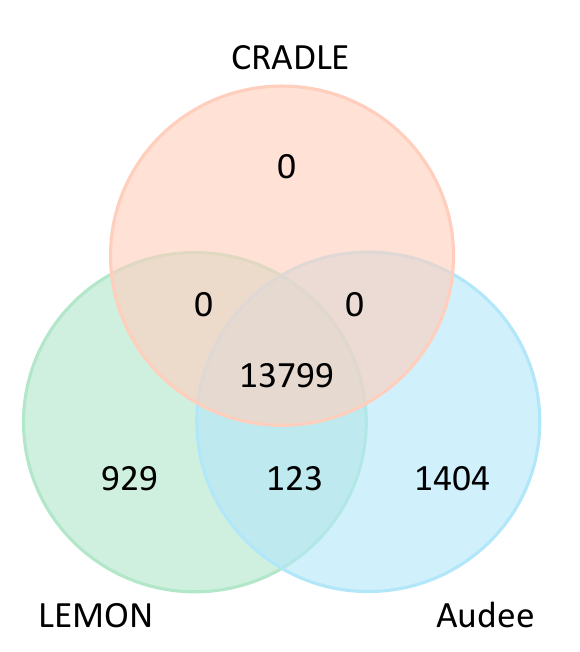}
}
\subfigure[\aadd{Function coverage}]{
\includegraphics[width=0.3\linewidth]{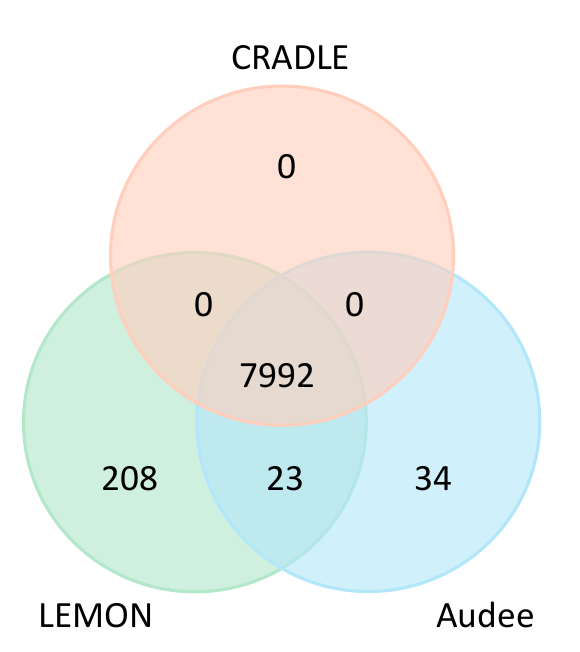}
}
\DeclareGraphicsExtensions.
\caption{\aadd{Number of overlapping elements covered by different approaches on PyTorch}}
\label{fig:vn_unique_cov_pt}
\end{figure}

We then compared the test coverage achieved by each of the three testing techniques.
Figure~\ref{fig:vn_unique_cov} \aadd{ and Figure~\ref{fig:vn_unique_cov_pt} show}\ddel{shows} the Venn diagrams to analyze the overlaps of their covered lines, branches, and functions.
We found that the number of unique lines, branches, and functions covered by each technique is small, especially compared with those that can be covered by all of them.
The results indicate that these techniques have the significant commonality in terms of test coverage.
In particular, both LEMON and Audee are on the basis of CRADLE, and according to Figure~\ref{fig:vn_unique_cov} \aadd{ and Figure~\ref{fig:vn_unique_cov_pt} }we found that their achieved test coverage mainly depends on the used pre-trained models and the coverage increments achieved by the mutated models by both LEMON and Audee are small.
\aadd{That means the diversity of the mutated models from the same pre-trained model is limited, and using more pre-trained models could increase the test diversity as well as the test coverage of the studied testing techniques.}
\ddel{That suggests that using more pre-trained models could facilitate to improve test coverage of the \del{current} \add{studied} testing techniques, and meanwhile}
\aadd{Moreover, }it is necessary to design new techniques with great diversity compared with these existing ones.

\find{The \del{current} \add{studied} DL framework testing techniques share a significant commonality in terms of test coverage and their coverage mainly depends on the used pre-trained models rather than mutated models by LEMON and Audee.}



\subsection{Implications}
According to the performance of existing state-of-the-art testing techniques (Section~\ref{sec:coverage}), there is still much room for improvement.
By further combining our findings, in this section we provide a series of actionable guidelines for future research on the detection and debugging of DL framework bugs.

\textbf{New mutation operators.}
Based on Findings 1 and 2, DL-involving root causes can result in a large percentage of bugs, and thus defining new mutation operators specific to their characteristics is helpful to efficiently explore whether DL frameworks can handle various cases involving them correctly.
New mutation operators can include:
1) \textit{type mutation}: many Type Confusion bugs are caused by type conversion, especially implicit type conversion, and thus we can \textit{add typecast for tensors} so that implicit type conversion may be triggered in tensor computation;
2) \textit{shape mutation}: we can create the scenarios, in which various tensor shapes can be involved to check whether they match, by \textit{inserting new layers with diverse shapes into different contexts};
3) \textit{environment mutation}: we can \textit{put a DL program into various environments for model building}, to test whether the used DL framework can stably support the training process.

\textbf{Test oracle improvement.}
Based on Findings 4, 5, 7, Crash and Incorrect Functionality are two most common symptoms for DL framework bugs, and thus designing effective test oracles with regard to them can cover a large percentage of bugs.
Regarding Crash, it has an explicit test oracle with error messages, but we still found many bug reporters complained that the error messages are ambiguous, which could affect the follow-up debugging process.
For example, among 55 bugs caused by Incorrect Exception Handling, 56.36\% are due to \textit{wrong exception messages}.
Hence, it is necessary for developers to refine error messages to make them precise and informative.
Regarding Incorrect Functionality, although differential testing on multiple DL frameworks has been adopted in the existing DL framework testing techniques by pre-defining a threshold for determining an inconsistency, it still cannot precisely identify Incorrect Functionality bugs due to inherent non-determinism in DL.
To reduce false positives and false negatives, a voting mechanism can be incorporated by integrating several test oracles, including differential testing on multiple versions of one DL framework as well as multiple environments, and metamorphic testing by constructing a group of equivalent tests.
Although integrating various test oracles may relieve the test oracle problem to some degree, new test oracles specific to such non-determinism definitely deserve more attention from the software engineering community.

\textbf{Component-targeted testing.}
In general, it is challenging to design a general testing technique that can effectively detect bugs occurring at various components, which can be demonstrated by \ddel{Findings 12 and 13}\aadd{Finding 11} to some degree (i.e., all these general testing techniques suffer from the low test coverage issue, especially on some components).
Hence, conducting component-targeted testing could be more practical.
According to Findings 9 and \ddel{12} \aadd{11}, we can assign the component of Operation Implementation the highest priority for designing targeted testing techniques, because this component involves the largest number of bugs but has little test coverage regardless of using existing testing techniques or the equipped test suite. Further analysis shows that only a small portion of operators are indeed covered by the existing approaches. However, since there are hundreds and even thousands of operators in a DL framework, the current testing is definitely insufficient. Testing tools targeting this component are urgently desired due to its core importance in the framework.
To achieve the targeted testing for the component of Operation Implementation, it is better to construct tests on the computational graph level, since it can more directly invoke and operate various operations to achieve high coverage compared with the widely-used model level by the existing testing techniques.

\textbf{Efficient reproduction.}
Based on Finding 6, over 55.10\% of bugs occur at the training process.
Since the training process involves heavy numerical computation based on a large amount of training data and inherent non-determinism, bug reproduction is unstable and time-consuming.
Therefore, efficient bug reproduction deserves much more attention.
One promising direction may be to shorten the training process by simplifying the model structure and reducing the amount of training data, which can still trigger the bug but with higher efficiency.
Here, adapting the idea of delta debugging~\cite{zeller2002simplifying} to both model and training data may be effective to achieve this goal.

\textbf{Build failure fixing.}
Based on Findings 3 and 8, Build Failure is common and has two highly relevant root causes, i.e., Misconfiguration and Environment Incompatibility.
Thus, there are hints for fixing building failures, making the design of automated methods feasible.
Indeed, there are some automated build failure fixing methods proposed for traditional software~\cite{hassan2018hirebuild,lou2019history}, but these methods tend to target the \textit{gradle} build framework~\cite{gradle.org}, which is different from the one depended by DL frameworks.
Moreover, as shown in Finding 3, Misconfiguration bugs in DL frameworks have different characteristics with those in traditional software \aadd{(i.e., there are a large number of configuration files/options for compilation, installation, and ensuring compatibility of DL frameworks due to their complex implementations involving multiple programming languages as well as the large number of dependent third-party libraries and hardware/software environments, leading to different proportions of Misconfiguration bugs)}.
Hence, not only investigating whether the existing methods still work on build failures of DL frameworks is valuable, but also designing new methods specific to the characteristics of DL frameworks is necessary.

\subsection{\tool{}: A Preliminary Application}
\label{sec:application}

\begin{table}[t]
\caption{Mutation Operators supported in \tool{}.}
\label{tab:mutator}
\begin{adjustbox}{max width=.99\columnwidth,center}
\begin{tabular}{l|c}
\toprule
\textbf{Mutation Operator} & \textbf{Brief Description}  \\ \midrule
\textit{tensor type mutation} &\makecell[l]{replacing the type of a tensor with another compatible type supported in TensorFlow.} \\\midrule
\textit{tensor shape mutation} &\makecell[l]{reshaping a tensor while keeping the data unchanged, e.g., changing the shape of \\a tensor from $3\times4$ to $2\times6$.
} 
\\\midrule
\textit{tensor structure mutation} &\makecell[l]{changing the structure of a tensor, e.g., changing the \texttt{tensor} to \texttt{sparse\_tensor} (that stores \\the non-zero values of the tensor and the corresponding coordinates of them) or \texttt{ragged\_tensor}\\ (that is the TensorFlow equivalent of nested variable-length lists).}   \\\midrule
\textit{tensor rotating} & \makecell[l]{rotating a tensor with a random angle $\theta$, where $\theta \in [30\degree, 270\degree]$ with the interval of $30\degree$. \\
Please note that we consider the tensors with the dimension no more than three.
}
\\\midrule
\textit{parameter mutation} &\makecell[l]{
changing the value of an API parameter to a special value, including the negation of \\the parameter value, \texttt{0}, \texttt{NaN}, and the maximum/minimum of the parameter.
}\\
\bottomrule
\end{tabular}
\end{adjustbox}
\end{table}

\add{
In this section, we demonstrate the usefulness of our findings with a preliminary proof-of-concept application \tool{}, which aims to generate tests for TensorFlow by mutating its equipped unit tests.
In the preliminary application, we selected TensorFlow as the representative due to its popularity.
\tool{} is designed based on two major findings:
(1) Bugs caused by tensors (especially tensor types and tensor shapes) are common;
\aadd{Specifically, the number of Tensor Type bugs is 100 (24 in TensorFlow, 19 in PyTorch, 22 in MXNet, 35 in DL4J) and the number of Tensor Shape Misalignment bugs is 122 (28 in TensorFlow, 33 in PyTorch, 27 in MXNet, 34 in DL4J).}
(2) The studied DL framework testing techniques (by generating DL models) achieve less coverage than the equipped unit tests on all the five components.
Hence, \tool{} takes the equipped unit tests in TensorFlow as the \aadd{initial pool of tests}\ddel{seeds} and designs five mutation operators on tensors (as presented in Table~\ref{tab:mutator}).
\aadd{These mutation operators can usually produce the exceptional cases of tensor type incompatibility or tensor shape misalignment, which can be helpful to more sufficiently test the ability of TensorFlow for handling these exceptional cases.}
Please note that for a given \ddel{seed}\aadd{test}, not all the mutation operators are applicable.
\aadd{This is because a test may not contain the elements required by each mutation operator (e.g., the mutation operator of negating the value of an API parameter cannot be applied when the test does not contain the APIs with numerical parameters).}

During the testing process with \tool{}, it first randomly selects \ddel{a seed}\aadd{a test from the pool} and then performs static analysis to determine a set of applicable mutation operators for this \ddel{seed}\aadd{test}.
\ddel{For example, for the mutation operator of replacing the parameter value with 0, \tool{} analyzes the type of the parameter to determine whether this operator is applicable.}
Further, it randomly selects an applicable mutation operator and then applies it to the \ddel{seed}\aadd{test} for generating a new test.
\aadd{That is, a new test is a mutant from the original test.}
\ddel{If the new test runs normally and does not detect a bug}\aadd{If the new test does not produce different outputs after normal executions on different versions of TensorFlow (we used differential testing on different versions of TensorFlow as the test oracle, which will be presented later)}, it will be put into the \ddel{seed}\aadd{test} pool for supporting high-order mutation.
The testing process will terminate until the given time budget is reached.
Indeed, it would be better for \tool{} to perform deeper analysis to improve the possibility of generating bug-triggering tests.
For example, it is more likely for the mutation operator of replacing the parameter value with 0 to generate a bug-triggering test by analyzing whether there is division operation inside the API.
In the future, we will incorporate more analysis to further improve \tool{}.
Please note that \tool{} generates tests based on the existing unit tests, while unit tests do not distinguish different stages and can test the APIs for various stages (including the training stage).
Therefore, \ddel{\tool{} can test various stages that the existing unit tests can test.}\aadd{All the stages tested by the existing unit tests, can be also tested by \tool{}.}
}
\del{
According to the findings, we design a prototype DL-framework testing tool, called \tool{}, and we have conducted a preliminary study to demonstrate the validity of our findings and the significance of our implications for guiding future research. Specifically, \tool{} is a mutation-based test generation tool. It defines a set of mutation operators targeting those DL-specific bugs based on our findings, e.g., bugs in Graph-Level Implementation and Operation Implementation levels, since they take a significant large portion of all kinds of bugs. Table~\ref{tab:mutator} presents the mutation operators defined in \tool{} and their detailed descriptions. 
When providing a set of unit tests as seeds, \tool{} will try to apply those mutation operators to them (if applicable) and generate new tests that can be used for detecting potential bugs in DL frameworks. 
}
\del{We have conducted our study on the latest TensorFlow framework, i.e., v2.8.0, which is widely used in both academic and industrial practice. The detailed procedure and experimental results are explained as follows.}

\textbf{Procedure:} 
\add{We conducted an experiment on TensorFlow to evaluate the effectiveness of \tool{}.
During the testing process, we applied each generated test by \tool{} to test four TensorFlow versions, i.e., v2.5.0, v2.6.0, v2.7.0, and v2.8.0, respectively.
That is, \tool{} adopts cross-version differential testing to determine whether the test detects a bug.
Initially, we collected the tests from the \textit{python} folder in TensorFlow as the \ddel{seeds}\aadd{initial test pool} for \tool{}.
In particular, some of these initial \ddel{seeds}\aadd{tests} can produce inconsistent results on the four versions due to the version incompatibility, which can incur noise to the testing process with \tool{}.
\aadd{Specifically, for each initial test, we ran it on the four versions of TensorFlow and recorded the results (i.e., the values of the variables observed by the assertions in the test), respectively.
If the results from the four versions are different or some versions crash on the test, we regarded that the test triggers an inconsistency.}
Hence, we discarded them from the initial \ddel{seed}\aadd{test} pool.
In total, there are 509 tests in the initial \ddel{seed}\aadd{test} pool.
Here, we set the time budget as 24 hours for running \tool{}.
When a test makes some of these versions under test crash or produces different results on these versions, we regard it as a potential bug and report it to the developers for manual investigation.
\aadd{In particular, when the results from the four versions are different, we determined the buggy version through the voting mechanism, which assumes that most of versions can produce correct results for the same test.}
}
\del{In the experiment, we employed differential testing over different versions of the TensorFlow framework to decide the correctness of testing outputs.
Specifically, we employed another three early released versions, i.e., v2.5.0, v2.6.0 and v2.7.0, as the references. 
Initially, we collected a set of test cases from the \textit{python} folder in TensorFlow framework as the \textit{seed pool} feeding to \tool{} for subsequent mutation, where we filtered out test cases that produced inconsistent results over different versions. 
Next, \tool{} would automatically generate new test cases by randomly selecting one test case each time from the \textit{seed pool} and applying applicable mutation operators, and then run these new test cases over different versions of TensorFlow. If inconsistent outputs or crashes were produced, \tool{} would report it to developers for manual inspection. Besides, the newly generated test cases would be added to the \textit{seed pool} for facilitating subsequent testing. This testing process was proceeded iteratively until a termination condition was reached. In our experiment, we set the maximal running time as 24 hours.}

\textbf{Results:} In total, \tool{} reported 9 tests that triggered potential bugs (producing inconsistent results on different versions). 
After manual inspection for identifying duplicates, \tool{} detected 6 unique bugs, \del{in} \add{of} which 3 bugs exist in the early releases and have been fixed in the latest version, while the other 3 bugs still exist in the latest version. 
\aadd{Specifically, 4 bugs were triggered on all the four versions, one bug was triggered on v2.5.0, and one bug was triggered on v2.5.0 and v2.6.0.}
We have reported them to the TensorFlow developers on GitHub, all of which have been successfully reproduced and confirmed by the developers.
Figure~\ref{fig:bug_exam2} shows a test generated by \tool{}, which triggers a bug in the latest version of TensorFlow. 
Specifically, \tool{} \del{applied the \textit{mutate\_type} operator to}
\add{mutates the type of} the variable \texttt{grads}, which represents the step size during gradient calculation with the Adadelta optimizer (i.e., \texttt{adadelta\_opt} in the example). 
By replacing the type \texttt{float32} with \texttt{float16} as shown in Line~\ref{mutate2}, the TensorFlow crashed \add{with the message of ``Aborted (core dumped)''} when it ran into Line~\ref{bug_point2} \add{due to the exception-handling issue for the type incompatibility.} \del{ due to type incompatibility and the incomplete type checking in the framework.} 
This bug was confirmed once we submitted it to the TensorFlow developers. 
\add{\ddel{In fact, it is as expected that \tool{} can effectively detect exception-handling bugs, since the mutation operators designed in \tool{} can usually produce the exceptional cases of tensor type incompatibility or tensor shape misalignment, which is helpful to more sufficiently test the ability of TensorFlow for handling these exceptional cases.}\aadd{As expected before, \tool{} can effectively detect exception-handling bugs due to the designed mutation operators.}
In the future, we will improve \tool{} to detect bugs with more diverse root causes by designing more mutation operators or incorporating advanced static analysis to avoid the cases of tensor type incompatibility and tensor shape misalignment during test generation.}
We have also released all the detected bugs in our project homepage. 
In summary, the preliminary experimental results demonstrate the effectiveness of \tool{} and \del{that our implications are valuable for future research} \add{the value of our implications}. 
In addition, the results also indicate that more effective bug detection techniques are desired.

\aadd{In particular, the idea of \tool{} is general since (1) its designed mutation operators are applied to tensors or API parameters, which also exist in other DL frameworks; (2) it treats the equipped unit tests as the initial pool of tests, and indeed other DL frameworks also contain equipped unit tests.
However, the current \tool{} tool cannot be directly applied to other DL frameworks.
This is because the equipped tests in different DL frameworks have different data structures and different formats, causing that the mutation operators implemented in \tool{} (specific to the equipped unit tests in TensorFlow) cannot be used to mutate the equipped unit tests in other DL frameworks.
Therefore, before applying \tool{} to other DL frameworks, we need to modify the implementation of the mutation operators to make them applicable to the equipped unit tests in other DL frameworks according to the corresponding data structures and formats.
Furthermore, we have reported the detected bugs by \tool{} to the TensorFlow developers on GitHub, all of which have been successfully reproduced and confirmed by the developers.
This indicates that these bugs were not detected by the existing testing tools before to some degree.
Indeed, the exception-handling bugs detected by \tool{} cannot be detected by the existing testing tools studied in Section~\ref{sec:coverage} (i.e., CRADLE, LEMON, and Audee), since they cannot produce exceptional cases of tensor type incompatibility or tensor shape misalignment.
In the future, we will try more testing tools to better understand the effectiveness difference among various testing tools.}

\begin{figure}
    \centering
    \begin{lstlisting}[style=Python,numbers=left]
  num_updates = 4
  for grad in [0.2,0.1,0.01]:
    for lr in [1.0,0.5,0.1]:
      var0 = variables.Variable([1.0,2.0], dtype=dtypes.float32)
      var1 = variables.Variable([3.0,4.0], dtype=dtypes.float32)
      grads = constant_op.constant([grad,grad],<@\fbox{{\scriptsize {\ttfamily dtype=dtypes.float16}}}@>) <@\label{mutate2} \textcolor{gray}{\# mutate ``float32'' to ``float16''}@>
      adadelta_opt = adadelta.AdadeltaOptimizer(learning_rate=lr,rho=0.95,epsilon=1e-08)
      if (not context.executing_eagerly()):
        adadelta_update = adadelta_opt.apply_gradients(zip([grads,grads], [var0,var1]))
        slot = ([None]*2)
        slot_update = ([None]*2)
      for step in range(num_updates):
         adadelta_opt.apply_gradients(zip([grads,grads], [var0,var1])) <@ {\label{bug_point2}} \textcolor{gray}{\# Running crashed here} @>
    \end{lstlisting}
    \caption{An example test case generated by \tool{} that triggers a bug}
    \label{fig:bug_exam2}
\end{figure}

\section{Threats to Validity}
\label{sec:dis}


The \textit{external threats to validity} mainly lie in our used data.
We systematically collected 1000 bugs of four DL frameworks as our study data, including collecting closed and merged pull requests, identifying bug-fixing pull requests via keyword searching, and conducting manual investigation following the existing work~\cite{dlcompiler,zhang2018empirical,islam2019comprehensive}.
Hence, a big confidence can be obtained regarding the high quality of our data.
As the most large-scale study on DL framework bugs, the generalizability of our study can be demonstrated to a large extent.
\add{Please note that same as many existing studies~\cite{dlcompiler,islam2019comprehensive,zhang2018empirical}, we cannot consider the importance of each studied bug, since the information about bug importance is not provided in the corresponding GitHub repositories.
In the future, we may study the DL framework bugs with the importance information to reduce this threat.}


The \textit{internal threats to validity} mainly lie in our manual labeling process. 
To mitigate the inaccuracy and subjectivity of each individual developer, two authors with over 4-year developing experience conducted the labeling process independently \add{through the general open-coding scheme~\cite{khandkar2009open} following many of the existing studies~\cite{islam2019comprehensive,zhang2018empirical,dlcompiler}}.\del{, and w}
We leveraged the Cohen's Kappa coefficient to measure the inter-rater agreement between them, where a coefficient as high as 95\% is reached, indicating a high agreement between them. 
\add{Besides, when coming across the inconsistencies between two authors in each labeling study, the two authors discussed with \textit{the third author (a senior developer)} until the bugs were labeled consistently, which can help further improve the reliability of the labels.
Indeed, during the training session, we also involved the third person for the discussion of inconsistencies, which can help reduce the risk that the two authors agree on the same (but wrong) thing.}

\add{The \textit{construct threats} to validity mainly lie in the selection of our studied DL framework testing techniques in Section~\ref{sec:coverage}.
Specifically, the findings (Findings \ddel{12} \aadd{11} and \ddel{13} \aadd{12}) obtained based on the three studied techniques may not represent the other DL framework testing techniques.
In the future, we will evaluate more techniques to further reduce this threat.}


\section{Related Work}
\label{sec:related_work}

The most related work to ours is the empirical study on TensorFlow bugs~\cite{jia2020empirical,JIA2021110935}.
This study is the only one on investigating DL framework bugs, but it is not enough to comprehensively understand bugs in the family of DL frameworks:
1) It investigates the bugs in only one DL framework (i.e., TensorFlow), while our study analyzed 1000 bugs of four popular and diverse DL frameworks.
That shows that our study is indeed large-scale and general (e.g., obtaining more general root-cause and symptom taxonomies), and facilitates the understanding of bugs across different DL frameworks.
2) It directly uses the folders organizing TensorFlow code as the component categories, which cannot be generalized to other DL frameworks.
However, our work proposes a general top-down five-level architecture for DL frameworks, and \del{analyzed} \add{analyzes} root cause distribution on each component, which facilitates the more fine-grained understanding of DL framework bugs.
3) Our study involves more study points, including studying bugs from some individual aspects (e.g., root causes) as well as associating different aspects for comprehensive analysis (e.g., associating root causes with DL framework components).
In particular, our study further associates our identified bug characteristics with existing testing and debugging practice for DL framework bugs, in order to dissect the current status in testing and debugging DL frameworks and then guide the direction of improving them.
Therefore, we believe that our study makes significantly novel contributions to understanding DL framework bugs comprehensively and further ensuring DL frameworks' quality by providing insightful guidelines for repairing DL related bugs~\cite{/publisher/Science China Press/journal/SCIENCE CHINA Information Sciences///10.1007/s11432-022-3580-5} . 

There are also some studies on investigating DL program bugs~\cite{zhang2018empirical,islam2019comprehensive,humbatova2020taxonomy,xie2019deephunter,ma2018deepmutation,tian2018deeptest}.
As explained in Sections~\ref{sec:intro} and~\ref{sec:background}, \del{DL program bugs are actually the incorrect usage of DL frameworks rather than the bugs in DL framework code.
The latter is the target of our work.}
\add{DL programs refer to the programs used for training DL models, which are implemented by invoking the APIs provided by DL frameworks, while DL frameworks implement the functionalities of those APIs.
Therefore, DL program bugs can be caused by incorrectly using the APIs provided by DL frameworks when implementing the DL programs, rather than the bugs inside the DL frameworks.
The latter is the target of our work.}
\add{Hence, in our study, there is no bug caused due to using incorrect versions of a DL framework.
However, since we analyzed 250 bugs for each DL framework, these bugs can involve different versions of the DL framework.}
In particular, the distribution and characteristics of DL framework bugs and DL program bugs are largely different. For example, there are many DL framework bugs being triggered during the installation process. On the contrary,
there is no such kind of bugs in DL programs. Similarly, the bug category of \textit{Incorrect Model Parameter} in DL programs~\cite{zhang2018empirical} does not appear in DL frameworks either. 
Furthermore, Garcia et al.~\cite{garcia2020comprehensive} studied the bugs of autonomous vehicles, which is a kind of DL-based applications and lies in the production level.
Shen et al.~\cite{dlcompiler} conducted an empirical study on DL compilers (e.g., TVM). 
Nejadgholi and Yang~\cite{Nejadgholi2019study} studied the oracle approximation assertions implemented in DL libraries.
Different from them, our work conducted a comprehensive study on DL framework bugs by investigating 1000 bugs from four DL frameworks.

Besides the studies on investigating DL (program and framework) bugs, there are also many studies focusing on traditional software bugs in the literature~\cite{lu2008learning,di2017comprehensive,DBLP:conf/esem/HanY16,ocariza2013empirical,wang2020empirical,MLBugs,hoang2019deepjit,li2021large}.
For example, Ocariza et al.~\cite{ocariza2013empirical} conducted a study on client-side JavaScript bugs.
Lu et al.~\cite{lu2008learning} investigated the characteristics of concurrency bugs, while Li et al.~\cite{li2021large} conducted a large-scale study on API misuses in traditional Java programs.
Different from them, our work targets DL framework bugs, which not only investigates the bug characteristics specific to DL frameworks, but also analyze the difference for the common bug characteristics (such as some common root causes) between DL frameworks and traditional software.
\add{In particular, 
a DL framework has the following typical characteristics, leading to four unique root causes compared with traditional software bugs.
First, it is the fundamental infrastructure of DL, which supports the construction and usage of DL models.
Therefore, many of DL framework bugs involve tensor types and tensor shapes (almost all the DL operations depend on tensors).
Second, it is the bridge between DL functionalities and various hardware, which implements some strategies to support DL functionalities on different hardware.
Therefore, many of DL framework bugs involve the environment incompatibility.
Third, DL is still a fast-growing area and thus DL frameworks have to be frequently updated to incorporate the rapid advancement in DL algorithms.
Therefore, many of DL framework bugs occur in the implementations of DL-specific algorithms.}

\add{Recently, many testing techniques for DL frameworks have been proposed, which can be mainly divided into two categories based on the data format during the generation of test cases. They are graph-level~\cite{wang2020deep,pham2019cradle,guo2020audee,shenDLFMutation,10.1145/3510003.3510092, 10.1145/3510003.3510165} and operator-level (API-level) test generation~\cite{duo,zhang2021predoo, 10.1145/3510003.3510041} techniques. 
In the first category, Pham et al.\cite{pham2019cradle} proposed CRADLE to detect DL framework bugs via differential testing. After that, LEMON~\cite{wang2020deep} and Audee~\cite{guo2020audee} were proposed and both of them generated tests via a set of mutation rules, but their mutation targets were different. Similarly,
the latest work EAGLE~\cite{10.1145/3510003.3510165} defined a set of equivalence transformation rules to facilitate this testing process.
Besides, Muffin~\cite{10.1145/3510003.3510092} aims at detecting DL framework bugs related to model training. It used a DAG-based algorithm to generate diverse DL models and measured the inconsistencies in training phase with multiple metrics.
For the second category,  Predoo~\cite{zhang2021predoo}, FreeFuzz~\cite{10.1145/3510003.3510041}, and DocTer\cite{xie2022docter} are three representative techniques. The major difference is the data used for test generation.
Predoo mutates the input tensor values in the original program, FreeFuzz leverages code snippets from different sources to aid the generation, while DocTer depends on the API constraints extracted from corresponding documents.
In this paper, we conducted a preliminary study over the three typical DL framework testing techniques (i.e., CRADLE, LEMON, and AUDEE), based on which we provide some guidance for future research and propose a simple DL framework testing technique. 
In the future, we plan to perform a more systematic review and comparison over all the existing techniques, and further improve the performance of DL framework testing techniques. 
}

\section{Conclusion}
\label{sec:conclusion}

In this work, we conducted the most large-scale empirical study on the characteristics (e.g., root causes, symptoms, and their correlations with DL-framework components) of DL framework bugs, where we manually analyzed 1,000 bugs from four popular DL frameworks. Through the comprehensive study and further analyzing the current status of existing DL framework testing techniques, we summarized \ddel{13} \aadd{12} major findings, based on which we provided a series of actionable implications for future studies on the detection and debugging of DL framework bugs. 
Finally, on the basis of those implications, we have developed a prototype DL-framework testing tool, called \tool{}, which was evaluated to be effective to find unknown DL framework bugs in a preliminary study, indicating the significance of those implications to guide feature research.

\begin{acks}
We thank the anonymous reviewers for their constructive suggestions to help improve the quality of this paper. This work was supported by the National Natural Science Foundation of China under Grant Nos. 62002256, 62232001, and 62202324.
\end{acks}

\bibliographystyle{ACM-Reference-Format}
\bibliography{dllibrary}


\begin{thebibliography}{81}


\ifx \showCODEN    \undefined \def \showCODEN     #1{\unskip}     \fi
\ifx \showDOI      \undefined \def \showDOI       #1{#1}\fi
\ifx \showISBNx    \undefined \def \showISBNx     #1{\unskip}     \fi
\ifx \showISBNxiii \undefined \def \showISBNxiii  #1{\unskip}     \fi
\ifx \showISSN     \undefined \def \showISSN      #1{\unskip}     \fi
\ifx \showLCCN     \undefined \def \showLCCN      #1{\unskip}     \fi
\ifx \shownote     \undefined \def \shownote      #1{#1}          \fi
\ifx \showarticletitle \undefined \def \showarticletitle #1{#1}   \fi
\ifx \showURL      \undefined \def \showURL       {\relax}        \fi
\providecommand\bibfield[2]{#2}
\providecommand\bibinfo[2]{#2}
\providecommand\natexlab[1]{#1}
\providecommand\showeprint[2][]{arXiv:#2}

\bibitem[cov(2021)]%
        {coverage_python}
 \bibinfo{year}{Accessed: 2021}\natexlab{}.
\newblock \bibinfo{title}{Coverage.py}.
\newblock
\newblock
\newblock
\shownote{\url{https://coverage.readthedocs.io/}}.


\bibitem[dl4(2021)]%
        {dl4j}
 \bibinfo{year}{Accessed: 2021}\natexlab{}.
\newblock \bibinfo{title}{Deeplearning4J}.
\newblock
\newblock
\newblock
\shownote{\url{https://deeplearning4j.org/}}.


\bibitem[gco(2021)]%
        {gcov_tool}
 \bibinfo{year}{Accessed: 2021}\natexlab{}.
\newblock \bibinfo{title}{Gcov}.
\newblock
\newblock
\newblock
\shownote{\url{https://gcc.gnu.org/onlinedocs/gcc/Gcov.html}}.


\bibitem[gra(2021)]%
        {gradle.org}
 \bibinfo{year}{Accessed: 2021}\natexlab{}.
\newblock \bibinfo{title}{Gradle}.
\newblock
\newblock
\newblock
\shownote{\url{https://gradle.org/}}.


\bibitem[mxn(2021)]%
        {mxnet}
 \bibinfo{year}{Accessed: 2021}\natexlab{}.
\newblock \bibinfo{title}{MXNet}.
\newblock
\newblock
\newblock
\shownote{\url{https://mxnet.apache.org}}.


\bibitem[new(2021a)]%
        {news1}
 \bibinfo{year}{Accessed: 2021}\natexlab{a}.
\newblock \bibinfo{title}{News}.
\newblock
\newblock
\newblock
\shownote{\url{https://www.vice.com/en_us/article/9kga85/uber-is-giving-up-on-self-driving-cars-in-california-after-deadly-crash}}.


\bibitem[new(2021b)]%
        {news2}
 \bibinfo{year}{Accessed: 2021}\natexlab{b}.
\newblock \bibinfo{title}{News}.
\newblock
\newblock
\newblock
\shownote{\url{https://www.newsweek.com/autonomous-tesla-crashes-parked-fire-truck-california-freeway-789177}}.


\bibitem[pyt(2021)]%
        {pytorch}
 \bibinfo{year}{Accessed: 2021}\natexlab{}.
\newblock \bibinfo{title}{PyTorch}.
\newblock
\newblock
\newblock
\shownote{\url{https://pytorch.org}}.


\bibitem[ten(2021)]%
        {tensorflow}
 \bibinfo{year}{Accessed: 2021}\natexlab{}.
\newblock \bibinfo{title}{TensorFlow}.
\newblock
\newblock
\newblock
\shownote{\url{https://www.tensorflow.org}}.


\bibitem[baz(2022)]%
        {bazel}
 \bibinfo{year}{Accessed: 2022}\natexlab{}.
\newblock \bibinfo{title}{Bazel}.
\newblock
\newblock
\newblock
\shownote{\url{https://bazel.build/}}.


\bibitem[caf(2022)]%
        {caffe}
 \bibinfo{year}{Accessed: 2022}\natexlab{}.
\newblock \bibinfo{title}{Caffe}.
\newblock
\newblock
\newblock
\shownote{\url{https://github.com/intel/caffe}}.


\bibitem[ker(2022)]%
        {keras}
 \bibinfo{year}{Accessed: 2022}\natexlab{}.
\newblock \bibinfo{title}{Keras}.
\newblock
\newblock
\newblock
\shownote{\url{https://github.com/keras-team/keras}}.


\bibitem[Abadi et~al\mbox{.}(2016)]%
        {DBLP:journals/corr/AbadiABBCCCDDDG16}
\bibfield{author}{\bibinfo{person}{Mart{\'{\i}}n Abadi},
  \bibinfo{person}{Ashish Agarwal}, \bibinfo{person}{Paul Barham},
  \bibinfo{person}{Eugene Brevdo}, \bibinfo{person}{Zhifeng Chen},
  \bibinfo{person}{Craig Citro}, \bibinfo{person}{Gregory~S. Corrado},
  \bibinfo{person}{Andy Davis}, \bibinfo{person}{Jeffrey Dean},
  \bibinfo{person}{Matthieu Devin}, \bibinfo{person}{Sanjay Ghemawat},
  \bibinfo{person}{Ian~J. Goodfellow}, \bibinfo{person}{Andrew Harp},
  \bibinfo{person}{Geoffrey Irving}, \bibinfo{person}{Michael Isard},
  \bibinfo{person}{Yangqing Jia}, \bibinfo{person}{Rafal J{\'{o}}zefowicz},
  \bibinfo{person}{Lukasz Kaiser}, \bibinfo{person}{Manjunath Kudlur},
  \bibinfo{person}{Josh Levenberg}, \bibinfo{person}{Dan Man{\'{e}}},
  \bibinfo{person}{Rajat Monga}, \bibinfo{person}{Sherry Moore},
  \bibinfo{person}{Derek~Gordon Murray}, \bibinfo{person}{Chris Olah},
  \bibinfo{person}{Mike Schuster}, \bibinfo{person}{Jonathon Shlens},
  \bibinfo{person}{Benoit Steiner}, \bibinfo{person}{Ilya Sutskever},
  \bibinfo{person}{Kunal Talwar}, \bibinfo{person}{Paul~A. Tucker},
  \bibinfo{person}{Vincent Vanhoucke}, \bibinfo{person}{Vijay Vasudevan},
  \bibinfo{person}{Fernanda~B. Vi{\'{e}}gas}, \bibinfo{person}{Oriol Vinyals},
  \bibinfo{person}{Pete Warden}, \bibinfo{person}{Martin Wattenberg},
  \bibinfo{person}{Martin Wicke}, \bibinfo{person}{Yuan Yu}, {and}
  \bibinfo{person}{Xiaoqiang Zheng}.} \bibinfo{year}{2016}\natexlab{}.
\newblock \showarticletitle{TensorFlow: Large-Scale Machine Learning on
  Heterogeneous Distributed Systems}.
\newblock \bibinfo{journal}{\emph{CoRR}}  \bibinfo{volume}{abs/1603.04467}
  (\bibinfo{year}{2016}).
\newblock
\showeprint[arXiv]{1603.04467}
\urldef\tempurl%
\url{http://arxiv.org/abs/1603.04467}
\showURL{%
\tempurl}


\bibitem[Amann et~al\mbox{.}(2016)]%
        {amann2016mubench}
\bibfield{author}{\bibinfo{person}{Sven Amann}, \bibinfo{person}{Sarah Nadi},
  \bibinfo{person}{Hoan~A Nguyen}, \bibinfo{person}{Tien~N Nguyen}, {and}
  \bibinfo{person}{Mira Mezini}.} \bibinfo{year}{2016}\natexlab{}.
\newblock \showarticletitle{MUBench: A benchmark for API-misuse detectors}. In
  \bibinfo{booktitle}{\emph{Proceedings of the 13th International Conference on
  Mining Software Repositories}}. \bibinfo{pages}{464--467}.
\newblock


\bibitem[Amann et~al\mbox{.}(2018)]%
        {amann2018systematic_api}
\bibfield{author}{\bibinfo{person}{Sven Amann}, \bibinfo{person}{Hoan~Anh
  Nguyen}, \bibinfo{person}{Sarah Nadi}, \bibinfo{person}{Tien~N Nguyen}, {and}
  \bibinfo{person}{Mira Mezini}.} \bibinfo{year}{2018}\natexlab{}.
\newblock \showarticletitle{A systematic evaluation of static api-misuse
  detectors}.
\newblock \bibinfo{journal}{\emph{IEEE Transactions on Software Engineering}}
  \bibinfo{volume}{45}, \bibinfo{number}{12} (\bibinfo{year}{2018}),
  \bibinfo{pages}{1170--1188}.
\newblock


\bibitem[Chen et~al\mbox{.}(2015)]%
        {chen2015deepdriving}
\bibfield{author}{\bibinfo{person}{Chenyi Chen}, \bibinfo{person}{Ari Seff},
  \bibinfo{person}{Alain Kornhauser}, {and} \bibinfo{person}{Jianxiong Xiao}.}
  \bibinfo{year}{2015}\natexlab{}.
\newblock \showarticletitle{Deepdriving: Learning affordance for direct
  perception in autonomous driving}. In \bibinfo{booktitle}{\emph{Proceedings
  of the IEEE International Conference on Computer Vision}}.
  \bibinfo{pages}{2722--2730}.
\newblock


\bibitem[Chen et~al\mbox{.}(2014)]%
        {10.1145/2627508.2627515}
\bibfield{author}{\bibinfo{person}{Fangwei Chen}, \bibinfo{person}{Lei Li},
  \bibinfo{person}{Jing Jiang}, {and} \bibinfo{person}{Li Zhang}.}
  \bibinfo{year}{2014}\natexlab{}.
\newblock \showarticletitle{Predicting the Number of Forks for Open Source
  Software Project}. In \bibinfo{booktitle}{\emph{Proceedings of the 2014 3rd
  International Workshop on Evidential Assessment of Software Technologies}}
  (Nanjing, China) \emph{(\bibinfo{series}{EAST 2014})}.
  \bibinfo{publisher}{Association for Computing Machinery},
  \bibinfo{address}{New York, NY, USA}, \bibinfo{pages}{40–47}.
\newblock
\showISBNx{9781450329651}
\urldef\tempurl%
\url{https://doi.org/10.1145/2627508.2627515}
\showDOI{\tempurl}


\bibitem[Chen et~al\mbox{.}(2020)]%
        {chen2020practical}
\bibfield{author}{\bibinfo{person}{Junjie Chen}, \bibinfo{person}{Zhuo Wu},
  \bibinfo{person}{Zan Wang}, \bibinfo{person}{Hanmo You},
  \bibinfo{person}{Lingming Zhang}, {and} \bibinfo{person}{Ming Yan}.}
  \bibinfo{year}{2020}\natexlab{}.
\newblock \showarticletitle{Practical accuracy estimation for efficient deep
  neural network testing}.
\newblock \bibinfo{journal}{\emph{ACM Transactions on Software Engineering and
  Methodology}} \bibinfo{volume}{29}, \bibinfo{number}{4}
  (\bibinfo{year}{2020}), \bibinfo{pages}{1--35}.
\newblock


\bibitem[Di~Franco et~al\mbox{.}(2017)]%
        {di2017comprehensive}
\bibfield{author}{\bibinfo{person}{Anthony Di~Franco}, \bibinfo{person}{Hui
  Guo}, {and} \bibinfo{person}{Cindy Rubio-Gonz{\'a}lez}.}
  \bibinfo{year}{2017}\natexlab{}.
\newblock \showarticletitle{A comprehensive study of real-world numerical bug
  characteristics}. In \bibinfo{booktitle}{\emph{Proceedings of 32nd IEEE/ACM
  International Conference on Automated Software Engineering}}.
  \bibinfo{pages}{509--519}.
\newblock


\bibitem[Du et~al\mbox{.}(2021)]%
        {9113719}
\bibfield{author}{\bibinfo{person}{Mengnan Du}, \bibinfo{person}{Fan Yang},
  \bibinfo{person}{Na Zou}, {and} \bibinfo{person}{Xia Hu}.}
  \bibinfo{year}{2021}\natexlab{}.
\newblock \showarticletitle{Fairness in Deep Learning: A Computational
  Perspective}.
\newblock \bibinfo{journal}{\emph{IEEE Intelligent Systems}}
  \bibinfo{volume}{36}, \bibinfo{number}{4} (\bibinfo{year}{2021}),
  \bibinfo{pages}{25--34}.
\newblock
\urldef\tempurl%
\url{https://doi.org/10.1109/MIS.2020.3000681}
\showDOI{\tempurl}


\bibitem[Ferreira et~al\mbox{.}(2019)]%
        {ferreira2019software}
\bibfield{author}{\bibinfo{person}{Fabio Ferreira}, \bibinfo{person}{Luciana
  Lourdes~Silva}, {and} \bibinfo{person}{Marco Tulio~Valente}.}
  \bibinfo{year}{2019}\natexlab{}.
\newblock \showarticletitle{Software engineering meets deep learning: A
  literature review}.
\newblock \bibinfo{journal}{\emph{arXiv e-prints}} (\bibinfo{year}{2019}),
  \bibinfo{pages}{arXiv--1909}.
\newblock


\bibitem[Garcia et~al\mbox{.}(2020)]%
        {garcia2020comprehensive}
\bibfield{author}{\bibinfo{person}{Joshua Garcia}, \bibinfo{person}{Yang Feng},
  \bibinfo{person}{Junjie Shen}, \bibinfo{person}{Sumaya Almanee},
  \bibinfo{person}{Yuan Xia}, {and} \bibinfo{person}{Qi~Alfred Chen}.}
  \bibinfo{year}{2020}\natexlab{}.
\newblock \showarticletitle{A comprehensive study of autonomous vehicle bugs}.
  In \bibinfo{booktitle}{\emph{Proceedings of the ACM/IEEE 42nd International
  Conference on Software Engineering}}. \bibinfo{pages}{385--396}.
\newblock


\bibitem[Goodfellow et~al\mbox{.}(2015)]%
        {DBLP:journals/corr/GoodfellowSS14}
\bibfield{author}{\bibinfo{person}{Ian~J. Goodfellow},
  \bibinfo{person}{Jonathon Shlens}, {and} \bibinfo{person}{Christian
  Szegedy}.} \bibinfo{year}{2015}\natexlab{}.
\newblock \showarticletitle{Explaining and Harnessing Adversarial Examples}. In
  \bibinfo{booktitle}{\emph{3rd International Conference on Learning
  Representations}}.
\newblock


\bibitem[Gu et~al\mbox{.}(2022)]%
        {10.1145/3510003.3510092}
\bibfield{author}{\bibinfo{person}{Jiazhen Gu}, \bibinfo{person}{Xuchuan Luo},
  \bibinfo{person}{Yangfan Zhou}, {and} \bibinfo{person}{Xin Wang}.}
  \bibinfo{year}{2022}\natexlab{}.
\newblock \showarticletitle{Muffin: Testing Deep Learning Libraries via Neural
  Architecture Fuzzing}. In \bibinfo{booktitle}{\emph{Proceedings of the 44th
  International Conference on Software Engineering}}
  \emph{(\bibinfo{series}{ICSE '22})}. \bibinfo{pages}{1418–1430}.
\newblock
\showISBNx{9781450392211}


\bibitem[Guo et~al\mbox{.}(2020)]%
        {guo2020audee}
\bibfield{author}{\bibinfo{person}{Qianyu Guo}, \bibinfo{person}{Xiaofei Xie},
  \bibinfo{person}{Yi Li}, \bibinfo{person}{Xiaoyu Zhang},
  \bibinfo{person}{Yang Liu}, \bibinfo{person}{Xiaohong Li}, {and}
  \bibinfo{person}{Chao Shen}.} \bibinfo{year}{2020}\natexlab{}.
\newblock \showarticletitle{Audee: Automated testing for deep learning
  frameworks}. In \bibinfo{booktitle}{\emph{2020 35th IEEE/ACM International
  Conference on Automated Software Engineering}}. \bibinfo{pages}{486--498}.
\newblock


\bibitem[Han et~al\mbox{.}(2020)]%
        {han2020empirical}
\bibfield{author}{\bibinfo{person}{Junxiao Han}, \bibinfo{person}{Shuiguang
  Deng}, \bibinfo{person}{David Lo}, \bibinfo{person}{Chen Zhi},
  \bibinfo{person}{Jianwei Yin}, {and} \bibinfo{person}{Xin Xia}.}
  \bibinfo{year}{2020}\natexlab{}.
\newblock \showarticletitle{An empirical study of the dependency networks of
  deep learning libraries}. In \bibinfo{booktitle}{\emph{2020 IEEE
  International Conference on Software Maintenance and Evolution (ICSME)}}.
  IEEE, \bibinfo{pages}{868--878}.
\newblock


\bibitem[Han and Yu(2016)]%
        {DBLP:conf/esem/HanY16}
\bibfield{author}{\bibinfo{person}{Xue Han} {and} \bibinfo{person}{Tingting
  Yu}.} \bibinfo{year}{2016}\natexlab{}.
\newblock \showarticletitle{An Empirical Study on Performance Bugs for Highly
  Configurable Software Systems}. In \bibinfo{booktitle}{\emph{Proceedings of
  the 10th ACM/IEEE International Symposium on Empirical Software Engineering
  and Measurement}}. \bibinfo{pages}{23:1--23:10}.
\newblock


\bibitem[Hapke and Nelson(2020)]%
        {hapke2020building}
\bibfield{author}{\bibinfo{person}{Hannes Hapke} {and}
  \bibinfo{person}{Catherine Nelson}.} \bibinfo{year}{2020}\natexlab{}.
\newblock \bibinfo{booktitle}{\emph{Building Machine Learning Pipelines}}.
\newblock \bibinfo{publisher}{O'Reilly Media}.
\newblock


\bibitem[Hassan and Wang(2018)]%
        {hassan2018hirebuild}
\bibfield{author}{\bibinfo{person}{Foyzul Hassan} {and}
  \bibinfo{person}{Xiaoyin Wang}.} \bibinfo{year}{2018}\natexlab{}.
\newblock \showarticletitle{Hirebuild: An automatic approach to history-driven
  repair of build scripts}. In \bibinfo{booktitle}{\emph{2018 IEEE/ACM 40th
  International Conference on Software Engineering (ICSE)}}. IEEE,
  \bibinfo{pages}{1078--1089}.
\newblock


\bibitem[Hoang et~al\mbox{.}(2019)]%
        {hoang2019deepjit}
\bibfield{author}{\bibinfo{person}{Thong Hoang}, \bibinfo{person}{Hoa~Khanh
  Dam}, \bibinfo{person}{Yasutaka Kamei}, \bibinfo{person}{David Lo}, {and}
  \bibinfo{person}{Naoyasu Ubayashi}.} \bibinfo{year}{2019}\natexlab{}.
\newblock \showarticletitle{DeepJIT: an end-to-end deep learning framework for
  just-in-time defect prediction}. In \bibinfo{booktitle}{\emph{2019 IEEE/ACM
  16th International Conference on Mining Software Repositories (MSR)}}. IEEE,
  \bibinfo{pages}{34--45}.
\newblock


\bibitem[Humbatova et~al\mbox{.}(2020)]%
        {humbatova2020taxonomy}
\bibfield{author}{\bibinfo{person}{Nargiz Humbatova}, \bibinfo{person}{Gunel
  Jahangirova}, \bibinfo{person}{Gabriele Bavota}, \bibinfo{person}{Vincenzo
  Riccio}, \bibinfo{person}{Andrea Stocco}, {and} \bibinfo{person}{Paolo
  Tonella}.} \bibinfo{year}{2020}\natexlab{}.
\newblock \showarticletitle{Taxonomy of real faults in deep learning systems}.
  In \bibinfo{booktitle}{\emph{Proceedings of the ACM/IEEE 42nd International
  Conference on Software Engineering}}. \bibinfo{pages}{1110--1121}.
\newblock


\bibitem[Islam et~al\mbox{.}(2019)]%
        {islam2019comprehensive}
\bibfield{author}{\bibinfo{person}{Md~Johirul Islam}, \bibinfo{person}{Giang
  Nguyen}, \bibinfo{person}{Rangeet Pan}, {and} \bibinfo{person}{Hridesh
  Rajan}.} \bibinfo{year}{2019}\natexlab{}.
\newblock \showarticletitle{A comprehensive study on deep learning bug
  characteristics}. In \bibinfo{booktitle}{\emph{Proceedings of the 2019 27th
  ACM Joint Meeting on European Software Engineering Conference and Symposium
  on the Foundations of Software Engineering}}. \bibinfo{pages}{510--520}.
\newblock


\bibitem[Jia et~al\mbox{.}(2020)]%
        {jia2020empirical}
\bibfield{author}{\bibinfo{person}{Li Jia}, \bibinfo{person}{Hao Zhong},
  \bibinfo{person}{Xiaoyin Wang}, \bibinfo{person}{Linpeng Huang}, {and}
  \bibinfo{person}{Xuansheng Lu}.} \bibinfo{year}{2020}\natexlab{}.
\newblock \showarticletitle{An Empirical Study on Bugs Inside TensorFlow}. In
  \bibinfo{booktitle}{\emph{International Conference on Database Systems for
  Advanced Applications}}. \bibinfo{pages}{604--620}.
\newblock


\bibitem[Jia et~al\mbox{.}(2021)]%
        {JIA2021110935}
\bibfield{author}{\bibinfo{person}{Li Jia}, \bibinfo{person}{Hao Zhong},
  \bibinfo{person}{Xiaoyin Wang}, \bibinfo{person}{Linpeng Huang}, {and}
  \bibinfo{person}{Xuansheng Lu}.} \bibinfo{year}{2021}\natexlab{}.
\newblock \showarticletitle{The symptoms, causes, and repairs of bugs inside a
  deep learning library}.
\newblock \bibinfo{journal}{\emph{Journal of Systems and Software}}
  \bibinfo{volume}{177} (\bibinfo{year}{2021}), \bibinfo{pages}{110935}.
\newblock
\showISSN{0164-1212}
\urldef\tempurl%
\url{https://doi.org/10.1016/j.jss.2021.110935}
\showDOI{\tempurl}


\bibitem[Julian et~al\mbox{.}(2016)]%
        {julian2016policy}
\bibfield{author}{\bibinfo{person}{Kyle~D Julian}, \bibinfo{person}{Jessica
  Lopez}, \bibinfo{person}{Jeffrey~S Brush}, \bibinfo{person}{Michael~P Owen},
  {and} \bibinfo{person}{Mykel~J Kochenderfer}.}
  \bibinfo{year}{2016}\natexlab{}.
\newblock \showarticletitle{Policy compression for aircraft collision avoidance
  systems}. In \bibinfo{booktitle}{\emph{2016 IEEE/AIAA 35th Digital Avionics
  Systems Conference}}. \bibinfo{pages}{1--10}.
\newblock


\bibitem[Kang et~al\mbox{.}(2021)]%
        {kang2021apirecx}
\bibfield{author}{\bibinfo{person}{Yuning Kang}, \bibinfo{person}{Zan Wang},
  \bibinfo{person}{Hongyu Zhang}, \bibinfo{person}{Junjie Chen}, {and}
  \bibinfo{person}{Hanmo You}.} \bibinfo{year}{2021}\natexlab{}.
\newblock \showarticletitle{Apirecx: Cross-library api recommendation via
  pre-trained language model}. In \bibinfo{booktitle}{\emph{Proceedings of the
  2021 Conference on Empirical Methods in Natural Language Processing}}.
  \bibinfo{pages}{3425--3436}.
\newblock


\bibitem[Khandkar(2009)]%
        {khandkar2009open}
\bibfield{author}{\bibinfo{person}{Shahedul~Huq Khandkar}.}
  \bibinfo{year}{2009}\natexlab{}.
\newblock \showarticletitle{Open coding}.
\newblock \bibinfo{journal}{\emph{University of Calgary}}  \bibinfo{volume}{23}
  (\bibinfo{year}{2009}), \bibinfo{pages}{2009}.
\newblock


\bibitem[Kim et~al\mbox{.}(2019)]%
        {DBLP:conf/icse/SADL}
\bibfield{author}{\bibinfo{person}{Jinhan Kim}, \bibinfo{person}{Robert Feldt},
  {and} \bibinfo{person}{Shin Yoo}.} \bibinfo{year}{2019}\natexlab{}.
\newblock \showarticletitle{Guiding deep learning system testing using surprise
  adequacy}. In \bibinfo{booktitle}{\emph{Proceedings of the 41st International
  Conference on Software Engineering}}. \bibinfo{pages}{1039--1049}.
\newblock


\bibitem[Kurakin et~al\mbox{.}(2017)]%
        {DBLP:conf/iclr/KurakinGB17a}
\bibfield{author}{\bibinfo{person}{Alexey Kurakin}, \bibinfo{person}{Ian~J.
  Goodfellow}, {and} \bibinfo{person}{Samy Bengio}.}
  \bibinfo{year}{2017}\natexlab{}.
\newblock \showarticletitle{Adversarial examples in the physical world}. In
  \bibinfo{booktitle}{\emph{5th International Conference on Learning
  Representations}}.
\newblock


\bibitem[Li et~al\mbox{.}(2021)]%
        {li2021large}
\bibfield{author}{\bibinfo{person}{Xia Li}, \bibinfo{person}{Jiajun Jiang},
  \bibinfo{person}{Samuel Benton}, \bibinfo{person}{Yingfei Xiong}, {and}
  \bibinfo{person}{Lingming Zhang}.} \bibinfo{year}{2021}\natexlab{}.
\newblock \showarticletitle{A Large-scale Study on API Misuses in the Wild}. In
  \bibinfo{booktitle}{\emph{2021 14th IEEE Conference on Software Testing,
  Verification and Validation (ICST)}}. \bibinfo{pages}{241--252}.
\newblock
\urldef\tempurl%
\url{https://doi.org/10.1109/ICST49551.2021.00034}
\showDOI{\tempurl}


\bibitem[Lou et~al\mbox{.}(2019)]%
        {lou2019history}
\bibfield{author}{\bibinfo{person}{Yiling Lou}, \bibinfo{person}{Junjie Chen},
  \bibinfo{person}{Lingming Zhang}, \bibinfo{person}{Dan Hao}, {and}
  \bibinfo{person}{Lu Zhang}.} \bibinfo{year}{2019}\natexlab{}.
\newblock \showarticletitle{History-driven build failure fixing: how far are
  we?}. In \bibinfo{booktitle}{\emph{Proceedings of the 28th ACM SIGSOFT
  International Symposium on Software Testing and Analysis}}.
  \bibinfo{pages}{43--54}.
\newblock


\bibitem[Lu et~al\mbox{.}(2008)]%
        {lu2008learning}
\bibfield{author}{\bibinfo{person}{Shan Lu}, \bibinfo{person}{Soyeon Park},
  \bibinfo{person}{Eunsoo Seo}, {and} \bibinfo{person}{Yuanyuan Zhou}.}
  \bibinfo{year}{2008}\natexlab{}.
\newblock \showarticletitle{Learning from mistakes: a comprehensive study on
  real world concurrency bug characteristics}. In
  \bibinfo{booktitle}{\emph{Proceedings of the 13th international conference on
  Architectural support for programming languages and operating systems}}.
  \bibinfo{pages}{329--339}.
\newblock


\bibitem[Lune and Berg(2017)]%
        {lune2017qualitative}
\bibfield{author}{\bibinfo{person}{Howard Lune} {and} \bibinfo{person}{Bruce~L
  Berg}.} \bibinfo{year}{2017}\natexlab{}.
\newblock \bibinfo{booktitle}{\emph{Qualitative research methods for the social
  sciences}}.
\newblock \bibinfo{publisher}{Pearson}.
\newblock


\bibitem[Ma et~al\mbox{.}(2018a)]%
        {DBLP:conf/kbse/deepgauge}
\bibfield{author}{\bibinfo{person}{Lei Ma}, \bibinfo{person}{Felix
  Juefei{-}Xu}, \bibinfo{person}{Fuyuan Zhang}, \bibinfo{person}{Jiyuan Sun},
  \bibinfo{person}{Minhui Xue}, \bibinfo{person}{Bo Li},
  \bibinfo{person}{Chunyang Chen}, \bibinfo{person}{Ting Su},
  \bibinfo{person}{Li Li}, \bibinfo{person}{Yang Liu}, \bibinfo{person}{Jianjun
  Zhao}, {and} \bibinfo{person}{Yadong Wang}.}
  \bibinfo{year}{2018}\natexlab{a}.
\newblock \showarticletitle{DeepGauge: multi-granularity testing criteria for
  deep learning systems}. In \bibinfo{booktitle}{\emph{Proceedings of the 33rd
  {ACM/IEEE} International Conference on Automated Software Engineering}}.
  \bibinfo{pages}{120--131}.
\newblock


\bibitem[Ma et~al\mbox{.}(2018b)]%
        {ma2018deepmutation}
\bibfield{author}{\bibinfo{person}{Lei Ma}, \bibinfo{person}{Fuyuan Zhang},
  \bibinfo{person}{Jiyuan Sun}, \bibinfo{person}{Minhui Xue},
  \bibinfo{person}{Bo Li}, \bibinfo{person}{Felix Juefei-Xu},
  \bibinfo{person}{Chao Xie}, \bibinfo{person}{Li Li}, \bibinfo{person}{Yang
  Liu}, \bibinfo{person}{Jianjun Zhao}, {et~al\mbox{.}}}
  \bibinfo{year}{2018}\natexlab{b}.
\newblock \showarticletitle{Deepmutation: Mutation testing of deep learning
  systems}. In \bibinfo{booktitle}{\emph{2018 IEEE 29th International Symposium
  on Software Reliability Engineering (ISSRE)}}. IEEE,
  \bibinfo{pages}{100--111}.
\newblock


\bibitem[Ma et~al\mbox{.}(2018c)]%
        {DBLP:journals/corr/abs-1806-07723}
\bibfield{author}{\bibinfo{person}{Lei Ma}, \bibinfo{person}{Fuyuan Zhang},
  \bibinfo{person}{Minhui Xue}, \bibinfo{person}{Bo Li}, \bibinfo{person}{Yang
  Liu}, \bibinfo{person}{Jianjun Zhao}, {and} \bibinfo{person}{Yadong Wang}.}
  \bibinfo{year}{2018}\natexlab{c}.
\newblock \showarticletitle{Combinatorial Testing for Deep Learning Systems}.
\newblock \bibinfo{journal}{\emph{CoRR}}  \bibinfo{volume}{abs/1806.07723}
  (\bibinfo{year}{2018}).
\newblock
\showeprint[arXiv]{1806.07723}


\bibitem[Nejadgholi and Yang(2019)]%
        {Nejadgholi2019study}
\bibfield{author}{\bibinfo{person}{Mahdi Nejadgholi} {and}
  \bibinfo{person}{Jinqiu Yang}.} \bibinfo{year}{2019}\natexlab{}.
\newblock \showarticletitle{A Study of Oracle Approximations in Testing Deep
  Learning Libraries}. In \bibinfo{booktitle}{\emph{2019 34th IEEE/ACM
  International Conference on Automated Software Engineering (ASE)}}.
  \bibinfo{pages}{785--796}.
\newblock
\urldef\tempurl%
\url{https://doi.org/10.1109/ASE.2019.00078}
\showDOI{\tempurl}


\bibitem[Ocariza et~al\mbox{.}(2013)]%
        {ocariza2013empirical}
\bibfield{author}{\bibinfo{person}{Frolin Ocariza}, \bibinfo{person}{Kartik
  Bajaj}, \bibinfo{person}{Karthik Pattabiraman}, {and} \bibinfo{person}{Ali
  Mesbah}.} \bibinfo{year}{2013}\natexlab{}.
\newblock \showarticletitle{An empirical study of client-side JavaScript bugs}.
  In \bibinfo{booktitle}{\emph{2013 ACM/IEEE International Symposium on
  Empirical Software Engineering and Measurement}}. \bibinfo{pages}{55--64}.
\newblock


\bibitem[Perez et~al\mbox{.}(2017)]%
        {perez2017prevalence}
\bibfield{author}{\bibinfo{person}{Alexandre Perez}, \bibinfo{person}{Rui
  Abreu}, {and} \bibinfo{person}{Marcelo D'Amorim}.}
  \bibinfo{year}{2017}\natexlab{}.
\newblock \showarticletitle{Prevalence of Single-Fault Fixes and Its Impact on
  Fault Localization}. In \bibinfo{booktitle}{\emph{2017 IEEE International
  Conference on Software Testing, Verification and Validation (ICST)}}.
  \bibinfo{pages}{12--22}.
\newblock


\bibitem[Pham et~al\mbox{.}(2019)]%
        {pham2019cradle}
\bibfield{author}{\bibinfo{person}{Hung~Viet Pham}, \bibinfo{person}{Thibaud
  Lutellier}, \bibinfo{person}{Weizhen Qi}, {and} \bibinfo{person}{Lin Tan}.}
  \bibinfo{year}{2019}\natexlab{}.
\newblock \showarticletitle{CRADLE: cross-backend validation to detect and
  localize bugs in deep learning libraries}. In \bibinfo{booktitle}{\emph{2019
  IEEE/ACM 41st International Conference on Software Engineering}}.
  \bibinfo{pages}{1027--1038}.
\newblock


\bibitem[Shen et~al\mbox{.}(2022)]%
        {shen2022natural}
\bibfield{author}{\bibinfo{person}{Qingchao Shen}, \bibinfo{person}{Junjie
  Chen}, \bibinfo{person}{Jie~M Zhang}, \bibinfo{person}{Haoyu Wang},
  \bibinfo{person}{Shuang Liu}, {and} \bibinfo{person}{Menghan Tian}.}
  \bibinfo{year}{2022}\natexlab{}.
\newblock \showarticletitle{Natural Test Generation for Precise Testing of
  Question Answering Software}. In \bibinfo{booktitle}{\emph{37th IEEE/ACM
  International Conference on Automated Software Engineering}}.
  \bibinfo{pages}{1--12}.
\newblock


\bibitem[Shen et~al\mbox{.}(2021a)]%
        {dlcompiler}
\bibfield{author}{\bibinfo{person}{Qingchao Shen}, \bibinfo{person}{Haoyang
  Ma}, \bibinfo{person}{Junjie Chen}, \bibinfo{person}{Yongqiang Tian},
  \bibinfo{person}{Shing-Chi Cheung}, {and} \bibinfo{person}{Xiang Chen}.}
  \bibinfo{year}{2021}\natexlab{a}.
\newblock \showarticletitle{A Comprehensive Study of Deep Learning Compiler
  Bugs}. In \bibinfo{booktitle}{\emph{Proceedings of the 29th ACM Joint
  European Software Engineering Conference and Symposium on the Foundations of
  Software Engineering}}.
\newblock
\newblock
\shownote{to appear}.


\bibitem[Shen et~al\mbox{.}(2021b)]%
        {shenDLFMutation}
\bibfield{author}{\bibinfo{person}{Xiangzhong Shen}, \bibinfo{person}{Jieyi
  Zhang}, \bibinfo{person}{Xiaonan Wang}, \bibinfo{person}{Hongfang Yu}, {and}
  \bibinfo{person}{Gang Sun}.} \bibinfo{year}{2021}\natexlab{b}.
\newblock \showarticletitle{Deep Learning Framework Fuzzing Based on Model
  Mutation}. In \bibinfo{booktitle}{\emph{2021 IEEE Sixth International
  Conference on Data Science in Cyberspace (DSC)}}. \bibinfo{pages}{375--380}.
\newblock
\urldef\tempurl%
\url{https://doi.org/10.1109/DSC53577.2021.00059}
\showDOI{\tempurl}


\bibitem[Sun et~al\mbox{.}(2016)]%
        {sun2016toward}
\bibfield{author}{\bibinfo{person}{Chengnian Sun}, \bibinfo{person}{Vu Le},
  \bibinfo{person}{Qirun Zhang}, {and} \bibinfo{person}{Zhendong Su}.}
  \bibinfo{year}{2016}\natexlab{}.
\newblock \showarticletitle{Toward understanding compiler bugs in GCC and
  LLVM}. In \bibinfo{booktitle}{\emph{Proceedings of the 25th International
  Symposium on Software Testing and Analysis}}. \bibinfo{pages}{294--305}.
\newblock


\bibitem[Tan et~al\mbox{.}(2014)]%
        {tan2014bug}
\bibfield{author}{\bibinfo{person}{Lin Tan}, \bibinfo{person}{Chen Liu},
  \bibinfo{person}{Zhenmin Li}, \bibinfo{person}{Xuanhui Wang},
  \bibinfo{person}{Yuanyuan Zhou}, {and} \bibinfo{person}{Chengxiang Zhai}.}
  \bibinfo{year}{2014}\natexlab{}.
\newblock \showarticletitle{Bug characteristics in open source software}.
\newblock \bibinfo{journal}{\emph{Empirical software engineering}}
  \bibinfo{volume}{19}, \bibinfo{number}{6} (\bibinfo{year}{2014}),
  \bibinfo{pages}{1665--1705}.
\newblock


\bibitem[Thung et~al\mbox{.}(2012)]%
        {MLBugs}
\bibfield{author}{\bibinfo{person}{Ferdian Thung}, \bibinfo{person}{Shaowei
  Wang}, \bibinfo{person}{David Lo}, {and} \bibinfo{person}{Lingxiao Jiang}.}
  \bibinfo{year}{2012}\natexlab{}.
\newblock \showarticletitle{An empirical study of bugs in machine learning
  systems}. In \bibinfo{booktitle}{\emph{Proceedings of 23rd International
  Symposium on Software Reliability Engineering}}. \bibinfo{pages}{271--280}.
\newblock


\bibitem[Tian et~al\mbox{.}(2018)]%
        {tian2018deeptest}
\bibfield{author}{\bibinfo{person}{Yuchi Tian}, \bibinfo{person}{Kexin Pei},
  \bibinfo{person}{Suman Jana}, {and} \bibinfo{person}{Baishakhi Ray}.}
  \bibinfo{year}{2018}\natexlab{}.
\newblock \showarticletitle{Deeptest: Automated testing of
  deep-neural-network-driven autonomous cars}. In
  \bibinfo{booktitle}{\emph{Proceedings of the 40th international conference on
  software engineering}}. \bibinfo{pages}{303--314}.
\newblock


\bibitem[Tian et~al\mbox{.}(2022)]%
        {tian2022learning}
\bibfield{author}{\bibinfo{person}{Zhao Tian}, \bibinfo{person}{Junjie Chen},
  \bibinfo{person}{Qihao Zhu}, \bibinfo{person}{Junjie Yang}, {and}
  \bibinfo{person}{Lingming Zhang}.} \bibinfo{year}{2022}\natexlab{}.
\newblock \showarticletitle{Learning to Construct Better Mutation Faults}. In
  \bibinfo{booktitle}{\emph{37th IEEE/ACM International Conference on Automated
  Software Engineering}}. \bibinfo{pages}{1--13}.
\newblock


\bibitem[Vieira et~al\mbox{.}(2010)]%
        {vieira2010cohen}
\bibfield{author}{\bibinfo{person}{Susana~M Vieira}, \bibinfo{person}{Uzay
  Kaymak}, {and} \bibinfo{person}{Jo{\~a}o~MC Sousa}.}
  \bibinfo{year}{2010}\natexlab{}.
\newblock \showarticletitle{Cohen's kappa coefficient as a performance measure
  for feature selection}. In \bibinfo{booktitle}{\emph{Proceedings of
  International Conference on Fuzzy Systems}}. \bibinfo{pages}{1--8}.
\newblock


\bibitem[Voas(1992)]%
        {pie}
\bibfield{author}{\bibinfo{person}{J.M. Voas}.}
  \bibinfo{year}{1992}\natexlab{}.
\newblock \showarticletitle{PIE: a dynamic failure-based technique}.
\newblock \bibinfo{journal}{\emph{IEEE Transactions on Software Engineering}}
  \bibinfo{volume}{18}, \bibinfo{number}{8} (\bibinfo{year}{1992}),
  \bibinfo{pages}{717--727}.
\newblock
\urldef\tempurl%
\url{https://doi.org/10.1109/32.153381}
\showDOI{\tempurl}


\bibitem[Wang et~al\mbox{.}(2022b)]%
        {wang2022empirical}
\bibfield{author}{\bibinfo{person}{Gan Wang}, \bibinfo{person}{Zan Wang},
  \bibinfo{person}{Junjie Chen}, \bibinfo{person}{Xiang Chen}, {and}
  \bibinfo{person}{Ming Yan}.} \bibinfo{year}{2022}\natexlab{b}.
\newblock \showarticletitle{An Empirical Study on Numerical Bugs in Deep
  Learning Programs}. In \bibinfo{booktitle}{\emph{37th IEEE/ACM International
  Conference on Automated Software Engineering}}. \bibinfo{pages}{1--5}.
\newblock


\bibitem[Wang et~al\mbox{.}(2022a)]%
        {10.1145/3510003.3510165}
\bibfield{author}{\bibinfo{person}{Jiannan Wang}, \bibinfo{person}{Thibaud
  Lutellier}, \bibinfo{person}{Shangshu Qian}, \bibinfo{person}{Hung~Viet
  Pham}, {and} \bibinfo{person}{Lin Tan}.} \bibinfo{year}{2022}\natexlab{a}.
\newblock \showarticletitle{EAGLE: Creating Equivalent Graphs to Test Deep
  Learning Libraries}. In \bibinfo{booktitle}{\emph{Proceedings of the 44th
  International Conference on Software Engineering}}
  \emph{(\bibinfo{series}{ICSE '22})}. \bibinfo{publisher}{Association for
  Computing Machinery}, \bibinfo{address}{New York, NY, USA},
  \bibinfo{pages}{798–810}.
\newblock
\showISBNx{9781450392211}
\urldef\tempurl%
\url{https://doi.org/10.1145/3510003.3510165}
\showDOI{\tempurl}


\bibitem[Wang et~al\mbox{.}(2020a)]%
        {wang2020empirical}
\bibfield{author}{\bibinfo{person}{Peipei Wang}, \bibinfo{person}{Chris Brown},
  \bibinfo{person}{Jamie~A Jennings}, {and} \bibinfo{person}{Kathryn~T
  Stolee}.} \bibinfo{year}{2020}\natexlab{a}.
\newblock \showarticletitle{An empirical study on regular expression bugs}. In
  \bibinfo{booktitle}{\emph{Proceedings of the 17th International Conference on
  Mining Software Repositories}}. \bibinfo{pages}{103--113}.
\newblock


\bibitem[Wang et~al\mbox{.}(2020b)]%
        {wang2020deep}
\bibfield{author}{\bibinfo{person}{Zan Wang}, \bibinfo{person}{Ming Yan},
  \bibinfo{person}{Junjie Chen}, \bibinfo{person}{Shuang Liu}, {and}
  \bibinfo{person}{Dongdi Zhang}.} \bibinfo{year}{2020}\natexlab{b}.
\newblock \showarticletitle{Deep learning library testing via effective model
  generation}. In \bibinfo{booktitle}{\emph{Proceedings of the 28th ACM Joint
  Meeting on European Software Engineering Conference and Symposium on the
  Foundations of Software Engineering}}. \bibinfo{pages}{788--799}.
\newblock


\bibitem[Wang et~al\mbox{.}(2021)]%
        {wang2021prioritizing}
\bibfield{author}{\bibinfo{person}{Zan Wang}, \bibinfo{person}{Hanmo You},
  \bibinfo{person}{Junjie Chen}, \bibinfo{person}{Yingyi Zhang},
  \bibinfo{person}{Xuyuan Dong}, {and} \bibinfo{person}{Wenbin Zhang}.}
  \bibinfo{year}{2021}\natexlab{}.
\newblock \showarticletitle{Prioritizing test inputs for deep neural networks
  via mutation analysis}. In \bibinfo{booktitle}{\emph{2021 IEEE/ACM 43rd
  International Conference on Software Engineering}}. IEEE,
  \bibinfo{pages}{397--409}.
\newblock


\bibitem[Wardat et~al\mbox{.}(2021)]%
        {wardat2021deeplocalize}
\bibfield{author}{\bibinfo{person}{Mohammad Wardat}, \bibinfo{person}{Wei Le},
  {and} \bibinfo{person}{Hridesh Rajan}.} \bibinfo{year}{2021}\natexlab{}.
\newblock \showarticletitle{DeepLocalize: Fault Localization for Deep Neural
  Networks}. In \bibinfo{booktitle}{\emph{2021 IEEE/ACM 43rd International
  Conference on Software Engineering}}. \bibinfo{pages}{251--262}.
\newblock


\bibitem[Wei et~al\mbox{.}(2022)]%
        {10.1145/3510003.3510041}
\bibfield{author}{\bibinfo{person}{Anjiang Wei}, \bibinfo{person}{Yinlin Deng},
  \bibinfo{person}{Chenyuan Yang}, {and} \bibinfo{person}{Lingming Zhang}.}
  \bibinfo{year}{2022}\natexlab{}.
\newblock \showarticletitle{Free Lunch for Testing: Fuzzing Deep-Learning
  Libraries from Open Source}. In \bibinfo{booktitle}{\emph{Proceedings of the
  44th International Conference on Software Engineering}} (Pittsburgh,
  Pennsylvania) \emph{(\bibinfo{series}{ICSE '22})}.
  \bibinfo{publisher}{Association for Computing Machinery},
  \bibinfo{address}{New York, NY, USA}, \bibinfo{pages}{995–1007}.
\newblock
\showISBNx{9781450392211}
\urldef\tempurl%
\url{https://doi.org/10.1145/3510003.3510041}
\showDOI{\tempurl}


\bibitem[Xie et~al\mbox{.}(2022)]%
        {xie2022docter}
\bibfield{author}{\bibinfo{person}{Danning Xie}, \bibinfo{person}{Yitong Li},
  \bibinfo{person}{Mijung Kim}, \bibinfo{person}{Hung~Viet Pham},
  \bibinfo{person}{Lin Tan}, \bibinfo{person}{Xiangyu Zhang}, {and}
  \bibinfo{person}{Michael~W Godfrey}.} \bibinfo{year}{2022}\natexlab{}.
\newblock \showarticletitle{Docter: Documentation-guided fuzzing for testing
  deep learning api functions}. In \bibinfo{booktitle}{\emph{Proceedings of the
  31st ACM SIGSOFT International Symposium on Software Testing and Analysis}}.
  \bibinfo{pages}{176--188}.
\newblock


\bibitem[Xie et~al\mbox{.}(2019)]%
        {xie2019deephunter}
\bibfield{author}{\bibinfo{person}{Xiaofei Xie}, \bibinfo{person}{Lei Ma},
  \bibinfo{person}{Felix Juefei-Xu}, \bibinfo{person}{Minhui Xue},
  \bibinfo{person}{Hongxu Chen}, \bibinfo{person}{Yang Liu},
  \bibinfo{person}{Jianjun Zhao}, \bibinfo{person}{Bo Li},
  \bibinfo{person}{Jianxiong Yin}, {and} \bibinfo{person}{Simon See}.}
  \bibinfo{year}{2019}\natexlab{}.
\newblock \showarticletitle{Deephunter: a coverage-guided fuzz testing
  framework for deep neural networks}. In \bibinfo{booktitle}{\emph{Proceedings
  of the 28th ACM SIGSOFT International Symposium on Software Testing and
  Analysis}}. \bibinfo{pages}{146--157}.
\newblock


\bibitem[Yan et~al\mbox{.}(2021)]%
        {yan2021exposing}
\bibfield{author}{\bibinfo{person}{Ming Yan}, \bibinfo{person}{Junjie Chen},
  \bibinfo{person}{Xiangyu Zhang}, \bibinfo{person}{Lin Tan},
  \bibinfo{person}{Gan Wang}, {and} \bibinfo{person}{Zan Wang}.}
  \bibinfo{year}{2021}\natexlab{}.
\newblock \showarticletitle{Exposing numerical bugs in deep learning via
  gradient back-propagation}. In \bibinfo{booktitle}{\emph{Proceedings of the
  29th ACM Joint Meeting on European Software Engineering Conference and
  Symposium on the Foundations of Software Engineering}}.
  \bibinfo{pages}{627--638}.
\newblock


\bibitem[Yang et~al\mbox{.}(2021)]%
        {yang2021semi}
\bibfield{author}{\bibinfo{person}{Lin Yang}, \bibinfo{person}{Junjie Chen},
  \bibinfo{person}{Zan Wang}, \bibinfo{person}{Weijing Wang},
  \bibinfo{person}{Jiajun Jiang}, \bibinfo{person}{Xuyuan Dong}, {and}
  \bibinfo{person}{Wenbin Zhang}.} \bibinfo{year}{2021}\natexlab{}.
\newblock \showarticletitle{Semi-supervised log-based anomaly detection via
  probabilistic label estimation}. In \bibinfo{booktitle}{\emph{2021 IEEE/ACM
  43rd International Conference on Software Engineering}}. IEEE,
  \bibinfo{pages}{1448--1460}.
\newblock


\bibitem[You et~al\mbox{.}(2023)]%
        {you2023DRFuzz}
\bibfield{author}{\bibinfo{person}{Hanmo You}, \bibinfo{person}{Zan Wang},
  \bibinfo{person}{Junjie Chen}, \bibinfo{person}{Shuang Liu}, {and}
  \bibinfo{person}{Shuochuan Li}.} \bibinfo{year}{2023}\natexlab{}.
\newblock \showarticletitle{Regression Fuzzing for Deep Learning Systems}. In
  \bibinfo{booktitle}{\emph{45th International Conference on Software
  Engineering}}.
\newblock
\newblock
\shownote{to appear}.


\bibitem[Zar(2005)]%
        {inbook}
\bibfield{author}{\bibinfo{person}{Jerrold Zar}.}
  \bibinfo{year}{2005}\natexlab{}.
\newblock \bibinfo{booktitle}{\emph{Spearman Rank Correlation}}.
  Vol.~\bibinfo{volume}{5}.
\newblock
\showISBNx{9780470011812}
\urldef\tempurl%
\url{https://doi.org/10.1002/0470011815.b2a15150}
\showDOI{\tempurl}


\bibitem[Zeller and Hildebrandt(2002)]%
        {zeller2002simplifying}
\bibfield{author}{\bibinfo{person}{Andreas Zeller} {and} \bibinfo{person}{Ralf
  Hildebrandt}.} \bibinfo{year}{2002}\natexlab{}.
\newblock \showarticletitle{Simplifying and isolating failure-inducing input}.
\newblock \bibinfo{journal}{\emph{IEEE Transactions on Software Engineering}}
  \bibinfo{volume}{28}, \bibinfo{number}{2} (\bibinfo{year}{2002}),
  \bibinfo{pages}{183--200}.
\newblock


\bibitem[Zhang et~al\mbox{.}(2018b)]%
        {zhang2018code_api}
\bibfield{author}{\bibinfo{person}{Tianyi Zhang}, \bibinfo{person}{Ganesha
  Upadhyaya}, \bibinfo{person}{Anastasia Reinhardt}, \bibinfo{person}{Hridesh
  Rajan}, {and} \bibinfo{person}{Miryung Kim}.}
  \bibinfo{year}{2018}\natexlab{b}.
\newblock \showarticletitle{Are code examples on an online Q\&A forum
  reliable?: a study of API misuse on stack overflow}. In
  \bibinfo{booktitle}{\emph{Proceedings of 40th IEEE/ACM International
  Conference on Software Engineering}}. \bibinfo{pages}{886--896}.
\newblock


\bibitem[Zhang et~al\mbox{.}(2021a)]%
        {duo}
\bibfield{author}{\bibinfo{person}{Xufan Zhang}, \bibinfo{person}{Jiawei Liu},
  \bibinfo{person}{Ning Sun}, \bibinfo{person}{Chunrong Fang},
  \bibinfo{person}{Jia Liu}, \bibinfo{person}{Jiang Wang},
  \bibinfo{person}{Dong Chai}, {and} \bibinfo{person}{Zhenyu Chen}.}
  \bibinfo{year}{2021}\natexlab{a}.
\newblock \showarticletitle{Duo: Differential Fuzzing for Deep Learning
  Operators}.
\newblock \bibinfo{journal}{\emph{IEEE Transactions on Reliability}}
  \bibinfo{volume}{70}, \bibinfo{number}{4} (\bibinfo{year}{2021}),
  \bibinfo{pages}{1671--1685}.
\newblock
\urldef\tempurl%
\url{https://doi.org/10.1109/TR.2021.3107165}
\showDOI{\tempurl}


\bibitem[Zhang et~al\mbox{.}(2021b)]%
        {zhang2021predoo}
\bibfield{author}{\bibinfo{person}{Xufan Zhang}, \bibinfo{person}{Ning Sun},
  \bibinfo{person}{Chunrong Fang}, \bibinfo{person}{Jiawei Liu},
  \bibinfo{person}{Jia Liu}, \bibinfo{person}{Dong Chai},
  \bibinfo{person}{Jiang Wang}, {and} \bibinfo{person}{Zhenyu Chen}.}
  \bibinfo{year}{2021}\natexlab{b}.
\newblock \showarticletitle{Predoo: precision testing of deep learning
  operators}. In \bibinfo{booktitle}{\emph{Proceedings of the 30th ACM SIGSOFT
  International Symposium on Software Testing and Analysis}}.
  \bibinfo{pages}{400--412}.
\newblock


\bibitem[Zhang et~al\mbox{.}(2021c)]%
        {DBLP:conf/icse/ZhangZMS21}
\bibfield{author}{\bibinfo{person}{Xiaoyu Zhang}, \bibinfo{person}{Juan Zhai},
  \bibinfo{person}{Shiqing Ma}, {and} \bibinfo{person}{Chao Shen}.}
  \bibinfo{year}{2021}\natexlab{c}.
\newblock \showarticletitle{{AUTOTRAINER:} An Automatic {DNN} Training Problem
  Detection and Repair System}. In \bibinfo{booktitle}{\emph{43rd {IEEE/ACM}
  International Conference on Software Engineering}}.
  \bibinfo{pages}{359--371}.
\newblock


\bibitem[Zhang et~al\mbox{.}(2018a)]%
        {zhang2018empirical}
\bibfield{author}{\bibinfo{person}{Yuhao Zhang}, \bibinfo{person}{Yifan Chen},
  \bibinfo{person}{Shing-Chi Cheung}, \bibinfo{person}{Yingfei Xiong}, {and}
  \bibinfo{person}{Lu Zhang}.} \bibinfo{year}{2018}\natexlab{a}.
\newblock \showarticletitle{An empirical study on TensorFlow program bugs}. In
  \bibinfo{booktitle}{\emph{Proceedings of the 27th ACM SIGSOFT International
  Symposium on Software Testing and Analysis}}. \bibinfo{pages}{129--140}.
\newblock


\bibitem[Zhang et~al\mbox{.}(2022)]%
        {DBLP:conf/kbse/ZhangWJYC22}
\bibfield{author}{\bibinfo{person}{Yingyi Zhang}, \bibinfo{person}{Zan Wang},
  \bibinfo{person}{Jiajun Jiang}, \bibinfo{person}{Hanmo You}, {and}
  \bibinfo{person}{Junjie Chen}.} \bibinfo{year}{2022}\natexlab{}.
\newblock \showarticletitle{Toward Improving the Robustness of Deep Learning
  Models via Model Transformation}. In \bibinfo{booktitle}{\emph{37th
  {IEEE/ACM} International Conference on Automated Software Engineering}}.
  \bibinfo{publisher}{{ACM}}, \bibinfo{pages}{104:1--104:13}.
\newblock


\bibitem[Zhong et~al\mbox{.}(2021)]%
        {zhong2021understanding}
\bibfield{author}{\bibinfo{person}{Ziyuan Zhong}, \bibinfo{person}{Yuchi Tian},
  {and} \bibinfo{person}{Baishakhi Ray}.} \bibinfo{year}{2021}\natexlab{}.
\newblock \showarticletitle{Understanding local robustness of deep neural
  networks under natural variations}. In
  \bibinfo{booktitle}{\emph{International Conference on Fundamental Approaches
  to Software Engineering}}. Springer, Cham, \bibinfo{pages}{313--337}.
\newblock


\end{thebibliography}










\end{document}